\documentclass{pasj01}
\usepackage{url}

\begin{document}
\SetRunningHead{Photo-$z$ WG}{Photo-$z$ for HSC-SSP}

\title{Photometric Redshifts for the Hyper Suprime-Cam Subaru Strategic Program Data Release 1}

\author{
  Masayuki Tanaka\altaffilmark{1},
  Jean Coupon\altaffilmark{2},
  Bau-Ching Hsieh\altaffilmark{3},
  Sogo Mineo\altaffilmark{1},
  Atsushi J. Nishizawa\altaffilmark{4},
  Joshua Speagle\altaffilmark{5},
  Hisanori Furusawa\altaffilmark{1},
  Satoshi Miyazaki\altaffilmark{1,6},
  Hitoshi Murayama\altaffilmark{7,8,9}
}
\altaffiltext{1}{National Astronomical Observatory of Japan, 2-21-1 Osawa, Mitaka, Tokyo 181-8588, Japan}
\altaffiltext{2}{Department of Astronomy, University of Geneva, ch. d’\'Ecogia 16, 1290 Versoix, Switzerland}
\altaffiltext{3}{Academia Sinica Institute of Astronomy and Astrophysics, P.O. Box 23-141, Taipei 10617, Taiwan}
\altaffiltext{4}{Institute for Advanced Research, Nagoya University Furocho, Chikusa-ku, Nagoya, 464-8602 Japan}
\altaffiltext{5}{Harvard University, 60 Garden St., Cambridge, MA 02138, USA}
\altaffiltext{6}{SOKENDAI(The Graduate University for Advanced Studies), Mitaka,
  Tokyo, 181-8588, Japan}
\altaffiltext{7}{Kavli Institute for the Physics and Mathematics of the Universe (Kavli IPMU, WPI), University of Tokyo, Chiba 277-8582, Japan}
\altaffiltext{8}{Department of Physics and Center for Japanese Studies, University of California, Berkeley, CA 94720, USA}
\altaffiltext{9}{Theoretical Physics Group, Lawrence Berkeley National Laboratory, MS 50A-5104, Berkeley, CA 94720}

\email{masayuki.tanaka@nao.ac.jp}


\maketitle
\newcommand{\commentblue}[1]{\textcolor{blue} {\textbf{#1}}}
\newcommand{\commentred}[1]{\textcolor{red} {\textbf{#1}}}
\newcommand{\bs}[1]{ {\boldsymbol{#1}}}

\begin{abstract}
  Photometric redshifts are a key component of many science objectives in the Hyper Suprime-Cam
  Subaru Strategic Program (HSC-SSP).  In this paper, we describe and compare the codes used to compute
  photometric redshifts for HSC-SSP, how we calibrate them, and the typical accuracy we achieve with 
  the HSC five-band photometry ($grizy$).  We introduce a new point estimator based on an improved loss function
  and demonstrate that it works better than other commonly used estimators.
  We find that our photo-$z$'s are most accurate at
  $0.2\lesssim z_{phot} \lesssim 1.5$, where we can straddle the 4000\AA{} break.  We achieve $\sigma(\Delta z_{phot}/(1+z_{phot}))\sim0.05$
  and an outlier rate of about 15\% for galaxies down to $i=25$ within this redshift range.
  If we limit to a brighter sample of $i<24$, we achieve $\sigma\sim0.04$ and $\sim8\%$ outliers.
  Our photo-$z$'s should thus enable many science cases for HSC-SSP.    We also characterize the accuracy of
  our redshift probability distribution function (PDF) and discover that some codes over/under-estimate
  the redshift uncertainties, which have implications for $N(z)$ reconstruction.
  Our photo-$z$ products for the entire area in the Public Data Release 1
  are publicly available, and
  both our catalog products (such as point estimates) and full PDFs can be retrieved from the data release site,
  \url{https://hsc-release.mtk.nao.ac.jp/}.
\end{abstract}

\section{Introduction}
\label{sec:introduction}

In the era of wide and deep imaging surveys,
the photometric redshift technique (hereafter photo-$z$,
see \cite{hildebrandt10} and references therein) has become
compulsory to uncover the large-scale distance and time
information of millions (soon billions) of galaxies.
While photo-$z$ algorithms and the photometry measurements have
improved significantly over the past two decades
\citep{coupon09,hildebrandt08,hildebrandt12,dahlen13,bonnett16},
the challenge of acquiring photo-$z$ estimates accurate enough to meet
the requirements of cosmology and
galaxy evolution studies continues to motivate the active development
of photometry extraction and photo-$z$ algorithms even today.

It is now clear that both template-fitting and machine-learning
methods are complementary and necessary
to compute meaningful photo-$z$'s. Template-fitting
methods \citep{arnouts1999,bmp2000,feldmann06,brammer2008,kotulla09}
use known galaxy spectral energy distributions (SED) and priors
\citep{benitez2000,ilbert06,tanaka15} to match the observed colors with predicted ones.
Such an approach currently represents the only way to
provide photo-$z$ estimates in regions of color/magnitude space where
no reference redshifts are available (but see also \cite{leistedt16}).
Machine learning methods
\citep{2003LNCS.2859..226T,collister04,lima08,wolf09,Carlilesetal:2010,singal11,brescia16}
are complementary as they provide efficient photo-$z$ estimates,
in terms of speed and precision, but require a training sample that is
a fair representation of the galaxy sample of interest, which is often
difficult to construct due to missing regions in the multi-color space.

Precise photo-$z$'s are needed to enable the selection of sharp,
non-overlapping redshift bins to ``slice'' the Universe.
For example, cosmic shear studies \citep{kilbinger13,hildebrandt17} suffer from
galaxies in adjacent redshift bins that dilute the cosmological signal
and increase the importance of systematic biases
such as the galaxy intrinsic alignments \citep{heymans13}.
For galaxy evolution studies,
it is often important to infer physical properties of galaxies such as
stellar mass in addition to redshifts.
It is thus crucial to minimize catastrophic photo-$z$ errors that lead
to erroneous physical parameters.

The accurate characterization of the true underlying redshift distribution
of a galaxy sample remains a major challenge in today's experiments.
With samples composed of hundreds of millions of galaxies,
systematic biases now largely dominate over statistical errors,
and gathering a complete and numerous calibration sample
has become increasingly pressing in the context of current
and planned large-scale imaging surveys.
Recently, significant progress has been made
in building fainter spectroscopic redshift (hereafter spec-$z$) samples,
e.g., DEEP2 \citep{davis03,cooper11,cooper12,newman13},
VVDS \citep{lefevre04,lefevre05,lefevre13},
VUDS \citep{tasca16}, and
3D-HST \citep{skelton14,momcheva16}.
These are complemented by larger but shallower surveys such as VIPERS \citep{garilli14},
SDSS \citep{alam15}, Wiggle-Z \citep{drinkwater10} and GAMA \citep{liske15}.
More complete but with lower redshift resolution samples are also available from
PRIMUS \citep{coil11,cool13} along with many-band photo-$z$'s from COSMOS \citep{laigle16}.
In parallel, the community
has developed new powerful tools to identify deficiencies in existing spec-$z$ samples
(see e.g. \cite{masters15}) in order to help focus resources on targeting specific
galaxy populations with the adequate instruments.

Still, additional effort is required to (1) improve photo-$z$ algorithms
to fully exploit the information provided by the calibration samples,
(2) gather and homogenize heterogeneous datasets,
and (3) fill in the underrepresented regions of color/magnitude space with reference redshifts
in order to calibrate all of the galaxies observed in the deepest photometric surveys.



These challenges are faced by all on-going and future large-scale photometric surveys,
such as the Kilo-Degree Survey (KiDS, \cite{dejong13}) 
which started in 2011 and whose aim is to map $1\,500$~deg$^2$
in four optical filters ($u, g, r, i$) at relatively shallow depths. 
At a similar depth but over a larger area, the Dark Energy Survey (DES, \cite{flaugher05}) 
is surveying $5\,000$~deg$^2$
in five filters ($g, r, i, z, Y$) in the southern sky since 2013.
In the future, Euclid \citep{laureijs11}, 
a space mission to be launched in 2020,
will observe $15\,000$~deg$^2$ in one optical ("vis")
and three near-infrared ($Y, J, H$) filters, 
complemented by optical multi-wavelength imaging data from the ground.
The Large Synoptic Survey Telescope (LSST, \cite{ivezic08})
will cover $20\,000$~deg$^2$ square degrees in 6 filters
($u, g, r, i, z, y$) over a period of 10 years, starting from 2022, with 
significantly deeper imaging data than the projects described above. 

Here we present the photo-$z$ results from 
the Hyper-Suprime-Cam Subaru Strategic Program (HSC-SSP, \cite{aihara17,aihara17b}),
a 300-night deep imaging survey dedicated to cosmology and
the study of galaxy formation and evolution.
The survey consists of  three components: a Wide layer 
($r\sim26$ at $5\sigma$ for point sources) over $1\,400$~deg$^2$,
a Deep layer ($r\sim27$) over  28~deg$^2$, and an UltraDeep layer ($r\sim28$) over 4~deg$^2$.
This paper presents the efforts led by the photo-$z$ team in HSC-SSP
to develop new photo-$z$ algorithms,
gather a state-of-the art reference redshift sample,
deal with an unprecedented amount of data, and release our products to the public.

The data presented in this study correspond to the Public Data Release 1  (PDR1)
and the S16A internal data release.
In Section~\ref{sec:training_validation_and_test_samples}, we describe the procedures
used to build a robust training sample and to validate our photo-$z$ estimates.
In Section~\ref{sec:method}, we present our photo-$z$ methods.
Section~\ref{sec:metric} defines our adopted performance metrics.
In Sections~\ref{sec:performance} and \ref{sec:pdf}, we characterize our photo-$z$
performance.  We give an overview of our photo-$z$ products included in the public
release in Section \ref{sec:products} and finally conclude in Section~\ref{sec:summary}.
As our previous internal photo-$z$ releases are often used in our science papers,
we briefly summarize our previous data products in Appendix~\ref{sec:previous}.
Unless otherwise stated, all the magnitudes are AB magnitudes.



\section{Training, Validation, and Test Samples}
\label{sec:training_validation_and_test_samples}

The HSC-SSP survey footprint has been designed in order to maximize the overlap with
other photometric and spectroscopic surveys, while keeping the survey geometry simple.
For photo-$z$ purposes, this means we can exploit a large number of public spectroscopic redshifts
in our survey fields and use them to calibrate our photo-$z$'s.
This section describes how we construct the training sample, and how we calibrate, validate
and test our photo-$z$ codes (details of the codes can be found in Section~\ref{sec:method}).

\subsection{Construction of the training sample}
\label{sec:construction_of_the_training_sample}


We first collect spectroscopic redshifts from the literature:
zCOSMOS DR3 \citep{lilly09},
UDSz \citep{bradshaw13,mclure13},
3D-HST \citep{skelton14,momcheva16},
FMOS-COSMOS \citep{silverman15},
VVDS \citep{lefevre13},
VIPERS PDR1 \citep{garilli14},
SDSS DR12 \citep{alam15},
GAMA DR2 \citep{liske15},
WiggleZ DR1 \citep{drinkwater10},
DEEP2 DR4 \citep{davis03,newman13}, and
PRIMUS DR1 \citep{coil11,cool13}.
As each of these surveys have its own flagging scheme to indicate redshift confidence, 
we homogenize them for selection of secure redshift.  The redshifts and flags are
fed to the HSC database and matched with the HSC objects.  This public spec-$z$ table
(described in detail on the spec-$z$ page at the data release site) 
is included in the PDR1 of HSC-SSP and made available to the community.

Our training data include $\sim$170k and 37k high-quality spec-$z$ and g/prism-$z$,
respectively, taken from the matched catalogs described above.  We supplement the
training data with 
$\sim170$k
COSMOS2015 many-band photo-$z$’s \citep{laigle16} along with a collection of private COSMOS spec-$z$’s 
(Mara Salvato, private communication) exclusively used for our photo-$z$ training (they are \emph{not} included in the PDR1).
Data are included in our training set if they meet the following quality cuts:

\textbf{Public spec-z data:}
\begin{enumerate}
	\item $0.01 < z < 9$ (no stars, quasars, or failures)
	\item $\sigma_z < 0.005(1+z)$ (error cut)
	\item SDSS/BOSS: $\textrm{\texttt{zWarning}} = 0$ (no apparent issues)
	\item DEEP2: $\textrm{\texttt{qFlag}} = 4$ ($>99.5\%$ confidence)
	\item PRIMUS: $\textrm{\texttt{qFlag}} = 4$ (very confident)
	\item VIPERS: $\textrm{\texttt{qFlag}} = 3-4$ ($>95\%$ confidence)
	\item VVDS: $\textrm{\texttt{qFlag}} = 3-4$ ($>95\%$ confidence)
	\item GAMA: $\textrm{\texttt{qFlag}} \geq 4$ (very confident)
	\item WiggleZ: $\textrm{\texttt{qFlag}} \geq 4$ (very confident)
	\item UDSz: $\textrm{\texttt{qFlag}} \geq 4$ (provisional catalog only includes $>95\%$ confidence)
	\item FMOS-COSMOS: $\textrm{\texttt{qFlag}} = 3-4$, $z > 0.01$,
	\texttt{flag\_star} is \texttt{False} ($>95\%$ confidence with no stars).
\end{enumerate}

\textbf{3DHST data:}
\begin{enumerate}
    \item \texttt{flag\_star} is \texttt{False} (no stars)
    \item $0 < z < 9$ (no stars, quasars, or redshift failures)
    \item $\max(z_{82}-z_{50},z_{50}-z_{18}) < 0.05(1+z)$ (1$\sigma$ redshift dispersion $<5\%$)
    \item $\max(z_{97.5}-z_{50},z_{50}-z_{2.5}) < 0.1(1+z)$ (2$\sigma$ redshift dispersion $<10\%$)
\end{enumerate}

\textbf{COSMOS data:}
\begin{enumerate}
    \item Spec-$z$:
      \begin{enumerate}
        \item $3 \leq \textrm{\texttt{qFlag}} < 6$ ($>99\%$ confidence)
        \item $0 < z < 7$ (no stars or quasars)
        \item For objects with repeat observations,
        $\sigma_z < 0.005(1+\langle z \rangle)$ (redshifts agree to within 0.5\%)
        \end{enumerate}
    \item Photo-$z$:
      \begin{enumerate}
        \item \texttt{flag\_capak} is \texttt{False} (no bad photometry)
        \item $\textrm{\texttt{type}} = 0$ (only galaxies)
        \item $\chi^2(\textrm{gal}) < \chi^2(\textrm{star})$ and
        $\chi^2(\textrm{gal})/N_{\textrm{bands}} < 5$ (fits are reasonable and better than stellar alternatives)
        \item \textrm{\texttt{$z_{\rm secondary}$}} $< 0$ (no secondary peaks)
        \item $\log M_* > 7.5$ (stellar mass recovery successful)
        \item $0 < z < 9$ (no stars, quasars, nor X-ray detected sources)
        \item $\max(z_{84}-z_{50},z_{50}-z_{16}) < 0.05(1+z)$ (1$\sigma$ redshift dispersion $<5\%$)
      \end{enumerate}
\end{enumerate}

Objects are subsequently matched directly to a set target UltraDeep/Deep catalogs selected using the following criteria:
\begin{enumerate}
    \item \texttt{detect\_is\_primary} is \texttt{True} (no duplicates)
    \item \texttt{[grizy]cmodel\_flux\_flags} is \texttt{False}
    \item \texttt{[grizy]flags\_pixel\_edge} is \texttt{False}
    \item \texttt{[grizy]flags\_pixel\_interpolated\_center} is \texttt{False}
    \item \texttt{[grizy]flags\_pixel\_saturated\_center} is \texttt{False}
    \item \texttt{[grizy]flags\_pixel\_cr\_center} is \texttt{False}
    \item \texttt{[grizy]flags\_pixel\_bad} is \texttt{False}
    \item \texttt{[grizy]centroid\_sdss\_flags} is \texttt{False}
\end{enumerate}
These are designed to maximize completeness while removing
objects with unreliable photometry.
Objects are iteratively matched to this
modified catalog within 1 arcsec at (1) UltraDeep, (2) Deep, and (3) Wide depths
in order to take advantage of higher-S/N data when available while avoiding possible duplicates.

The following quantities are then selected and/or computed:
\begin{enumerate}
\item \textbf{Identifiers}: \texttt{ID}, (\texttt{ra,dec}), and (\texttt{tract,patch}) coordinates.
\item \textbf{Fluxes}: \texttt{PSF} fluxes, \texttt{cmodel} fluxes, \texttt{cmodel\_exp} fluxes,
\texttt{cmodel\_dev} fluxes, and PSF-matched aperture fluxes
with target 1.1 arcsec PSF and 1.5 arcsec apertures taken from the afterburner
run (\texttt{afterburner} fluxes; \cite{aihara17}).
\item \textbf{Shapes}: \texttt{sdss\_shape} parameters.
\item \textbf{Miscellaneous}: \texttt{merge} measurement flags, attenuation
estimates (\texttt{a\_[grizy]}), and \texttt{extendedness} measurements.
\item \textbf{Redshift}: redshift, 1$\sigma$ error, parent survey (SDSS, etc.),
and redshift type (spectroscopic, g/prism, or many-band photometric).
\item \textbf{Depth}: flag for UltraDeep, Deep, and Wide photometry.
\item \textbf{Emulated errors}: emulated wide-depth photometric errors.
  These are relevant for objects from the Deep and UltraDeep layers, which have smaller photometric uncertainties
  than Wide due to the deeper depths.
  They are computed
  independently for each flux type (\texttt{PSF}, \texttt{cmodel}, etc.) by assigning signal-to-noise (S/N) values
from a $grizy$ nearest-neighbor search to a Wide catalog of $\sim$ 500k objects
selected to mimic the overall survey sample.
\item \textbf{Weights}: Color/magnitude weights are computed using a generalization of \citet{lima08}'s nearest neighbor approach.
  The color-magnitude distribution of the training sample is different from that of target sample from the Wide layer.
  The weights are computed so that the training sample can reproduce the color-magnitude distribution of the target galaxies.
  See Speagle et al. (in prep.) for more details.
\end{enumerate}

\begin{figure*}
  \begin{center}
    \includegraphics[width=16cm]{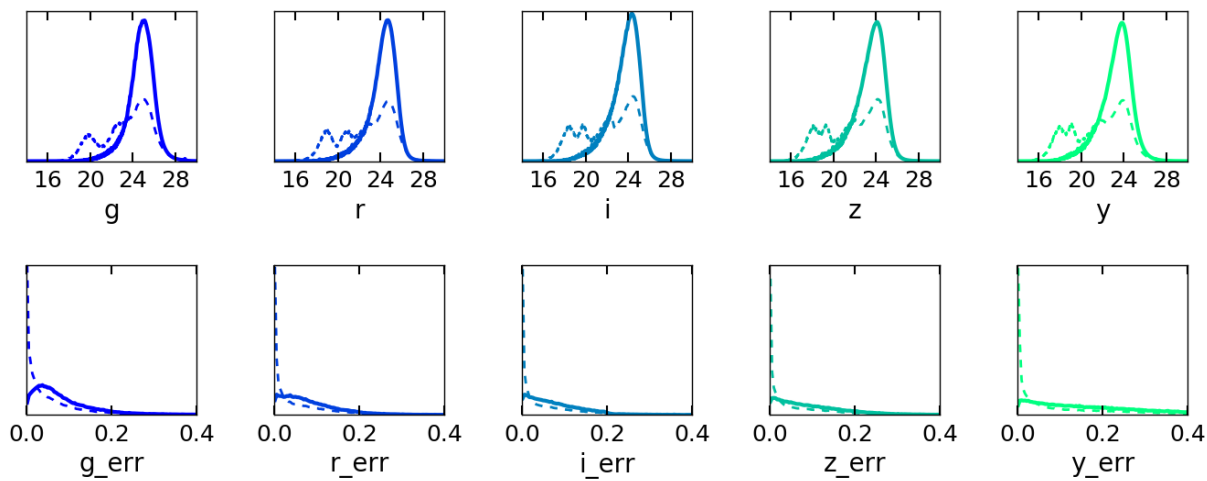}
  \end{center}
  \caption{
    The original (dashed) and re-weighted (solid) normalized number densities
    of our training sample as a function of $grizy$ (left-to-right) 
    magnitude (top) and error (bottom).
    Note that we use asinh magnitudes (i.e., Luptitudes) here.
    Our color-magnitude weights are able to
    effectively correct for biases in our original training sample to better
    mimic the HSC-SSP Wide data.
 }
 \label{fig:hsc_pz_1}
\end{figure*}

\begin{figure}
  \begin{center}
    \includegraphics[width=7cm]{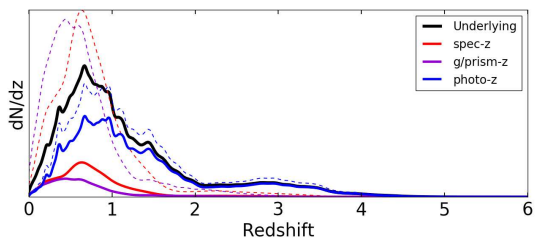}
  \end{center}
  \caption{
    The re-weighted, normalized redshift number density for our training sample.
    The full distribution is shown in solid black while the spec-$z$, g/prism-$z$, 
    and many-band photo-$z$ components are shown in solid red, purple, and blue, respectively.
    The dashed lines show these same components re-normalized to the full sample in order 
    to better highlight their differences. We can see that most of the substructure in the
    redshift distribution of our training sample comes from the many-band COSMOS photo-$z$'s,
    which also contribute almost all of our high-$z$ sources.
  }
 \label{fig:hsc_pz_3}
\end{figure}

\noindent
The underlying and re-weighted magnitude and redshift distributions of our training 
sample are shown in Figures \ref{fig:hsc_pz_1} and \ref{fig:hsc_pz_3}, respectively.

As described here, our training sample consists of various redshift measurements (spec, g/prism,
and many-band photo-$z$'s).  We use all of them as the `truth' throughout the paper, but some of the redshifts
(especially the many-band photo-$z$'s) may be erroneous. We thus urge caution when interpreting the absolute numbers in our adopted
metrics.  We refer to these 'true' redshifts as reference redshifts ($z_{ref}$) throughout the paper.

\subsection{Training and validation procedures}
\label{ssec:training_validation_proc}

The training sample is split into $k=5$ randomized 'folds'.  Because each fold
has a relatively large number of objects ($\sim$ 75k), most of us employ
a simple hold-out validation to train and validate our photo-$z$ methods.
To be specific,
we use folds $k=1-3$ to train our codes and $k=4$ to validate them, with the last fold ($k=5$) reserved for testing
(see below).
The only exception here is \texttt{FRANKEN-Z}, which uses cross-validation
(i.e., it used 5 rotating folds for training and validation).
Throughout the paper,
all statistics are computed using the color-magnitude weights described in the previous section.

\subsection{Test samples}
\label{ssec:test_samples}

We reserve a test sample from the training sample in order to evaluate the performance
of our codes.  Most of us use the subsample of the training sample described in  the previous section
(the 5th fold). For \texttt{FRANKEN-Z}, we use one of the 5-fold cross-validation runs.
We use the test sample to evaluate our performance at the UltraDeep depth.
This is reasonable because a significant fraction of the objects in the test sample come from UltraDeep COSMOS,
especially at faint magnitudes.

For the Wide-depth performance evaluation, we stack a subsample of the COSMOS UltraDeep data to
the Wide depth in all the bands.
We have computed the emulated Wide-depth photometric uncertainties as described earlier, but
they turn out to be problematic in a few cases.  Some of our codes use multiple photometry
techniques (e.g., \texttt{EPHOR} uses exponential and de Vaucouleur fluxes from CModel), but because
the measurements are done using the same pixels, these measurements are strongly correlated.
The random flux perturbation is no longer valid and we find that the resultant
photo-$z$'s have weird features.  We thus resort to the COSMOS Wide-depth stacks.
Thanks to a large number of visits available in the field,
we could generate stacks with three different seeing FWHMs (0.5, 0.7 and 1.0 arcsec), which we will
later use to evaluate the seeing dependence.  We call these stacks the Wide-depth 'best', 'median',
and 'worst' seeing stacks.  Because we use only a small subsample of the UltraDeep COSMOS data
(typically 1/10 of all the visits), it is reasonable to assume that the photometry is quasi-independent
from the training sample.  But, again, we urge caution when interpreting the absolute numbers.
We note that the Wide-depth stacks have the same $N(z)$ distribution as the training sample,
leading to some drawbacks we will discuss in Section \ref{ssec:dndz}.
The COSMOS wide-depth stacks are included in the public data
release \citep{aihara17} and can be exploited by the community.

We note that the current Wide-depth stacks have a known issue that the $i$-band
in the median seeing stack is slightly shallower than the Wide-depth (15min in total
as opposed to the nominal exposure of 20min).  But, for the purpose of photo-$z$ analyses in this paper,
we do not suffer significantly from this issue because we limit ourselves to
relatively bright magnitudes of $i<25$.  Also, the slightly shallower depth only in one
of the five bands does not have a major impact on the overall photo-$z$ performance.

\section{Methods}
\label{sec:method}

As we reviewed in the introduction section, each photo-$z$ technique has pros and cons.
For HSC-SSP, we use all the template fitting, empirical fitting, and machine-learning techniques
to cover the wide range of scientific applications.  We describe each of our code in this Section.

\subsection{DEmP}
\label{ssec:demp}

The Direct Empirical Photometric code (DEmP; \cite{hsieh14}) is the successor of the empirical
quadratic polynomial photometric redshift fitting code \citep{hsieh05} applied to the Red-Sequence
Cluster Survey data. DEmP is designed to minimize major issues of conventional empirical-fitting methods,
e.g., how to choose a proper form of the fitting functions, and biased results due to the population
distribution of the training set, by introducing two techniques: regional polynomial fitting and uniformly
weighted training set. The former is to perform fitting for each input galaxy using a subset of
the training set galaxies with photometry and colors closest to those of the input galaxy, and the latter
is to resample the training set to produce a flat population distribution. However,
we find that using a uniformly weighted training set does not
improve the overall photo-$z$ quality. This is because the number density of this training set is
sufficiently high thanks to the many-band photo-$z$'s from COSMOS; the subset of the training set used in the regional polynomial fitting consists of
galaxies with very similar magnitudes and colors, which reduces the bias caused by the population
distribution of the training set. Therefore, we use only the regional polynomial fitting to derive
the HSC photo-$z$s.

The probability distribution of photo-$z$ for each galaxy is generated using Monte Carlo
technique and the bootstraping method. We use Monte Carlo technique to generate 500 data sets based on
the photometry and uncertainties of the input galaxies to account for the effects due to photometric
uncertainties. We then bootstrap the training set for each input galaxy 500 times for each of
the Monte Carlo generated data set, to estimate the sampling effect in the training set. More details
are described in \citet{hsieh14}. We use the PSF-matched aperture photometry (a.k.a. the \texttt{afterburner}
photometry; \cite{aihara17}) to derive photo-$z$s for all the primary objects
even with only one-band detection.

\subsection{Ephor}
\label{ssec:ephor}
Extended Photometric redshift (EPHOR) is a publicly available,
neural network photo-$z$ code\footnote{\url{https://hsc-release.mtk.nao.ac.jp/doc/index.php/photometric-redshifts/}}.
We use a feedforward neural network that has an input layer ($\mathbf{x}_0$),
a series of hidden layers ($\mathbf{x}_i$ for $i = 1,\ldots,n$) and an output layer
($\mathbf{y}$).  Variables with the bold typeface are horizontal vectors.

We feed the neural network with two model fluxes;
de Vaucouleur flux and exponential flux in each band.
These are derived as part of the CModel photometry \citep{bosch17}.
The fluxes $f_k$ are normalized before being fed to the neural network:
\begin{eqnarray}
  (\mathbf{x}_0)_k
  &=& \mathop{\mathrm{arsinh}}\left( \frac{f_k - \mu_k}{\sigma_k} \right)
,
\end{eqnarray}
in which $\mu_k$ is the median of $f_k$ over the training dataset
(training as opposed to validation and test), and $\sigma_k$ is
the interquartile range, non-normalized, of the training dataset.
$\mathop{\mathrm{arsinh}}$ is applied so that unusually large fluxes
will not ruin the neural network.

The hidden layers employ softplus ($\mathop{\mathrm{softplus}}(x) = \ln(1 + \mathrm{e}^x)$)
as the activation function:
\begin{eqnarray}
  \mathbf{x}_i
  &=& \mathop{\mathrm{softplus}}(\mathbf{x}_{i-1} W_i + \mathbf{b}_i)
    \ \ \ \ \mbox{for $i = 1,\ldots,n$}
,
\end{eqnarray}

\noindent
where $W_i$ is a weight matrix and $\mathbf{b}_i$ is a bias vector, both of which are
determined in the training.  The $\mathop{\mathrm{softplus}}$ activation function is
applied to the argument vector elementwise.  The neural network performs slightly better
with softplus than with the rectifier $f(x) = \max(0,x)$.

The output layer is softmax: $\mathbf{y} = \sigma(\mathbf{x}_n)$, or
\begin{eqnarray}
  y_k
  &=& \Big( \sigma(\mathbf{x}_n) \Big)_k
  = \frac{\mathrm{e}^{(\mathbf{x}_n)_{k}}}{\sum_{\ell}
\mathrm{e}^{(\mathbf{x}_n)_{\ell}}}
.
\end{eqnarray}
We split the range of redshifts at equal intervals $z_0 < z_1 < \cdots < z_{d}$,
and equate $y_k$ with the probability of the redshift being within the $k$-th bin
$[z_{k-1}, z_{k})$.  We train the neural network by means of ADAM \citep{2014arXiv1412.6980K}
so that the cross entropy defined below is minimized:
\begin{eqnarray}
  H
  &=& \left\langle -\sum_{k=1}^{d} y'_k \ln y_k \right\rangle
,
\end{eqnarray}
in which the average is taken from the training dataset,
and $\mathbf{y}' = (0 \cdots 0\ 1\ 0 \cdots 0)$ is a one-hot vector for a sample:
\begin{eqnarray}
  y'_k
  &=& \left\{\begin{array}{ll}
        1 & \mbox{if the sample's redshift is in $[z_{k-1}, z_{k})$}
    ,\\ 0 & \mbox{otherwise}
    .
    \end{array}\right.
\end{eqnarray}

The default setup of \texttt{EPHOR} is to use the two model fluxes in each filter.
But, we also run the code using the PSF-matched aperture photometry (one flux in each band) and we refer to
the photo-$z$'s as \texttt{EPHOR\_AB}.  \texttt{AB} stands for afterburner.

\subsection{FRANKEN-Z}
\label{ssec:frankenz}

Flexible Regression over Associated Neighbors with Kernel dEnsity estimatioN
for Redshifts (FRANKEN-Z) is a hybrid approach that combines the data-driven
nature of machine learning with the statistical rigor of posterior-driven
(i.e. template-fitting) approaches. Using machine learning, FRANKEN-Z attempts to
approximate the `flux projection' from a set of unknown target objects
to a corresponding set of training objects in the presence of observational
errors within both datasets. The corresponding mapping to redshift is then computed by
stacking each training object's posterior-weighted redshift kernel
density estimate (KDE). This constitutes a generalization of typical
template-fitting approaches to the machine-learning regime.

For the HSC-SSP PDR1, we approximated the associated flux projection
using a collection of an object's nearest neighbors in magnitude space. We incorporated
observational errors by selecting object neighbors to be
the union of the 10 nearest-neighbors in
magnitude space computed using the PSF-matched photometry over
25 Monte Carlo realizations. The log-likelihoods for each object $i$ given
training object $j$ were then computed using the associated fluxes via

\begin{equation}
  -2\ln P(i|j) = \sum_{b} \frac{(F_{i,b}-F_{j,b})^2}{\sigma^2_{i,b}+\sigma^2_{j,b}} - n(i,j),
\end{equation}

\noindent
where the sum is taken over all bands indexed by $b$ and $n(i,j)$ is the number
of bands where both $i$ and $j$ are observed.

Because our nearest-neighbor search is in flux rather than redshift, our results
are (somewhat) more robust to domain mismatches between the training/target datasets.
We thus assume our prior is uniform over our training data such that our
posterior is directly proportional to our likelihood. The redshift PDF $P(z|i)$ then 
constitutes a posterior-weighted sum

\begin{equation}
P(z|i) = \sum_j P(z,j|i) = \sum_j P(z|j) P(j|i) \propto \sum_j P(z|j) P(i|j)
\end{equation}

\noindent
where $P(j|i)$ is the posterior and $P(i|j)$ is again the likelihood.

We note that the full code is still under active development and is more flexible
than the early version utilized here. It can be found at
\url{https://github.com/joshspeagle/frankenz}. See
Speagle et al. (in prep.) for additional details.

\subsection{MLZ}
\label{ssec:mlz}

SOMz is a part of the public photo-$z$ code, MLZ, which enables us to
estimate photometric redshift with {\it Self-Organizing Map} (SOM).
The SOM algorithm itself is an unsupervised machine learning method and
is widely used to classify a given dataset into small segments with
similar properties.
For photo-$z$ measurements, we first apply the SOM to the training set
and assign a redshift to each segment by computing the mean redshift of the
galaxies in that segment. Then, we find the closest segment for every
photometric objects to assign a redshift. Monte-Carlo and bootstrap
resampling enables us to produce the probability distribution of every
galaxy.

We describe each step in more detail.
First we prepare the random map of $N_{\rm pix}$ defined on a
2-dimensional sphere, where pixel is defined by Healpix pixelization
with $N_{\rm pix}=12\times N_{\rm side}^2$. The $p$-th pixel has a
vector $w_{pi}$ describing the object properties, e.g. 5-band
magnitudes,
where subscript $p$ runs from 1 to $N_{\rm pix}$, and
$i=1,2,\cdots, N_{\rm att}$ with $N_{\rm att}$ being the number of properties
to characterize objects, i.e. $k$-th galaxy has data vector
${\boldmath x}_k = \{x_{1k}, x_{2k}, \cdots, x_{N_{\rm att} k} \}$.
Here we utilize 5 band magnitudes of CModel photometry, and 10 colors
derived from those magnitudes. During the optimization, we find that
colors from afterburner photometry in addition to CModel magnitudes
and colors, slightly improve the photo-z performance. Therefore, we
characterize objects with 5+10+10 attributes with their measurement
errors.
Not all the attributes are independent and there are covariances between them.
We ignore the covariances for now and
leave it to our future work to evaluate their effects on photo-$z$'s.

As an initial condition of the map, we set the vector value in each
pixel randomly drawn from the data vector.
The Euclidean distance between a given galaxy and pixel is defined as
\begin{equation}
  d(p, k)
  =
  \sqrt{\sum_i \frac{(w_{pi}-x_{ik})^2}{\sigma_{ik}^2}}.
\end{equation}
Then we look for the nearest pixel for the given galaxy.
For the nearest pixel of the $k$-th galaxy $\hat{p}(k)$,
$d(\hat{p},k) \leq d(p,k)$ holds for any $p$.
The weight vectors of the nearest pixel and the vicinity of the nearest
pixels are iteratively updated as,
\begin{equation}
  \bs{w}_p(t+1)
  =
  \bs{w}_p(t) +
  \alpha(t)
  \exp\left[-\frac{\bs{\gamma}^2(p,\hat{p})}{2\sigma^2(t)} \right],
\end{equation}
where $\gamma(p, \hat{p})$ is the angular distance between pixel $p$
and the nearest pixel $\hat{p}$ and $\alpha, \sigma$ are
monotonically decreasing functions with time $t$. The time $t$
increases by 1 after we use one galaxy. After the pixels are updated
using all galaxies, the same processes are iteratively applied
except for setting the initial map to be random. We iterate this
for $N_{\rm ite}$ times.
In order to obtain a reliable redshift probability distribution function,
$P(z)$, we make a perturbed
catalog using both bootstrap resampling and Monte-Carlo methods.
For the latter, we perturb all the magnitudes and colors according to
their measurement errors. As a result, we have
$N_{\rm boot} \times N_{\rm MC}$ samples to derive our final $P(z)$.

As described in
Section~\ref{ssec:training_validation_proc},
we optimize the hyper-parameters using fold 1-3
and evaluate the performance with fold 4.
We note that the optimization is performed in terms of minimizing the
$\sigma_{conv}$ instead of \textit{loss} function introduced in
Section~\ref{ssec:metric}. That might partly be the reason why the
MLZ performed worse than other machine-learning codes, as we discuss later.
The hyper-parameters include $N_{\rm pix}$, $N_{\rm att}$,
$N_{\rm ite}$, $N_{\rm boot}$ and $N_{\rm MC}$. Given the reasonable
timescale to compute a large number of objects, we find out the
optimal hyper-parameter set as
$N_{\rm pix}=16$, $N_{\rm att}=5+10+10=25$, $N_{\rm ite}=200$,
$N_{\rm boot}=24$ and $N_{\rm MC}=16$. Except for the $N_{\rm att}$,
the increase of those parameters do not significantly improve our
results.

\subsection{NNPZ}
\label{ssec:nnpz}

Nearest Neighbors P(z) (NNPZ) redshifts are computed following the method introduced by
\citet{cunha09}. The principle of the method is explained in their Section~2.2 and can be
summarized as finding the nearest neighbors around an unknown object 
in the Euclidian color/magnitude space from a reference
sample and using the reference redshift histogram as the PDF. There exists, however,
a number of differences between the original method and the one applied here:
\begin{itemize}
\item $i$, $g-r$, $r-i$, $i-z$, $z-y$ color/magnitude attributes (CModel photometry) are used.
\item The reference sample is the weighted training sample (fold 1-3) as described in Section~\ref{sec:training_validation_and_test_samples}.
\item The neighbors are weighted according to the inverse Euclidean distance in the color/magnitude space.
\item To avoid giving too much weight to a neighbor with low signal-to-noise photometry that 
accidentally lies very close to the target object, the neighbors with large photometric errors are down-weighted. To do so, we first compute a photometric-error estimate as the sum of the photometric errors in all bands from both the unknown and neighbor objects, and we take the inverse of the photometric-error estimate as the weight. 
\end{itemize}

The final weight for each neighbor is the product of the reference, distance and photometric-error weights. The final object $P(z)$ is thus the weighted histogram of the neighbors. We also record the neighbor redshifts and weights in additional output tables.
We note that the choice of maximum number of neighbors, here 50, has little impact owing to the weighting
scheme in color/magnitude space.  We do not produce a $P(z)$ when the CModel measurement has failed in any of the bands.

\subsection{Mizuki}
\label{ssec:mizuki}

Finally, we use a template fitting-code {\sc mizuki} \citep{tanaka15}.
This code differs from classical template fitting codes in a few respects.  It uses a set of
templates generated with the \citet{bruzual03} stellar population synthesis code assuming
a \citet{chabrier03} IMF and \citet{calzetti00} dust attenuation curve.  Emission
lines are added to the templates assuming solar metallicity \citep{inoue11}.

There are pros and cons in using stellar population synthesis models.
One disadvantage of using theoretical templates is that they deliver less accurate photometric
redshifts than empirical templates because empirical templates often fit the observed SEDs of
galaxies better.  However, we correct for this template mismatch
by applying a template error function \citep{brammer2008}, which comes in two terms both as a function of rest-frame
wavelength.  One is a systematic flux correction applied to the templates to reduce the mismatch
and the other is template flux uncertainty to properly weight (un)reliable parts of SEDs.
This template error
function can be derived from the data by comparing the best-fit model fluxes and the observed fluxes
of objects.  We use the training sample (fold 1-3) to generate the template error function.

A big advantage of using theoretical templates is that we know the physical properties of
galaxies such as SFR and stellar mass for each template.  We apply a set of Bayesian priors
on the physical properties and let the priors depend on redshift.  Refer to \citet{tanaka15}
for details of the priors, but they are all observationally motivated.  What these priors
effectively do is (1) to keep the template parameters within realistic ranges to reduce the
degeneracy in the multi-color space and also (2) to let
templates evolve with redshift in an observationally motivated way.  Both template error
function and the physical priors improve photometric redshifts.
An improvement to the original code is that the $N(z|mag)$ prior is extended to multi-color
space and it now uses $N(z|g-i,i-y,i)$.  We make grids in the two-color magnitude space
and pre-compute $N(z)$ in each grid using the training sample (fold 1-3).  There are some redshift
spikes in the COSMOS field and we apply Gaussian smoothing with $\sigma_z=0.05$ 
in each grid to largely smear out the COSMOS-specific features.
In addition to redshifts, we compute stellar mass, SFR, and extinction fully marginalized
over all the other parameters, which can be useful for galaxy science.
Appendix~\ref{sec:phys_mizuki} compares stellar mass and SFRs from the code against
an external multi-wavelength survey.

In addition to galaxy templates, we also include QSO/AGN templates and stellar templates.
The QSO/AGN templates are generated by combining the type-1 QSO spectrum from \citet{polletta07}
and young galaxy templates from \citet{bruzual03} assuming $\tau = 1$Gyr, age$<$ 2Gyr, and
$0 < \tau_V < 2$,
where $\tau$ is an exponential decay timescale of star formation history,
age is time since the onset of star formation, and
$\tau_V$ is the optical depth (attenuation) in the $V$-band.
The relative fractions of the QSO and galaxy components are 0.5:1, 1:1, 2:1, and 4:1.
These hybrid templates are similar to those presented in \citet{salvato09}.
For the stellar templates, we use BaSeL 3.1 stellar library \citep{westera02}.
These QSO and stellar templates are used to give relative probabilities of objects being
galaxy, QSO, or star.  At this point, this functionality of the code is still preliminary,
and for simplicity, we use stellar and QSO templates for compact sources
(we use the standard extendedness parameter from the pipeline down to $i\sim24$ to
identify compact objects and all the fainter objects are assumed to be extended;
see \cite{aihara17} for details).
Only the galaxy templates are used for extended sources.

One important caveat in this release is that the code is trained using an old version of the training sample
with erroneous weights (the one described in Section \ref{sec:construction_of_the_training_sample}
but without the \texttt{centroid\_sdss} flag cut).  We unfortunately did not have time to
re-train the code with the new training sample.  This might be part of the reason why
the code performs worse than the other codes.

\section{Metrics and Their Definitions}
\label{sec:metric}

\subsection{Metrics to characterize photo-$z$}
\label{ssec:metric}
There are a few standard quantities used to characterize photo-$z$ accuracy.  However,
as their definitions are not always the same in the literature, we explicitly define them here for
this paper.  We also introduce new quantities.

\begin{itemize}

\item {\bf Bias:}
  Photo-$z$'s may systematically be off from spectroscopic redshifts
  and we call this systematic offset bias.  We compute a systematic bias
  in $\Delta z=(z_{\rm phot}-z_{\rm ref})/(1+z_{\rm ref})$ by applying the biweight
  statistics \citep{beers90}.
  The biweight is a robust statistical method to estimate the center and
  dispersion of a data sample by applying a weight function to down-weight
  outliers, which we often have in photo-$z$'s.
  We iteratively apply 3$\sigma$ clipping for 3 times to further reduce outliers.

\item {\bf Dispersion:}
  In the literature, dispersion is often computed as

  \begin{equation}
    \sigma_{\rm conv}=1.48\times{\rm MAD}(\Delta z),
  \end{equation}

  \noindent
  where MAD is the median absolute deviation.  Note that this definition
  does not account for the systematic bias.  In addition to this conventional
  definition, we also measure the dispersion by accounting for the bias
  using the biweight statistics.  We iteratively apply a $3\sigma$ clipping
  as done for bias to measure the dispersion around the central value.
  We denote the conventional dispersion
  and the biweight dispersion as $\sigma_{\rm conv}$ and $\sigma$, respectively.

\item {\bf Outlier rate:}
  The conventional definition is

  \begin{equation}
    f_{\rm outlier,conv}=\frac{N\left(|\Delta z|>0.15\right)}{N_{\rm total}},
  \end{equation}

  \noindent
  where outliers are defined as $|\Delta z|>0.15$.
  Again, this definition does not account for the systematic bias.
  The threshold of 0.15 is an arbitrary value but is probably reasonable for
  photo-$z$'s with several bands.  It is clearly too large for those with many bands.
  Together with this conventional one, we also define outliers as those
  $2\sigma$ away from the central value (these $\sigma$ and center are
  from biweight; see above). This $2\sigma$ is an arbitrary choice, but
  it is motivated to match reasonably well with the conventional
  one for several band photo-$z$'s.  We will denote the $\sigma$-based
  outlier fraction as $f_{\rm outlier}$ and the conventional one as
  $f_{\rm outlier,conv}$.

\item {\bf Loss function:}
  It can be cumbersome to use multiple indicators to characterize the photo-$z$
  accuracy.  Here we define a simple loss function to remedy the complexity and
  help us capture the photo-$z$ accuracy with a single number.
  We define a loss function as

  \begin{equation}
    L(\Delta z)=1-\frac{1}{1+\left(\frac{\Delta z}{\gamma}\right)^2}.
  \end{equation}

  This is an 'inverted' Lorentz function.  The loss is zero when $\Delta z=0$
  and continuously increases with larger $\Delta z$.  Thus, this can be considered as
  a continuous form of the outlier rate defined above.  The loss also increases with
  the photo-$z$ bias because
  a systematic bias means non-zero $\Delta z$ for most objects.
  The loss also increases with dispersion because a larger dispersion means larger $\Delta z$.
  Therefore, it effectively combines the three popular metrics into a single number.
  In order to keep a rough consistency with the conventional outlier definition, we adopt $\gamma=0.15$.

\end{itemize}

\begin{table*}[hptb]
  \begin{center}
    \begin{tabular}{ccccccc}
      \hline
      Point Estimator  &  bias  &  $\sigma_{conv}$ & $f_{outlier,conv}$  & $\sigma$  & $f_{outlier}$ & $loss$\\
      \hline
      mean & $-0.003$ & $0.075$ & $0.227$ & $0.078$ & $0.218$ & $0.260$\\
      mode & $-0.002$ & $0.067$ & $0.213$ & $0.064$ & $0.240$ & $0.244$\\
      median & $-0.001$ & $0.066$ & $0.199$ & $0.064$ & $0.226$ & $0.236$\\
      best & $-0.003$ & $0.064$ & $0.197$ & $0.061$ & $0.229$ & $0.233$\\
      \hline
    \end{tabular}
  \end{center}
  \caption{
    Photo-$z$ performance using mean, mode, median, and best estimators.  The numbers are for
    MLZ, but all the other codes show the same trend.
  }
  \label{tab:point_stat}
\end{table*}

\subsection{Optimal point estimates and photo-$z$ risk parameter}
\label{ssec:zp_best}

Our photo-$z$ methods do not output a point redshift directly,
but instead infer a redshift PDF, $P(z)$.  We want to use the full $P(z)$ for science,
but it is often useful to reduce the PDF to a point estimate, $z_{\mathrm{phot}}$.
There are several ways to do it; the mean, median or mode of $P(z)$, for example.
To obtain the ``best'' point estimate, however, we take the {\it minimum risk} strategy ---
we define a ``risk'' parameter as a function of redshift
and choose the point where the risk is minimized as the best point estimate.

The loss function $L(\Delta z)$ defined above is a function of $z_{\mathrm{phot}}$ and $z_{\mathrm{ref}}$,
and can be viewed as a loss arising from $z_{\mathrm{phot}}$ being different from $z_{\mathrm{ref}}$.
The expected amount of loss for a point estimate $z_{\rm phot}$ can be estimated as

\begin{eqnarray}
  \label{eq:risk_function}
  R(z_{\mathrm{phot}})
  &=& \int \mathrm{d}z P(z) L\left( \frac{z_{\rm phot}-z}{1+z}\right)
.
\end{eqnarray}

\noindent
The integral $R(z_{\mathrm{phot}})$ depends only on $z_{\mathrm{phot}}$ and represents
the expected loss for a given choice of $z_{\mathrm{phot}}$ as the point estimate.
We employ $R(z_{\mathrm{phot}})$ as the ``risk'' function.
The risk $R(z_{\mathrm{phot}})$ can be roughly interpreted as the probability
of the inferred redshift $z_{\mathrm{phot}}$ being an outlier:
the loss $L(\Delta z)$ is approximately $0$ if the guess $z_{\mathrm{phot}}$
is close to the true answer $z_{\mathrm{ref}}$, and it is approximately $1$
if the guess $z_{\mathrm{phot}}$ differs largely from the true answer $z_{\mathrm{ref}}$.

As mentioned above, we take the minimum risk strategy to choose a point estimate
$z_{\mathrm{phot}}$ at which the risk $R(z_{\mathrm{phot}})$ is minimum,
which we call the \textit{best} point estimate $z_{\mathrm{best}}$:
\begin{eqnarray}
    z_{\mathrm{best}} = \mathop{\mathrm{argmin}}(R(z_{\mathrm{phot}}))
.
\end{eqnarray}
This minimal point has no closed-form solution and must be searched for numerically.

In addition to $z_{best}$, we also compute $z_{mean}$, $z_{mode}$, and $z_{median}$ and
make comparisons between them in the next section, where we demonstrate that
$z_{best}$ indeed performs best.  Equally important to the point
estimate is the reliability of the point estimate, and we naturally use the risk
parameter, $R(z_{phot})$, for this\footnote{
  In the catalog database at the data release site, this parameter is named \texttt{photoz\_risk}.
}.
We compute the risk parameter for each point estimate (e.g., $R(z_{mean})$). To facilitate comparisons
to previous work, we also compute the commonly used
estimator of redshift confidence, $C(z)$, defined as

\begin{equation}
  \label{eq:photoz_conf}
  C(z_{phot})=\int_{z_{phot}-0.03}^{z_{phot}+0.03} P(z)dz,
\end{equation}

\noindent
where $z_{phot}$ is a point estimate such as median and best.  This is primarily to
keep consistency with previous studies, since we will show later that $R(z_{phot})$ is
a better estimator of photo-$z$ reliability.

\section{Performance Evaluation Using Point Estimates}
\label{sec:performance}

We now characterize the performance of our photo-$z$'s.
We first evaluate how well the 'best' point estimator works compared to
other popular statistics.  We then move on to show our photo-$z$ accuracy
at the Wide depth, 
followed by discussions on the depth and seeing dependence of the accuracy.
We focus on the point estimator to characterize our photo-$z$
performance in this section.  We present PDF-based tests in Section \ref{sec:pdf}.

\subsection{The 'best' point estimate}
\label{ssec:best_point_estimate}

One of
the most popular point estimators used for photo-$z$ is the median, which is defined as
the redshift at which the integrated probability equals 0.5.
The mode of PDF is also frequently used.
We compare the mean,
mode, median, and best redshifts using the COSMOS Wide-depth median seeing stack (see the next
section for details) for \texttt{MLZ} in Table \ref{tab:point_stat}.
We use all galaxies with $i<25$ here.
The best estimator gives the smallest scatter and lowest outlier rate\footnote{
  $f_{outlier}$ is larger for best than for median, but it is due to reduced scatter ($\sigma$).
  Recall that $f_{outlier}$ is defined as $2\sigma$ outliers.
} compared to the other estimators.
The best estimator tends to introduce a small negative bias, but the bias is
not sufficiently large to prevent most scientific applications.
The other photo-$z$ codes show the same trend.  Based on this result,
we will use the best estimator in what follows and denote the best redshift as $z_{phot}$
for simplicity.

\subsection{Photo-$z$ performance at the Wide-depth}
\label{ssec:photoz_performance_at_the_wide_depth}

\begin{figure*}
  \begin{center}
    \includegraphics[width=7cm]{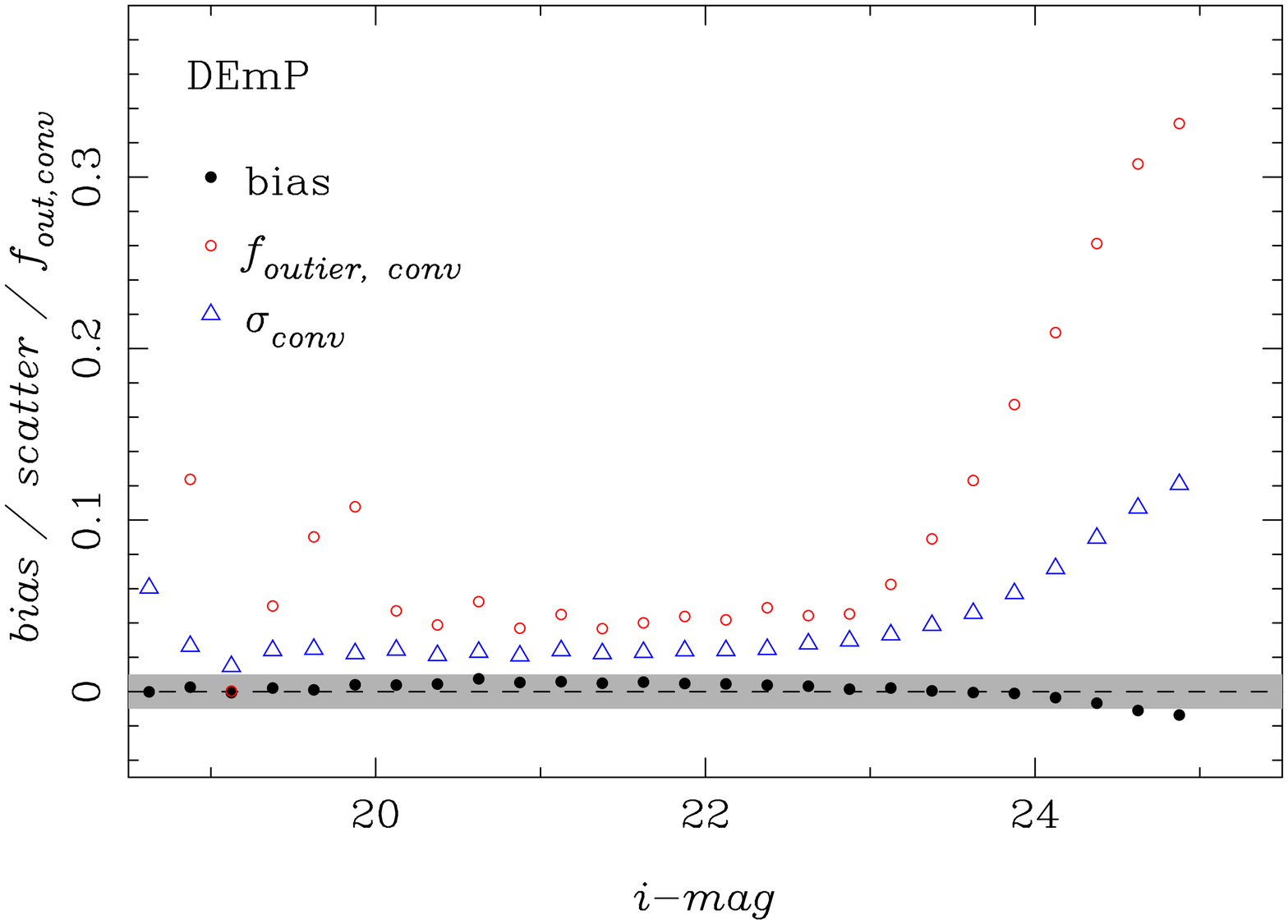}\hspace{0.5cm}
    \includegraphics[width=7cm]{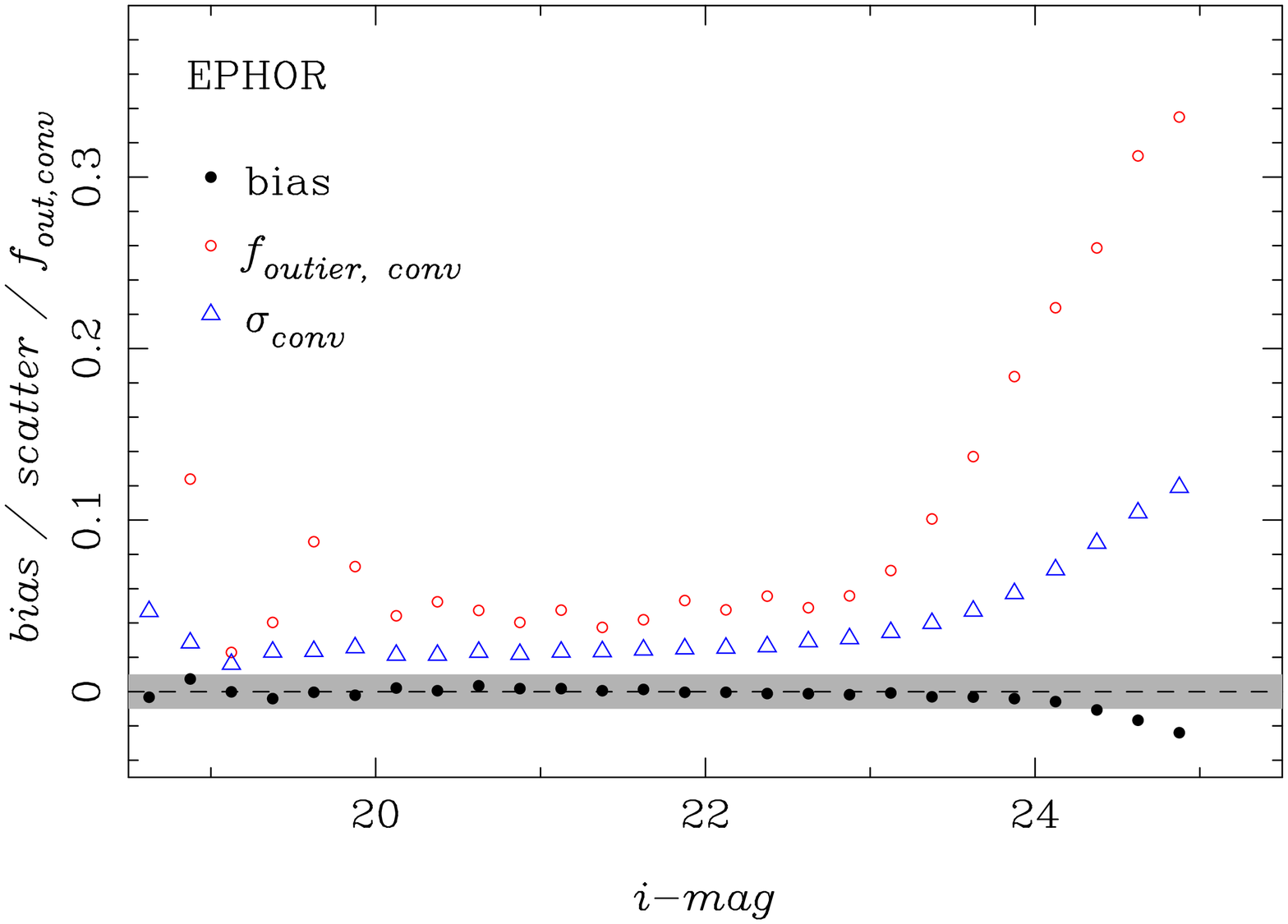}\\\vspace{0.5cm}
    \includegraphics[width=7cm]{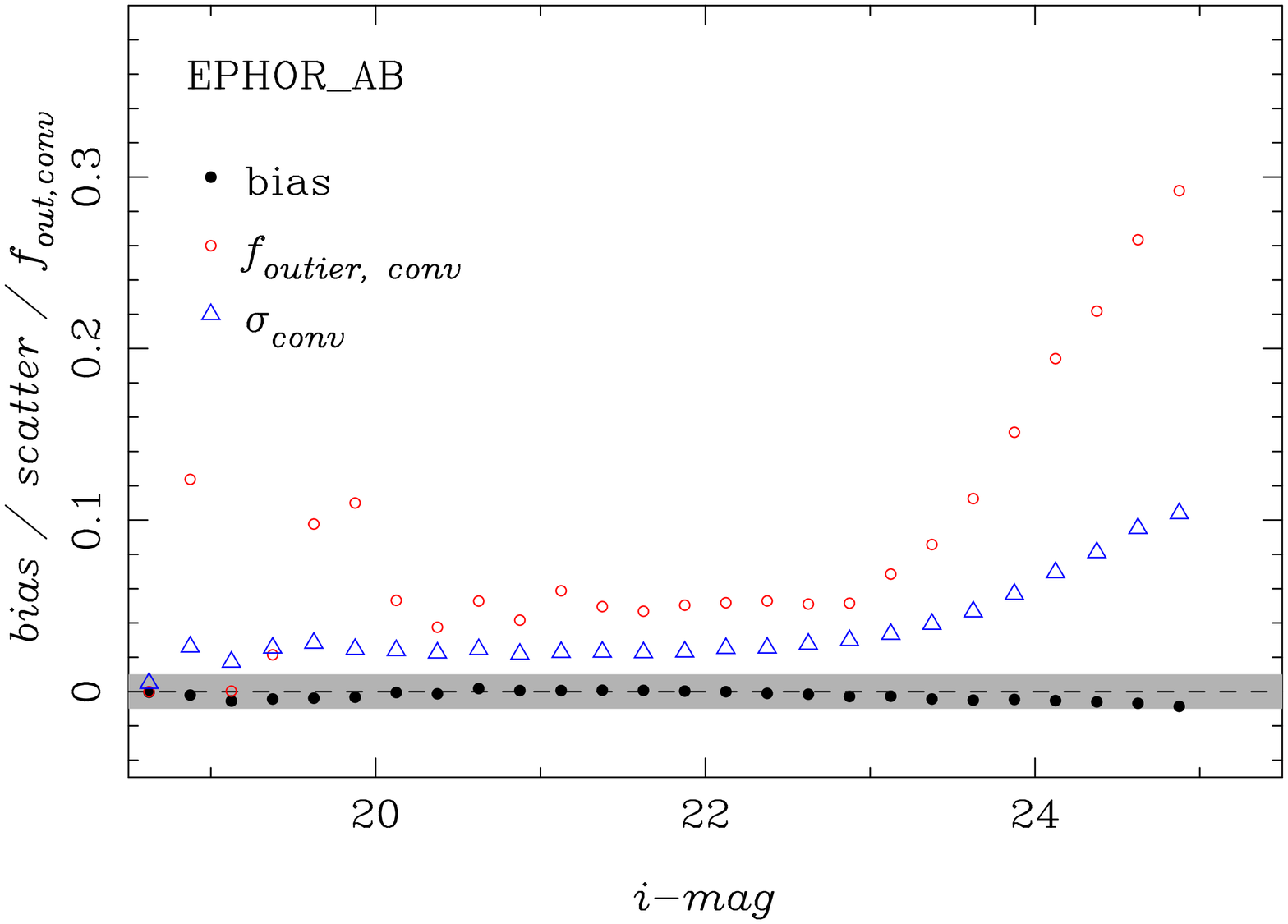}\hspace{0.5cm}
    \includegraphics[width=7cm]{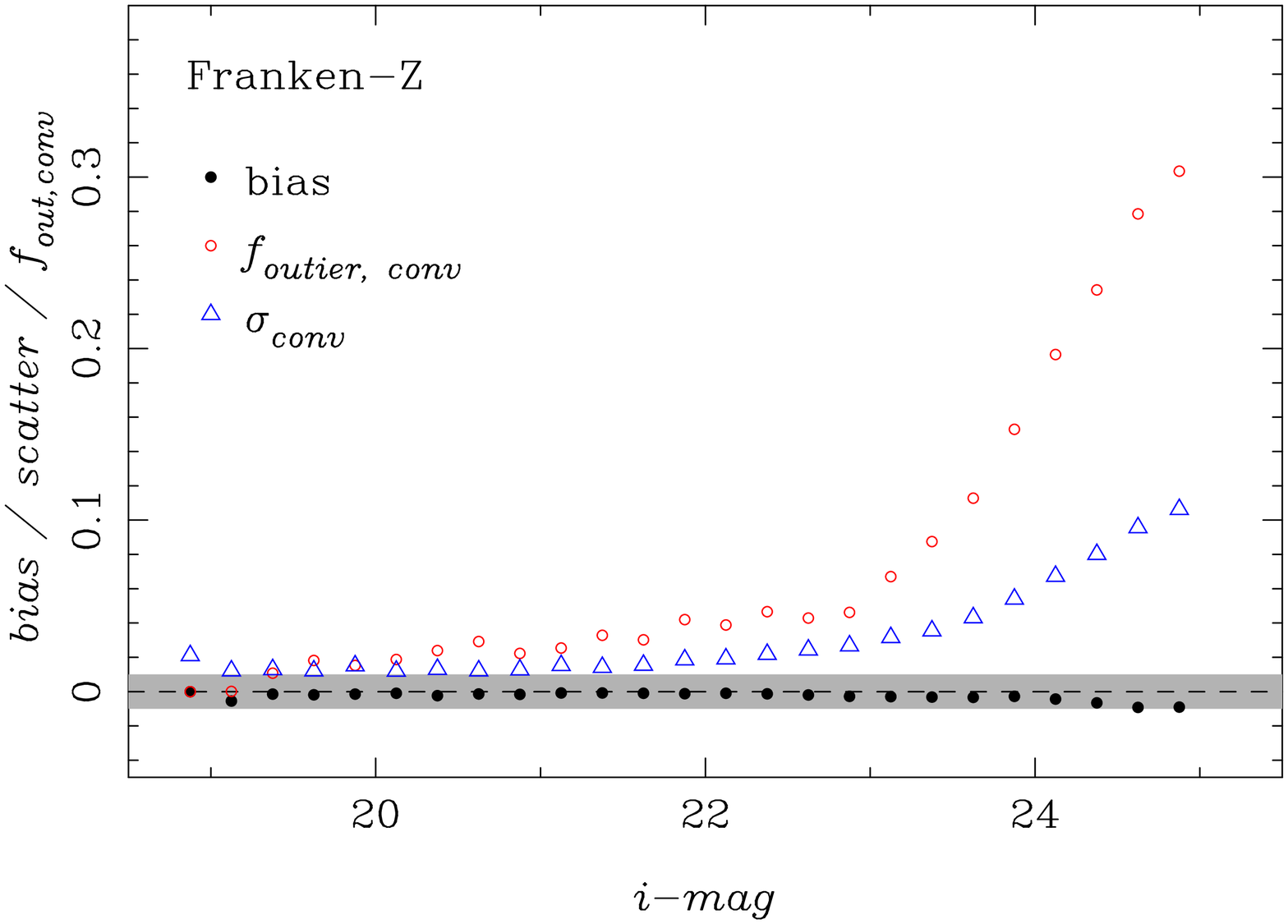}\\\vspace{0.5cm}
    \includegraphics[width=7cm]{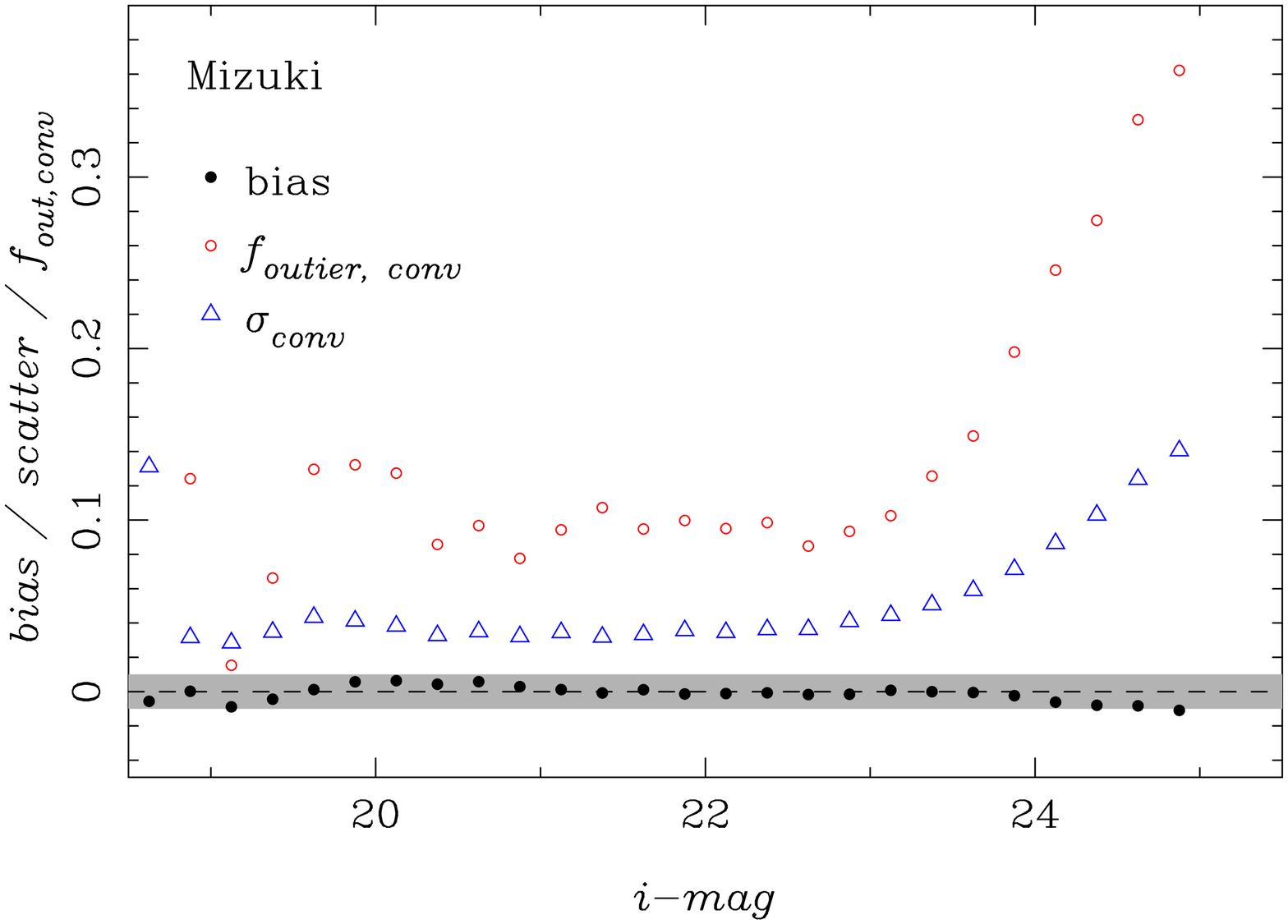}\hspace{0.5cm}
    \includegraphics[width=7cm]{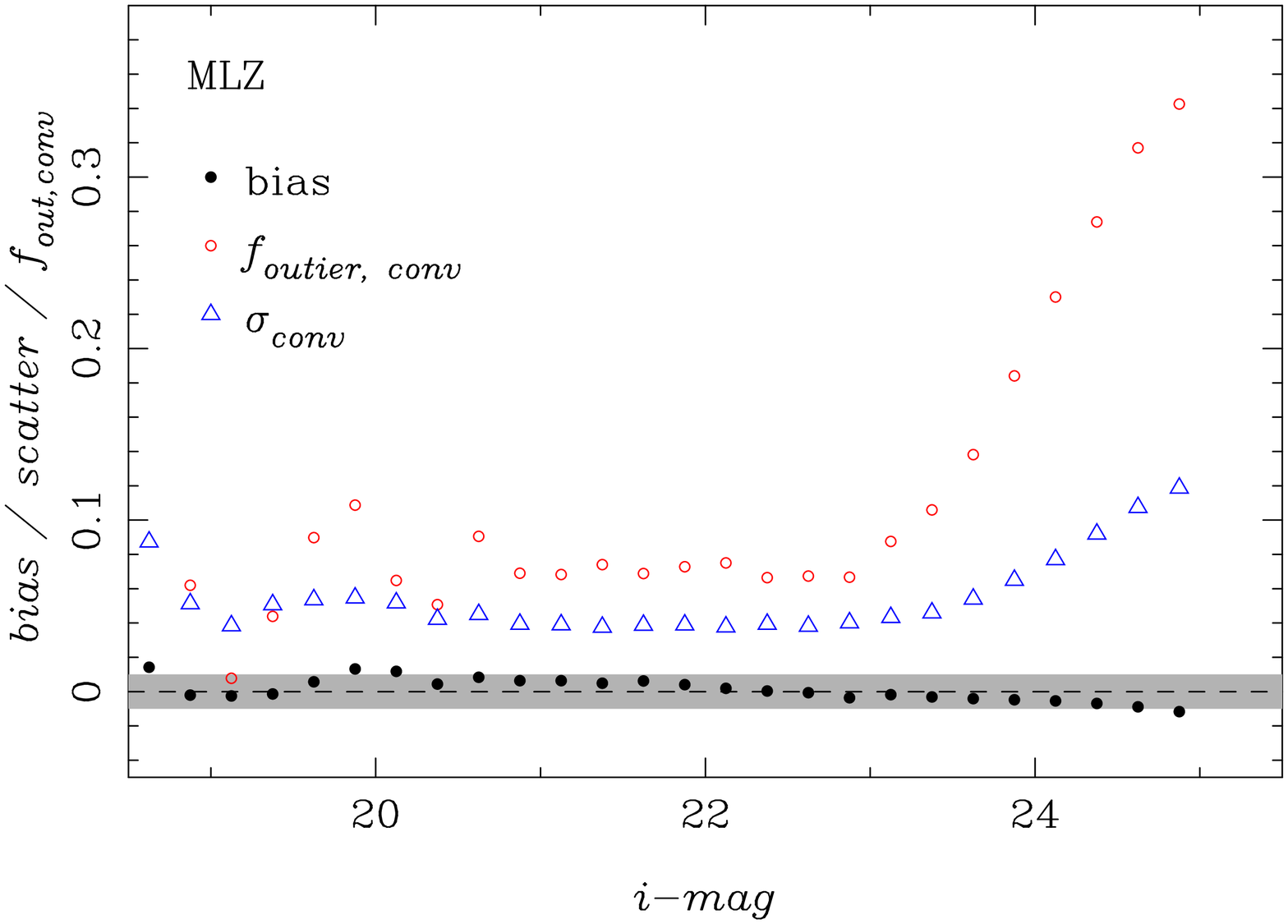}\\\vspace{0.5cm}
    \includegraphics[width=7cm]{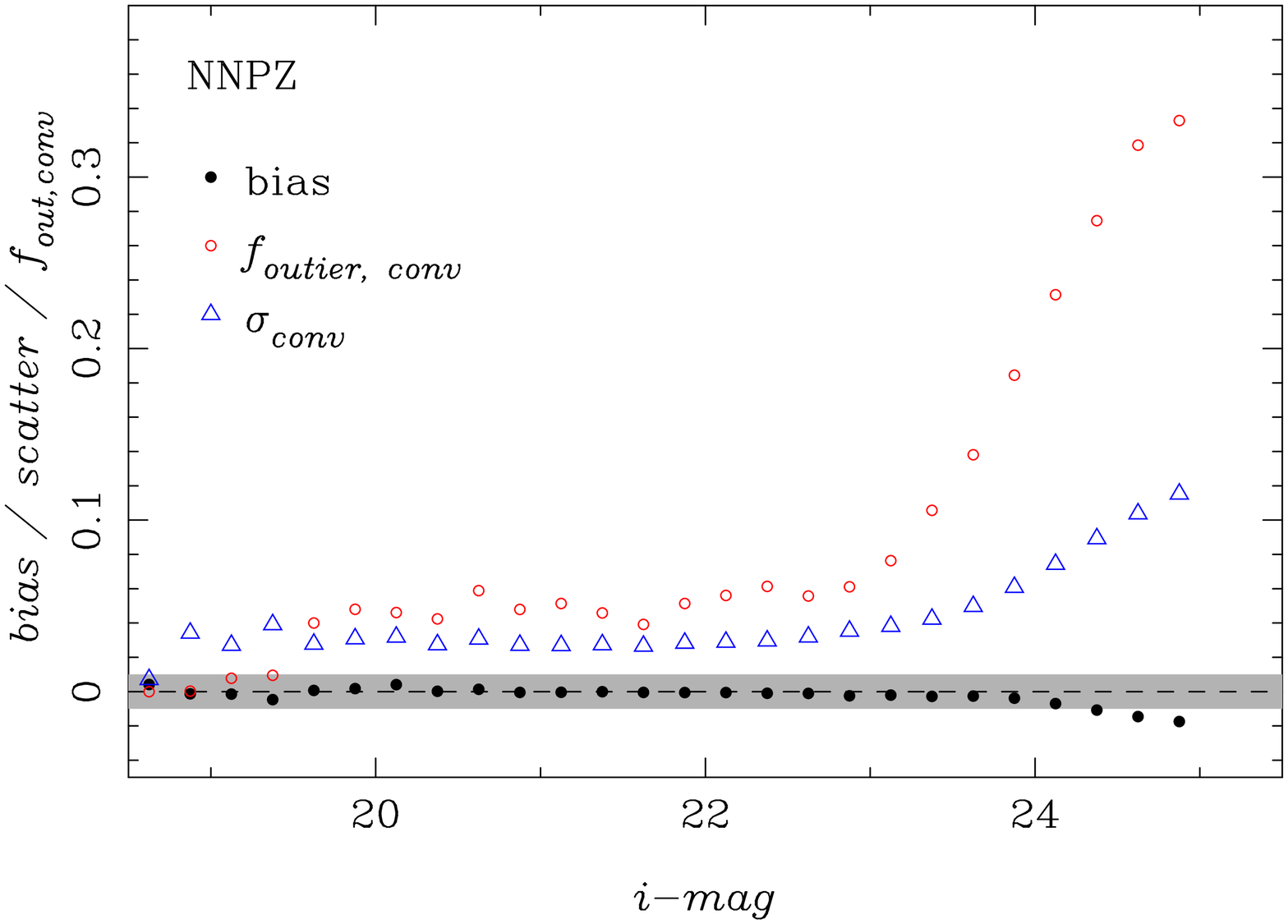}
  \end{center}
  \caption{
    Bias, $f_{outlier,conv}$ and $\sigma_{conv}$ plotted against $i$-band magnitude.
    The different panels are for different codes as indicated by the label on
    the top-left corner of each panel.  The gray shades show $\pm0.01$ range,
    which will be useful for bias.  The symbols are explained in the panels.
    Note that these plots are based on the COSMOS Wide-depth median stack
    and include objects in COSMOS only.
 }
 \label{fig:stat_mag}
\end{figure*}

We characterize the photo-$z$ performance at the Wide-depth, representative depth
of the HSC survey as a whole, using the metrics defined in Section \ref{sec:metric}.
We compare $z_{phot}$ with $z_{ref}$ for the COSMOS Wide-depth median seeing stack.
Recall that this is a subsample of UltraDeep COSMOS data stacked to the depth of
the Wide survey and all the filters have 0.7 arcsec seeing.
Fig.~\ref{fig:stat_mag} shows the bias, scatter and outlier fraction as a function of
$i$-band magnitude for all the codes.  More statistics are summarized in Table~\ref{tab:stat_mag}.
Most of the reference redshifts at faint magnitudes come from
the COSMOS photo-$z$ catalog and they are not very accurate at $i\gtrsim25$.
We thus cut at $i=25$ and characterize the performance at brighter mags.
Once again, not all the COSMOS photo-$z$'s are correct and we inherit the systematic uncertainty
from COSMOS.  The absolute numbers of the statistics shown in the figure should thus be
taken with caution.

Before we compare the codes, it is important to note that we observe a sign of
of over-fitting in \texttt{FRANKEN-Z} even though the COSMOS Wide-depth
stacks were not explicitly included during the training. This is likely due to 
\texttt{FRANKEN-Z}'s sensitivity to both the redshift PDF and error distributions 
in both datasets, which makes it more sensitive to our assumption that the two
datasets are quasi-independent (see section \ref{sec:pdf} for details).
It is thus unfair to compare its performance directly with the other codes, since
it is likely to be overly optimistic.

The photo-$z$ accuracy is a strong function of magnitude, but it is relatively
flat down to $i\sim23$ for all the codes.  The scatter and outlier rate are
about 0.03 and 5\%, respectively, at this bright mags.  This flat performance is likely because most
objects within this magnitude range are located at $z\lesssim1.5$, where we can
obtain fairly good photo-$z$'s with the $grizy$ photometry (see below).
At fainter mags,
the fraction of $z>1.5$ objects increases and these high redshift galaxies 
drive the poor performance at faint mags.  It is encouraging that the bias is
still within $\sim1\%$ at all mags for most codes.

{\scriptsize
  \begin{longtable}{llcccccc}
    \caption{
      Photo-$z$ statistics for all the codes as a function of magnitude.  The number are for all
      galaxies down to $i=25$.
    }
    \label{tab:stat_mag}
    \hline
    Code   &  mag.  &  bias  &  $\sigma_{conv}$ & $f_{outlier,conv}$ & $\sigma$ & $f_{outlier}$ & $<L(\Delta z))>$\\
    \endfirsthead
    \endhead
    \hline
    \endfoot
    \hline
    \endlastfoot
    \hline
     & $18.50 - 18.75$ & $-0.000$ & $0.060$ & $0.430$ & $0.023$ & $0.463$ & $0.373$\\
 & $18.75 - 19.00$ & $+0.003$ & $0.026$ & $0.124$ & $0.025$ & $0.210$ & $0.124$\\
 & $19.00 - 19.25$ & $-0.001$ & $0.015$ & $0.000$ & $0.017$ & $0.145$ & $0.027$\\
 & $19.25 - 19.50$ & $+0.002$ & $0.024$ & $0.050$ & $0.021$ & $0.156$ & $0.070$\\
 & $19.50 - 19.75$ & $+0.001$ & $0.025$ & $0.090$ & $0.023$ & $0.209$ & $0.101$\\
 & $19.75 - 20.00$ & $+0.004$ & $0.022$ & $0.108$ & $0.020$ & $0.245$ & $0.112$\\
 & $20.00 - 20.25$ & $+0.004$ & $0.024$ & $0.047$ & $0.023$ & $0.191$ & $0.079$\\
 & $20.25 - 20.50$ & $+0.004$ & $0.021$ & $0.039$ & $0.019$ & $0.178$ & $0.061$\\
 & $20.50 - 20.75$ & $+0.007$ & $0.023$ & $0.052$ & $0.021$ & $0.186$ & $0.077$\\
 & $20.75 - 21.00$ & $+0.005$ & $0.021$ & $0.037$ & $0.020$ & $0.166$ & $0.058$\\
 & $21.00 - 21.25$ & $+0.006$ & $0.024$ & $0.045$ & $0.022$ & $0.171$ & $0.072$\\
 & $21.25 - 21.50$ & $+0.005$ & $0.022$ & $0.037$ & $0.021$ & $0.182$ & $0.066$\\
	DEmP	& $21.50 - 21.75$ & $+0.006$ & $0.023$ & $0.040$ & $0.021$ & $0.156$ & $0.064$\\
 & $21.75 - 22.00$ & $+0.005$ & $0.024$ & $0.044$ & $0.024$ & $0.165$ & $0.073$\\
 & $22.00 - 22.25$ & $+0.005$ & $0.024$ & $0.042$ & $0.023$ & $0.163$ & $0.072$\\
 & $22.25 - 22.50$ & $+0.004$ & $0.025$ & $0.049$ & $0.024$ & $0.166$ & $0.078$\\
 & $22.50 - 22.75$ & $+0.003$ & $0.028$ & $0.044$ & $0.027$ & $0.153$ & $0.078$\\
 & $22.75 - 23.00$ & $+0.001$ & $0.030$ & $0.045$ & $0.030$ & $0.150$ & $0.082$\\
 & $23.00 - 23.25$ & $+0.002$ & $0.033$ & $0.063$ & $0.033$ & $0.160$ & $0.103$\\
 & $23.25 - 23.50$ & $+0.000$ & $0.039$ & $0.089$ & $0.038$ & $0.173$ & $0.128$\\
 & $23.50 - 23.75$ & $-0.001$ & $0.046$ & $0.123$ & $0.044$ & $0.201$ & $0.164$\\
 & $23.75 - 24.00$ & $-0.001$ & $0.057$ & $0.167$ & $0.054$ & $0.217$ & $0.208$\\
 & $24.00 - 24.25$ & $-0.004$ & $0.072$ & $0.209$ & $0.070$ & $0.222$ & $0.251$\\
 & $24.25 - 24.50$ & $-0.007$ & $0.090$ & $0.261$ & $0.089$ & $0.227$ & $0.297$\\
 & $24.50 - 24.75$ & $-0.011$ & $0.107$ & $0.308$ & $0.110$ & $0.222$ & $0.335$\\
 & $24.75 - 25.00$ & $-0.014$ & $0.121$ & $0.331$ & $0.127$ & $0.208$ & $0.357$\\
\hline
 & $18.50 - 18.75$ & $-0.003$ & $0.047$ & $0.430$ & $0.020$ & $0.437$ & $0.422$\\
 & $18.75 - 19.00$ & $+0.007$ & $0.028$ & $0.124$ & $0.017$ & $0.271$ & $0.139$\\
 & $19.00 - 19.25$ & $-0.000$ & $0.016$ & $0.023$ & $0.018$ & $0.133$ & $0.037$\\
 & $19.25 - 19.50$ & $-0.004$ & $0.023$ & $0.040$ & $0.024$ & $0.201$ & $0.077$\\
 & $19.50 - 19.75$ & $-0.000$ & $0.024$ & $0.087$ & $0.021$ & $0.176$ & $0.092$\\
 & $19.75 - 20.00$ & $-0.002$ & $0.026$ & $0.073$ & $0.024$ & $0.191$ & $0.099$\\
 & $20.00 - 20.25$ & $+0.002$ & $0.021$ & $0.044$ & $0.022$ & $0.205$ & $0.078$\\
 & $20.25 - 20.50$ & $+0.001$ & $0.021$ & $0.052$ & $0.020$ & $0.164$ & $0.071$\\
 & $20.50 - 20.75$ & $+0.003$ & $0.023$ & $0.047$ & $0.023$ & $0.157$ & $0.069$\\
 & $20.75 - 21.00$ & $+0.002$ & $0.022$ & $0.040$ & $0.021$ & $0.157$ & $0.065$\\
 & $21.00 - 21.25$ & $+0.002$ & $0.023$ & $0.048$ & $0.023$ & $0.157$ & $0.071$\\
 & $21.25 - 21.50$ & $+0.001$ & $0.023$ & $0.037$ & $0.024$ & $0.144$ & $0.064$\\
	EPHOR	& $21.50 - 21.75$ & $+0.001$ & $0.024$ & $0.042$ & $0.024$ & $0.145$ & $0.066$\\
 & $21.75 - 22.00$ & $-0.000$ & $0.025$ & $0.053$ & $0.026$ & $0.165$ & $0.078$\\
 & $22.00 - 22.25$ & $-0.000$ & $0.025$ & $0.048$ & $0.025$ & $0.168$ & $0.079$\\
 & $22.25 - 22.50$ & $-0.001$ & $0.026$ & $0.056$ & $0.026$ & $0.172$ & $0.084$\\
 & $22.50 - 22.75$ & $-0.001$ & $0.029$ & $0.049$ & $0.029$ & $0.153$ & $0.083$\\
 & $22.75 - 23.00$ & $-0.002$ & $0.031$ & $0.056$ & $0.031$ & $0.159$ & $0.092$\\
 & $23.00 - 23.25$ & $-0.001$ & $0.035$ & $0.071$ & $0.034$ & $0.164$ & $0.109$\\
 & $23.25 - 23.50$ & $-0.003$ & $0.040$ & $0.101$ & $0.037$ & $0.183$ & $0.137$\\
 & $23.50 - 23.75$ & $-0.003$ & $0.047$ & $0.137$ & $0.044$ & $0.211$ & $0.175$\\
 & $23.75 - 24.00$ & $-0.004$ & $0.057$ & $0.184$ & $0.052$ & $0.235$ & $0.217$\\
 & $24.00 - 24.25$ & $-0.006$ & $0.071$ & $0.224$ & $0.066$ & $0.243$ & $0.259$\\
 & $24.25 - 24.50$ & $-0.011$ & $0.087$ & $0.259$ & $0.085$ & $0.235$ & $0.293$\\
 & $24.50 - 24.75$ & $-0.017$ & $0.104$ & $0.312$ & $0.110$ & $0.226$ & $0.333$\\
 & $24.75 - 25.00$ & $-0.024$ & $0.119$ & $0.335$ & $0.125$ & $0.210$ & $0.356$\\
\hline
 & $18.50 - 18.75$ & $-0.000$ & $0.005$ & $0.000$ & $0.008$ & $0.325$ & $0.014$\\
 & $18.75 - 19.00$ & $-0.002$ & $0.026$ & $0.124$ & $0.026$ & $0.192$ & $0.140$\\
 & $19.00 - 19.25$ & $-0.005$ & $0.017$ & $0.000$ & $0.019$ & $0.123$ & $0.035$\\
 & $19.25 - 19.50$ & $-0.004$ & $0.026$ & $0.021$ & $0.021$ & $0.105$ & $0.046$\\
 & $19.50 - 19.75$ & $-0.004$ & $0.028$ & $0.098$ & $0.023$ & $0.169$ & $0.100$\\
 & $19.75 - 20.00$ & $-0.003$ & $0.025$ & $0.110$ & $0.021$ & $0.229$ & $0.121$\\
 & $20.00 - 20.25$ & $-0.001$ & $0.024$ & $0.053$ & $0.026$ & $0.174$ & $0.082$\\
 & $20.25 - 20.50$ & $-0.001$ & $0.023$ & $0.038$ & $0.021$ & $0.147$ & $0.061$\\
 & $20.50 - 20.75$ & $+0.002$ & $0.025$ & $0.053$ & $0.022$ & $0.163$ & $0.076$\\
 & $20.75 - 21.00$ & $+0.001$ & $0.022$ & $0.042$ & $0.022$ & $0.143$ & $0.059$\\
 & $21.00 - 21.25$ & $+0.001$ & $0.023$ & $0.059$ & $0.022$ & $0.178$ & $0.080$\\
 & $21.25 - 21.50$ & $+0.001$ & $0.023$ & $0.050$ & $0.022$ & $0.169$ & $0.072$\\
	EPHOR\_AB	& $21.50 - 21.75$ & $+0.001$ & $0.023$ & $0.047$ & $0.022$ & $0.153$ & $0.066$\\
 & $21.75 - 22.00$ & $+0.000$ & $0.023$ & $0.050$ & $0.024$ & $0.165$ & $0.075$\\
 & $22.00 - 22.25$ & $-0.000$ & $0.025$ & $0.052$ & $0.024$ & $0.157$ & $0.078$\\
 & $22.25 - 22.50$ & $-0.001$ & $0.026$ & $0.053$ & $0.025$ & $0.162$ & $0.078$\\
 & $22.50 - 22.75$ & $-0.002$ & $0.028$ & $0.051$ & $0.027$ & $0.147$ & $0.080$\\
 & $22.75 - 23.00$ & $-0.003$ & $0.030$ & $0.052$ & $0.030$ & $0.148$ & $0.085$\\
 & $23.00 - 23.25$ & $-0.003$ & $0.033$ & $0.069$ & $0.032$ & $0.155$ & $0.102$\\
 & $23.25 - 23.50$ & $-0.004$ & $0.039$ & $0.086$ & $0.037$ & $0.164$ & $0.123$\\
 & $23.50 - 23.75$ & $-0.005$ & $0.047$ & $0.113$ & $0.044$ & $0.182$ & $0.155$\\
 & $23.75 - 24.00$ & $-0.005$ & $0.057$ & $0.151$ & $0.052$ & $0.200$ & $0.194$\\
 & $24.00 - 24.25$ & $-0.005$ & $0.069$ & $0.194$ & $0.068$ & $0.209$ & $0.237$\\
 & $24.25 - 24.50$ & $-0.006$ & $0.081$ & $0.222$ & $0.081$ & $0.206$ & $0.264$\\
 & $24.50 - 24.75$ & $-0.007$ & $0.095$ & $0.263$ & $0.097$ & $0.215$ & $0.303$\\
 & $24.75 - 25.00$ & $-0.009$ & $0.104$ & $0.292$ & $0.110$ & $0.212$ & $0.326$\\
\hline
 & $18.75 - 19.00$ & $-0.000$ & $0.021$ & $0.000$ & $0.028$ & $0.120$ & $0.039$\\
 & $19.00 - 19.25$ & $-0.005$ & $0.012$ & $0.000$ & $0.010$ & $0.228$ & $0.032$\\
 & $19.25 - 19.50$ & $-0.001$ & $0.013$ & $0.011$ & $0.015$ & $0.189$ & $0.036$\\
 & $19.50 - 19.75$ & $-0.002$ & $0.012$ & $0.018$ & $0.015$ & $0.182$ & $0.038$\\
 & $19.75 - 20.00$ & $-0.001$ & $0.015$ & $0.015$ & $0.017$ & $0.145$ & $0.043$\\
 & $20.00 - 20.25$ & $-0.001$ & $0.012$ & $0.019$ & $0.013$ & $0.199$ & $0.039$\\
 & $20.25 - 20.50$ & $-0.002$ & $0.013$ & $0.024$ & $0.014$ & $0.200$ & $0.043$\\
 & $20.50 - 20.75$ & $-0.001$ & $0.012$ & $0.029$ & $0.013$ & $0.195$ & $0.045$\\
 & $20.75 - 21.00$ & $-0.002$ & $0.013$ & $0.022$ & $0.014$ & $0.162$ & $0.038$\\
 & $21.00 - 21.25$ & $-0.001$ & $0.015$ & $0.025$ & $0.016$ & $0.156$ & $0.046$\\
 & $21.25 - 21.50$ & $-0.001$ & $0.014$ & $0.033$ & $0.015$ & $0.178$ & $0.046$\\
 & $21.50 - 21.75$ & $-0.001$ & $0.015$ & $0.030$ & $0.016$ & $0.173$ & $0.048$\\
	Franken-Z	& $21.75 - 22.00$ & $-0.001$ & $0.019$ & $0.042$ & $0.019$ & $0.173$ & $0.060$\\
 & $22.00 - 22.25$ & $-0.001$ & $0.019$ & $0.039$ & $0.020$ & $0.174$ & $0.062$\\
 & $22.25 - 22.50$ & $-0.001$ & $0.022$ & $0.047$ & $0.022$ & $0.173$ & $0.069$\\
 & $22.50 - 22.75$ & $-0.002$ & $0.024$ & $0.043$ & $0.025$ & $0.152$ & $0.069$\\
 & $22.75 - 23.00$ & $-0.003$ & $0.027$ & $0.046$ & $0.027$ & $0.158$ & $0.079$\\
 & $23.00 - 23.25$ & $-0.003$ & $0.032$ & $0.067$ & $0.031$ & $0.172$ & $0.103$\\
 & $23.25 - 23.50$ & $-0.003$ & $0.035$ & $0.088$ & $0.034$ & $0.180$ & $0.122$\\
 & $23.50 - 23.75$ & $-0.003$ & $0.043$ & $0.113$ & $0.041$ & $0.197$ & $0.154$\\
 & $23.75 - 24.00$ & $-0.003$ & $0.054$ & $0.153$ & $0.050$ & $0.214$ & $0.195$\\
 & $24.00 - 24.25$ & $-0.004$ & $0.067$ & $0.197$ & $0.064$ & $0.224$ & $0.239$\\
 & $24.25 - 24.50$ & $-0.007$ & $0.080$ & $0.234$ & $0.079$ & $0.223$ & $0.273$\\
 & $24.50 - 24.75$ & $-0.009$ & $0.096$ & $0.279$ & $0.096$ & $0.226$ & $0.310$\\
 & $24.75 - 25.00$ & $-0.009$ & $0.106$ & $0.303$ & $0.112$ & $0.217$ & $0.332$\\
\hline
 & $18.50 - 18.75$ & $-0.006$ & $0.131$ & $0.430$ & $0.051$ & $0.464$ & $0.481$\\
 & $18.75 - 19.00$ & $+0.000$ & $0.032$ & $0.124$ & $0.036$ & $0.278$ & $0.190$\\
 & $19.00 - 19.25$ & $-0.009$ & $0.028$ & $0.015$ & $0.036$ & $0.115$ & $0.075$\\
 & $19.25 - 19.50$ & $-0.004$ & $0.035$ & $0.066$ & $0.036$ & $0.161$ & $0.112$\\
 & $19.50 - 19.75$ & $+0.001$ & $0.043$ & $0.130$ & $0.037$ & $0.228$ & $0.178$\\
 & $19.75 - 20.00$ & $+0.006$ & $0.041$ & $0.132$ & $0.041$ & $0.237$ & $0.185$\\
 & $20.00 - 20.25$ & $+0.006$ & $0.038$ & $0.127$ & $0.040$ & $0.234$ & $0.171$\\
 & $20.25 - 20.50$ & $+0.004$ & $0.033$ & $0.086$ & $0.030$ & $0.210$ & $0.122$\\
 & $20.50 - 20.75$ & $+0.006$ & $0.035$ & $0.097$ & $0.032$ & $0.218$ & $0.126$\\
 & $20.75 - 21.00$ & $+0.003$ & $0.032$ & $0.078$ & $0.033$ & $0.179$ & $0.110$\\
 & $21.00 - 21.25$ & $+0.001$ & $0.035$ & $0.094$ & $0.034$ & $0.174$ & $0.127$\\
 & $21.25 - 21.50$ & $-0.001$ & $0.032$ & $0.107$ & $0.029$ & $0.208$ & $0.134$\\
	Mizuki	& $21.50 - 21.75$ & $+0.001$ & $0.033$ & $0.095$ & $0.030$ & $0.166$ & $0.121$\\
 & $21.75 - 22.00$ & $-0.001$ & $0.036$ & $0.100$ & $0.033$ & $0.184$ & $0.132$\\
 & $22.00 - 22.25$ & $-0.001$ & $0.035$ & $0.095$ & $0.032$ & $0.187$ & $0.126$\\
 & $22.25 - 22.50$ & $-0.001$ & $0.036$ & $0.099$ & $0.034$ & $0.184$ & $0.131$\\
 & $22.50 - 22.75$ & $-0.002$ & $0.036$ & $0.085$ & $0.035$ & $0.172$ & $0.123$\\
 & $22.75 - 23.00$ & $-0.002$ & $0.041$ & $0.093$ & $0.039$ & $0.171$ & $0.136$\\
 & $23.00 - 23.25$ & $+0.001$ & $0.045$ & $0.103$ & $0.042$ & $0.174$ & $0.146$\\
 & $23.25 - 23.50$ & $-0.000$ & $0.051$ & $0.126$ & $0.048$ & $0.186$ & $0.170$\\
 & $23.50 - 23.75$ & $-0.001$ & $0.059$ & $0.149$ & $0.058$ & $0.192$ & $0.197$\\
 & $23.75 - 24.00$ & $-0.002$ & $0.071$ & $0.198$ & $0.070$ & $0.212$ & $0.244$\\
 & $24.00 - 24.25$ & $-0.006$ & $0.086$ & $0.246$ & $0.086$ & $0.217$ & $0.286$\\
 & $24.25 - 24.50$ & $-0.008$ & $0.103$ & $0.275$ & $0.102$ & $0.212$ & $0.316$\\
 & $24.50 - 24.75$ & $-0.008$ & $0.124$ & $0.334$ & $0.121$ & $0.228$ & $0.364$\\
 & $24.75 - 25.00$ & $-0.011$ & $0.141$ & $0.362$ & $0.137$ & $0.222$ & $0.391$\\
\hline
 & $18.50 - 18.75$ & $+0.014$ & $0.087$ & $0.430$ & $0.424$ & $0.000$ & $0.435$\\
 & $18.75 - 19.00$ & $-0.002$ & $0.051$ & $0.062$ & $0.050$ & $0.078$ & $0.132$\\
 & $19.00 - 19.25$ & $-0.002$ & $0.038$ & $0.008$ & $0.046$ & $0.062$ & $0.076$\\
 & $19.25 - 19.50$ & $-0.001$ & $0.051$ & $0.044$ & $0.046$ & $0.091$ & $0.107$\\
 & $19.50 - 19.75$ & $+0.006$ & $0.054$ & $0.090$ & $0.049$ & $0.154$ & $0.153$\\
 & $19.75 - 20.00$ & $+0.013$ & $0.055$ & $0.109$ & $0.053$ & $0.164$ & $0.170$\\
 & $20.00 - 20.25$ & $+0.012$ & $0.052$ & $0.065$ & $0.056$ & $0.127$ & $0.142$\\
 & $20.25 - 20.50$ & $+0.004$ & $0.042$ & $0.051$ & $0.045$ & $0.138$ & $0.111$\\
 & $20.50 - 20.75$ & $+0.008$ & $0.045$ & $0.091$ & $0.041$ & $0.159$ & $0.125$\\
 & $20.75 - 21.00$ & $+0.006$ & $0.039$ & $0.069$ & $0.043$ & $0.151$ & $0.116$\\
 & $21.00 - 21.25$ & $+0.006$ & $0.039$ & $0.068$ & $0.041$ & $0.146$ & $0.116$\\
 & $21.25 - 21.50$ & $+0.005$ & $0.038$ & $0.074$ & $0.037$ & $0.170$ & $0.111$\\
	MLZ	& $21.50 - 21.75$ & $+0.006$ & $0.039$ & $0.069$ & $0.037$ & $0.135$ & $0.106$\\
 & $21.75 - 22.00$ & $+0.004$ & $0.039$ & $0.073$ & $0.037$ & $0.155$ & $0.112$\\
 & $22.00 - 22.25$ & $+0.002$ & $0.038$ & $0.075$ & $0.038$ & $0.163$ & $0.115$\\
 & $22.25 - 22.50$ & $+0.000$ & $0.039$ & $0.066$ & $0.039$ & $0.147$ & $0.109$\\
 & $22.50 - 22.75$ & $-0.001$ & $0.038$ & $0.067$ & $0.039$ & $0.143$ & $0.110$\\
 & $22.75 - 23.00$ & $-0.004$ & $0.040$ & $0.067$ & $0.040$ & $0.148$ & $0.112$\\
 & $23.00 - 23.25$ & $-0.002$ & $0.043$ & $0.088$ & $0.042$ & $0.163$ & $0.130$\\
 & $23.25 - 23.50$ & $-0.003$ & $0.046$ & $0.106$ & $0.044$ & $0.175$ & $0.148$\\
 & $23.50 - 23.75$ & $-0.004$ & $0.054$ & $0.138$ & $0.052$ & $0.194$ & $0.183$\\
 & $23.75 - 24.00$ & $-0.005$ & $0.065$ & $0.184$ & $0.062$ & $0.214$ & $0.225$\\
 & $24.00 - 24.25$ & $-0.005$ & $0.077$ & $0.230$ & $0.075$ & $0.229$ & $0.265$\\
 & $24.25 - 24.50$ & $-0.007$ & $0.092$ & $0.274$ & $0.095$ & $0.229$ & $0.306$\\
 & $24.50 - 24.75$ & $-0.009$ & $0.107$ & $0.317$ & $0.110$ & $0.234$ & $0.341$\\
 & $24.75 - 25.00$ & $-0.012$ & $0.119$ & $0.343$ & $0.126$ & $0.228$ & $0.364$\\
\hline
 & $18.50 - 18.75$ & $+0.004$ & $0.007$ & $0.000$ & $0.009$ & $0.379$ & $0.020$\\
 & $18.75 - 19.00$ & $-0.001$ & $0.034$ & $0.000$ & $0.031$ & $0.079$ & $0.055$\\
 & $19.00 - 19.25$ & $-0.001$ & $0.027$ & $0.008$ & $0.032$ & $0.094$ & $0.059$\\
 & $19.25 - 19.50$ & $-0.005$ & $0.039$ & $0.010$ & $0.039$ & $0.079$ & $0.070$\\
 & $19.50 - 19.75$ & $+0.001$ & $0.028$ & $0.040$ & $0.029$ & $0.127$ & $0.073$\\
 & $19.75 - 20.00$ & $+0.002$ & $0.031$ & $0.048$ & $0.031$ & $0.150$ & $0.084$\\
 & $20.00 - 20.25$ & $+0.004$ & $0.032$ & $0.046$ & $0.034$ & $0.144$ & $0.094$\\
 & $20.25 - 20.50$ & $+0.000$ & $0.027$ & $0.042$ & $0.027$ & $0.164$ & $0.075$\\
 & $20.50 - 20.75$ & $+0.001$ & $0.031$ & $0.059$ & $0.030$ & $0.166$ & $0.089$\\
 & $20.75 - 21.00$ & $-0.000$ & $0.027$ & $0.048$ & $0.027$ & $0.161$ & $0.078$\\
 & $21.00 - 21.25$ & $-0.000$ & $0.027$ & $0.051$ & $0.027$ & $0.162$ & $0.081$\\
 & $21.25 - 21.50$ & $-0.000$ & $0.027$ & $0.046$ & $0.028$ & $0.148$ & $0.072$\\
	NNPZ	& $21.50 - 21.75$ & $-0.000$ & $0.027$ & $0.039$ & $0.026$ & $0.149$ & $0.071$\\
 & $21.75 - 22.00$ & $-0.001$ & $0.028$ & $0.051$ & $0.028$ & $0.153$ & $0.083$\\
 & $22.00 - 22.25$ & $-0.001$ & $0.029$ & $0.056$ & $0.028$ & $0.163$ & $0.088$\\
 & $22.25 - 22.50$ & $-0.001$ & $0.030$ & $0.061$ & $0.030$ & $0.161$ & $0.092$\\
 & $22.50 - 22.75$ & $-0.001$ & $0.032$ & $0.056$ & $0.031$ & $0.148$ & $0.091$\\
 & $22.75 - 23.00$ & $-0.002$ & $0.035$ & $0.061$ & $0.035$ & $0.151$ & $0.101$\\
 & $23.00 - 23.25$ & $-0.002$ & $0.038$ & $0.076$ & $0.038$ & $0.164$ & $0.119$\\
 & $23.25 - 23.50$ & $-0.003$ & $0.042$ & $0.106$ & $0.040$ & $0.187$ & $0.145$\\
 & $23.50 - 23.75$ & $-0.003$ & $0.050$ & $0.138$ & $0.048$ & $0.208$ & $0.178$\\
 & $23.75 - 24.00$ & $-0.004$ & $0.061$ & $0.185$ & $0.057$ & $0.227$ & $0.220$\\
 & $24.00 - 24.25$ & $-0.007$ & $0.074$ & $0.231$ & $0.071$ & $0.239$ & $0.264$\\
 & $24.25 - 24.50$ & $-0.011$ & $0.089$ & $0.275$ & $0.090$ & $0.239$ & $0.303$\\
 & $24.50 - 24.75$ & $-0.015$ & $0.104$ & $0.319$ & $0.112$ & $0.225$ & $0.337$\\
 & $24.75 - 25.00$ & $-0.017$ & $0.115$ & $0.333$ & $0.122$ & $0.219$ & $0.354$\\
\hline

    \hline
  \end{longtable}
}

\begin{figure*}
  \begin{center}
    \includegraphics[width=7cm]{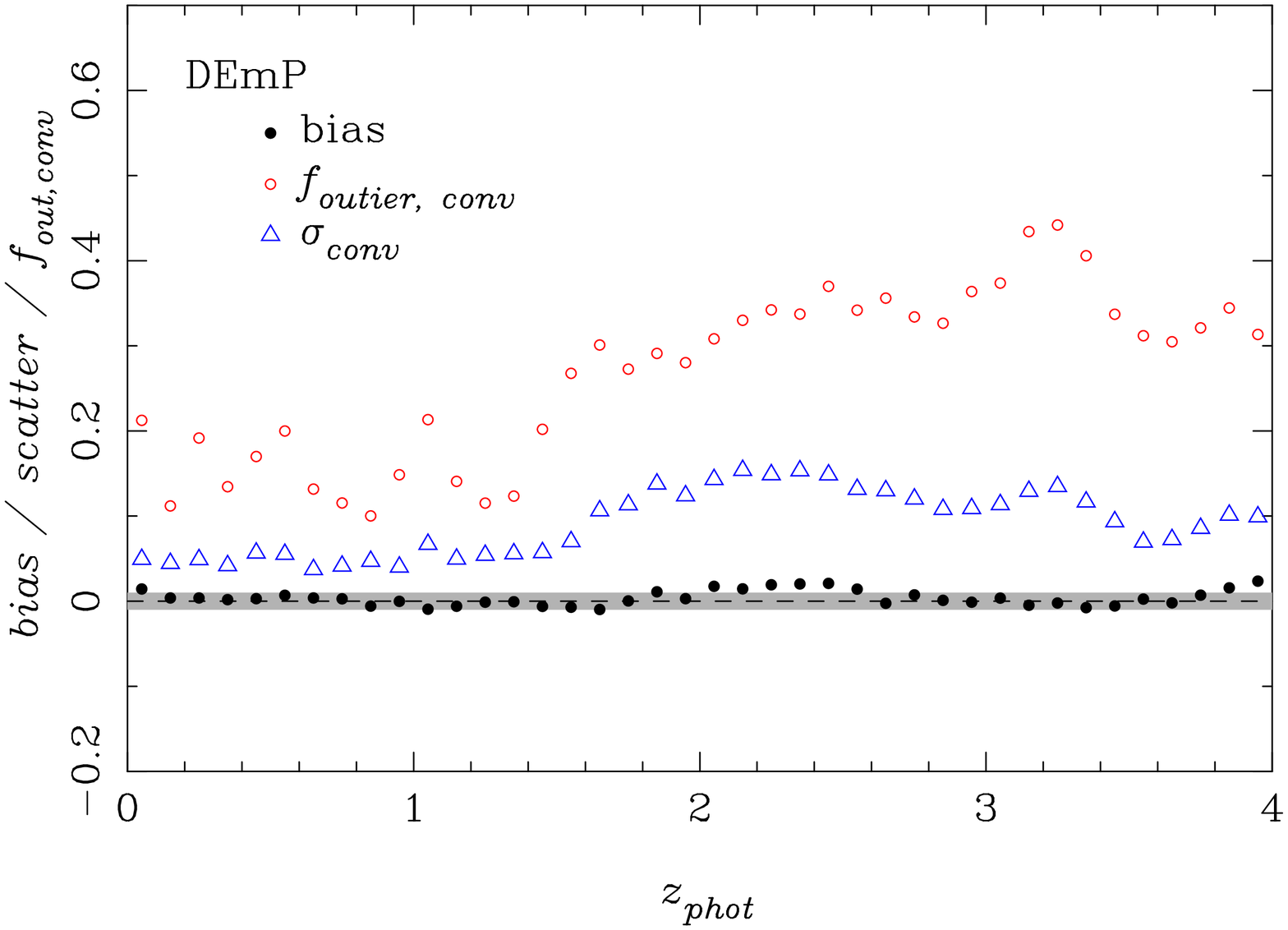}\hspace{0.5cm}
    \includegraphics[width=7cm]{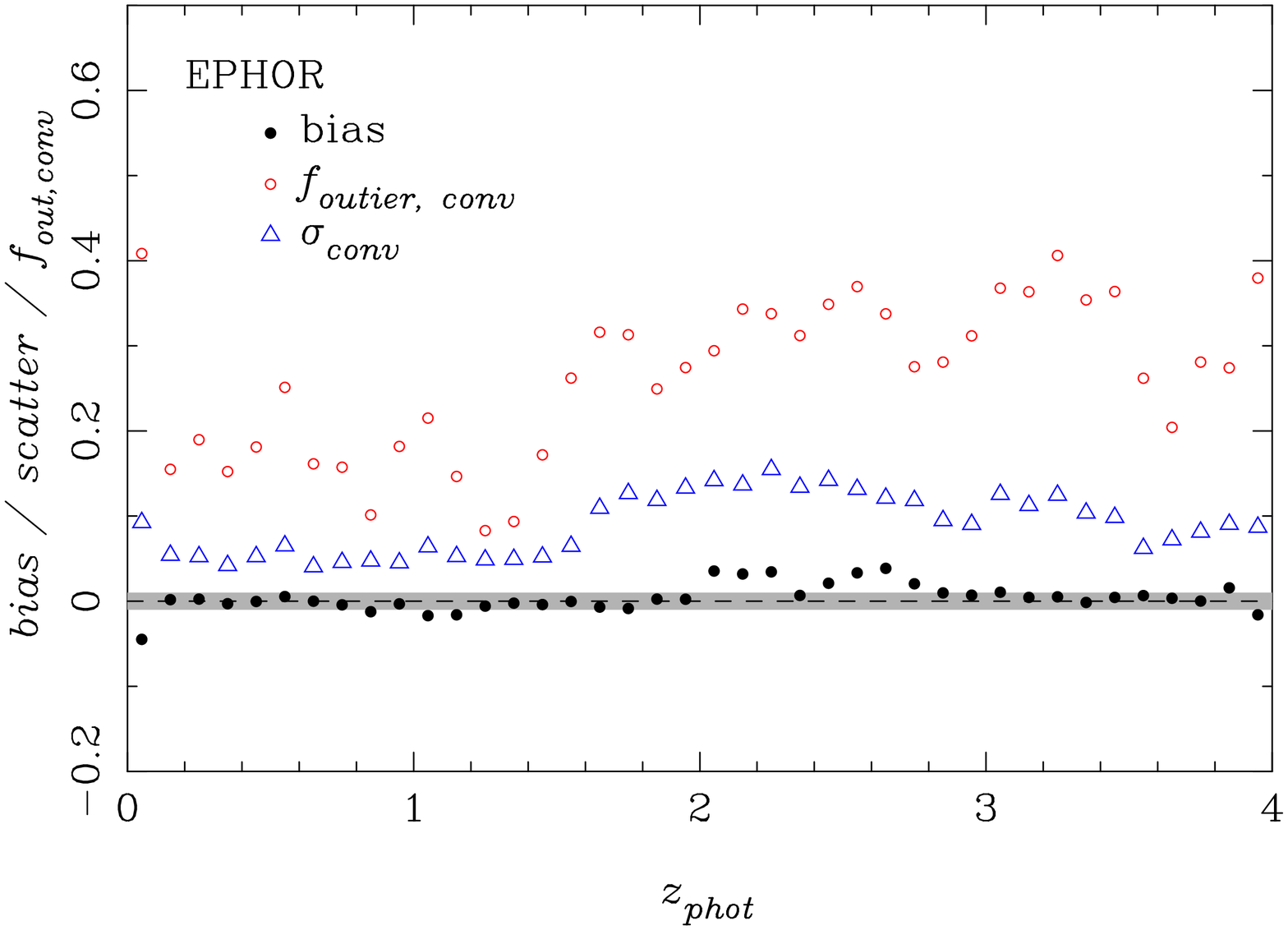}\\\vspace{0.5cm}
    \includegraphics[width=7cm]{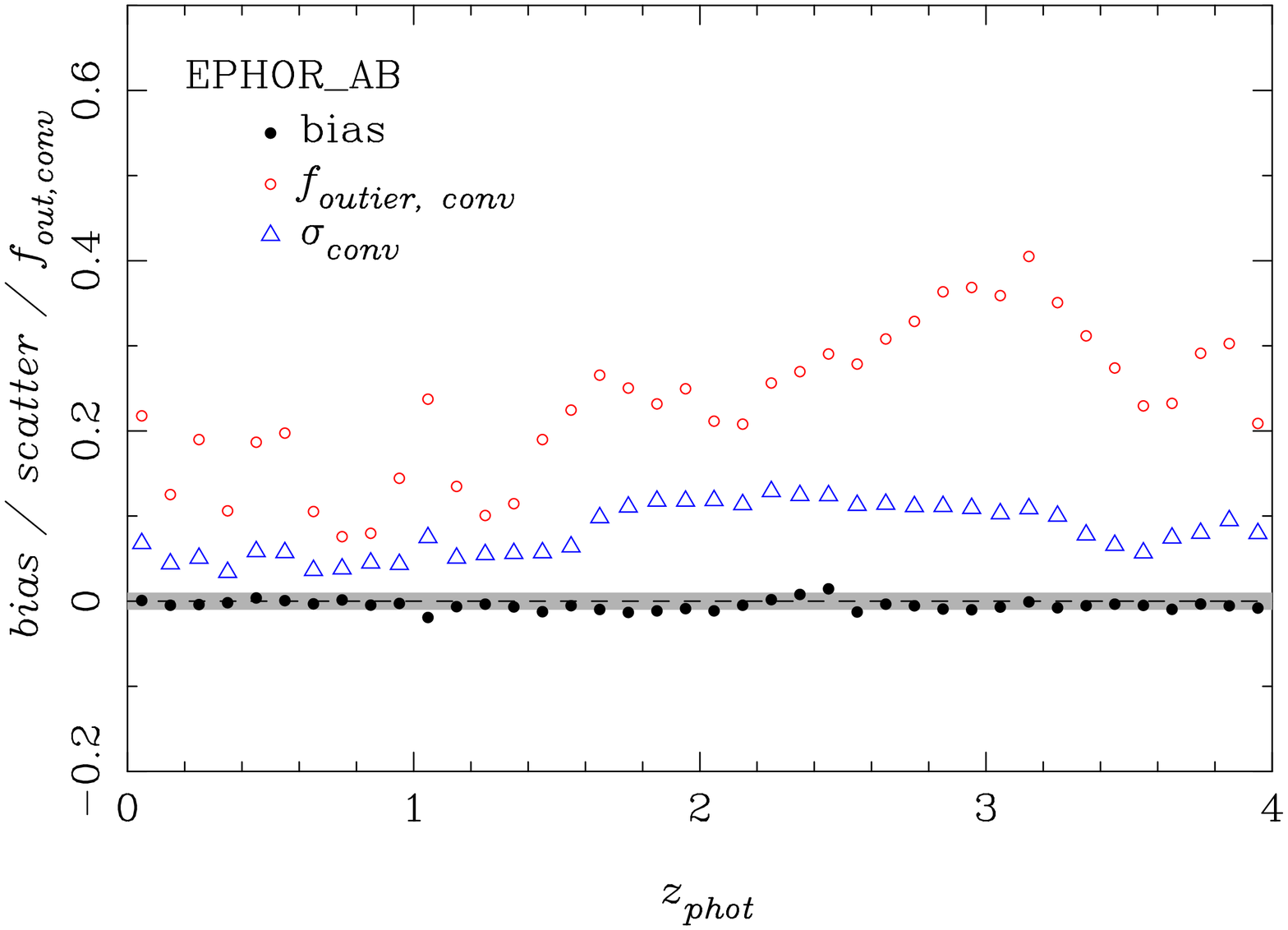}\hspace{0.5cm}
    \includegraphics[width=7cm]{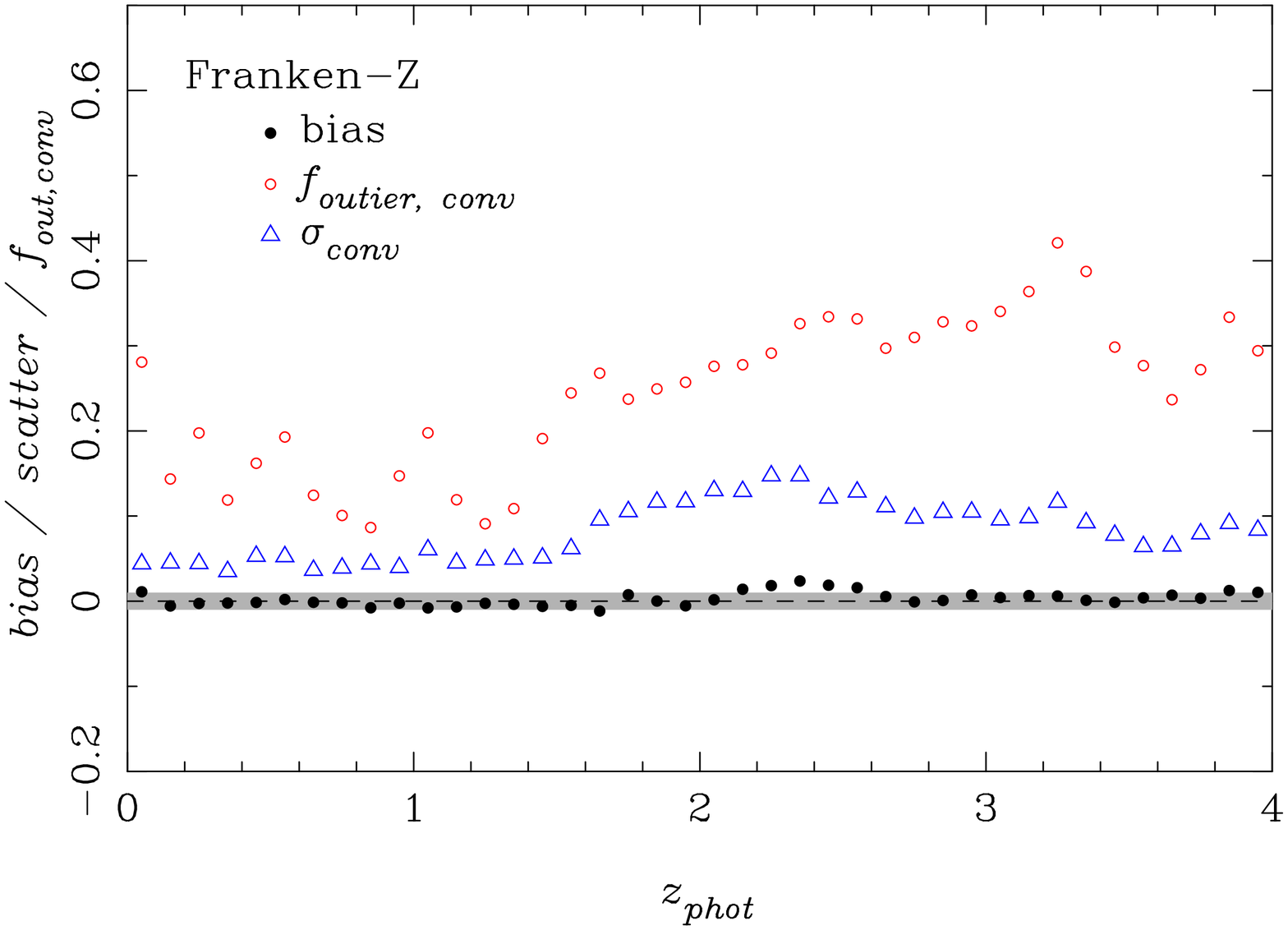}\\\vspace{0.5cm}
    \includegraphics[width=7cm]{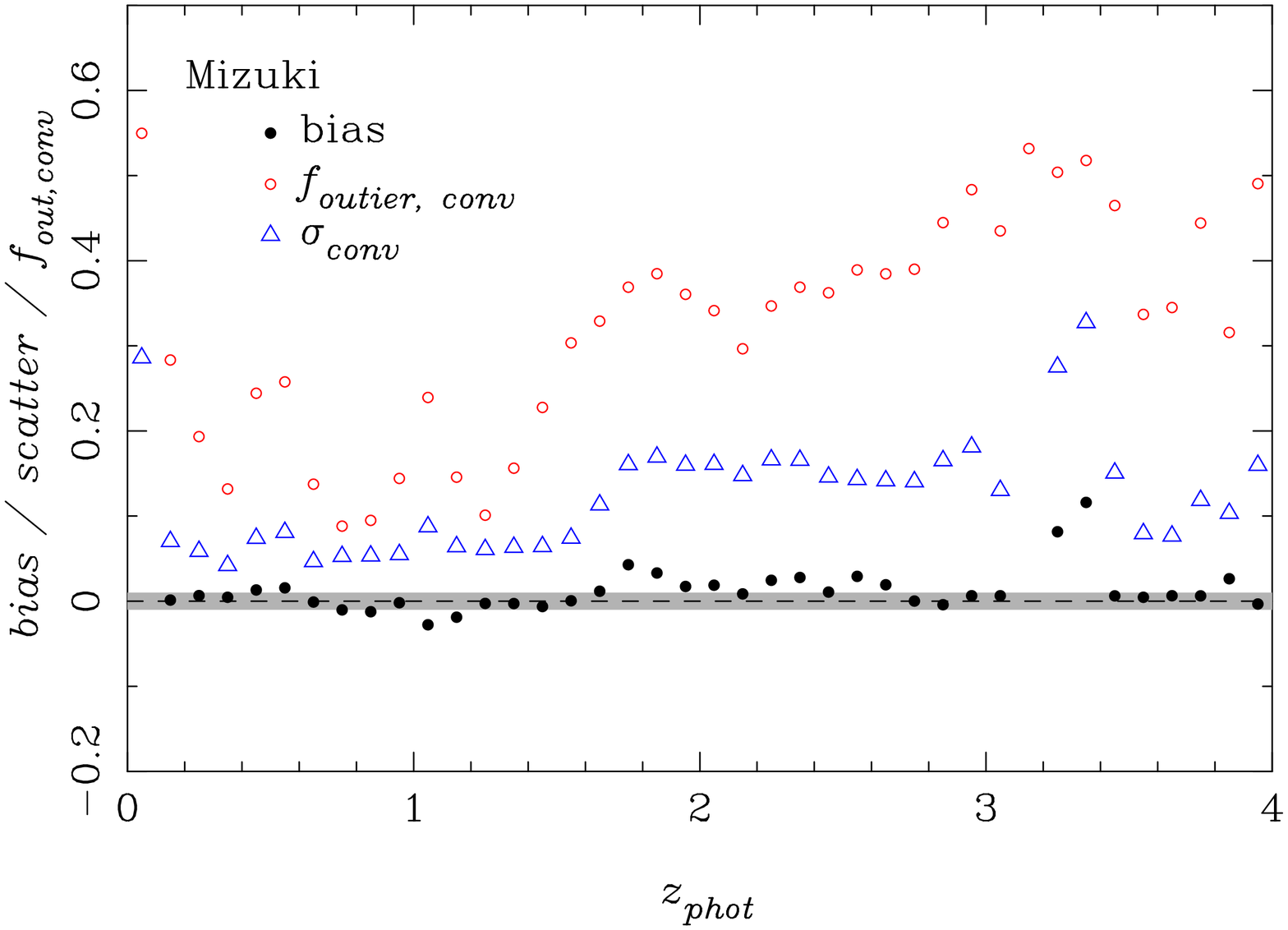}\hspace{0.5cm}
    \includegraphics[width=7cm]{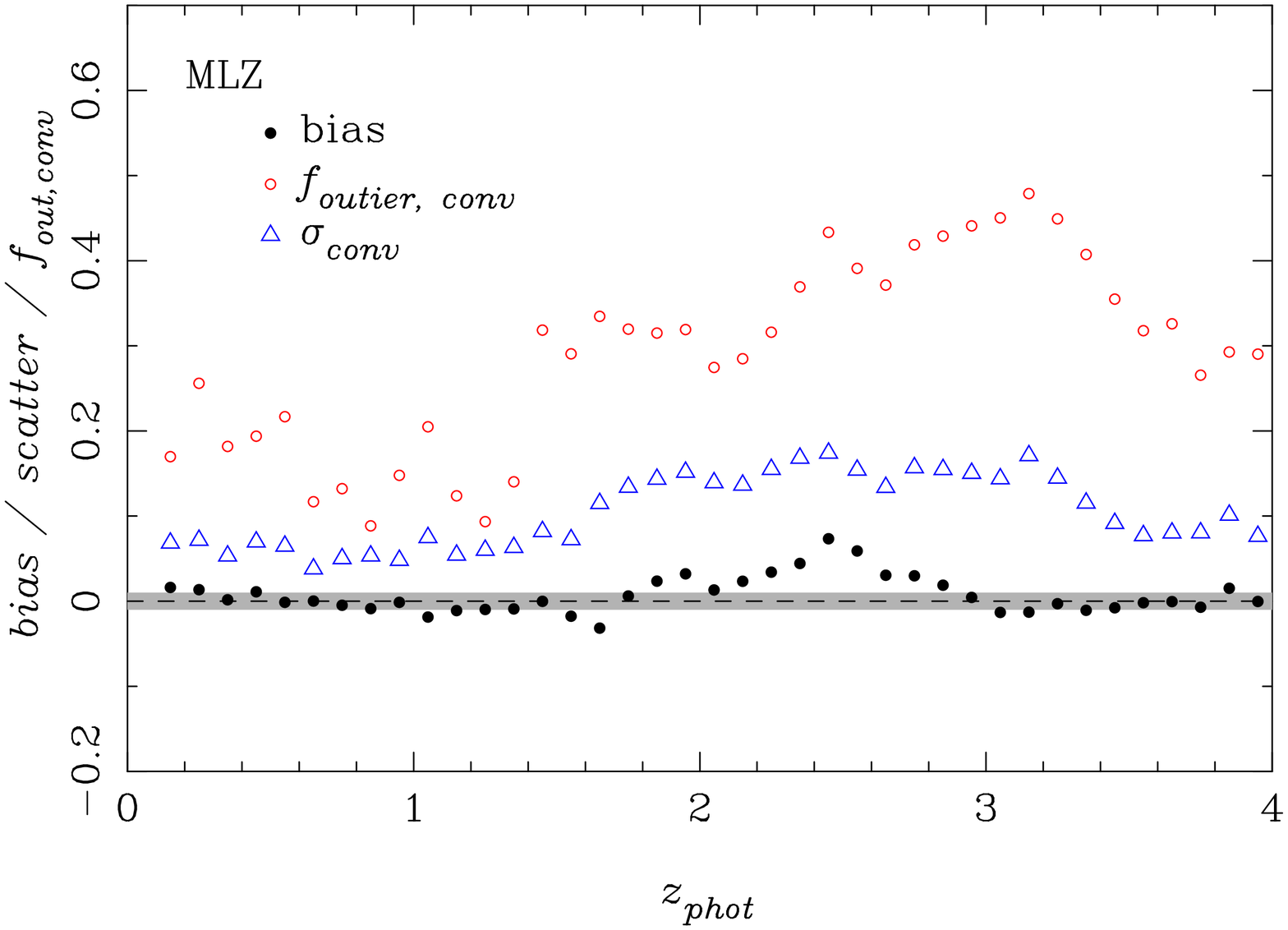}\\\vspace{0.5cm}
    \includegraphics[width=7cm]{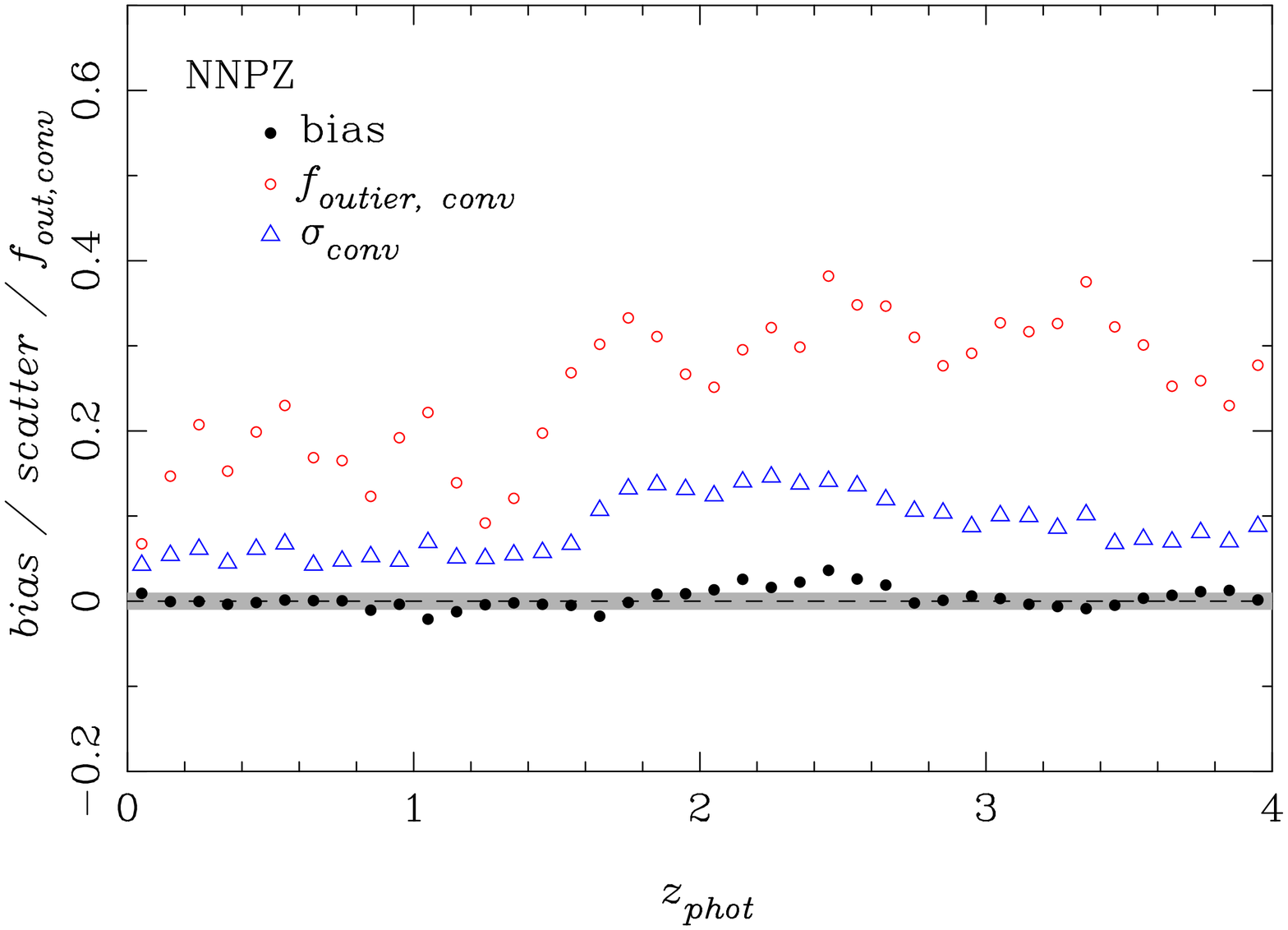}
  \end{center}
  \caption{
    Same as Fig.~\ref{fig:stat_mag} but as a function of $z_{phot}$.
 }
 \label{fig:stat_zphot}
\end{figure*}

Fig.~\ref{fig:stat_zphot} shows the same metrics but as a function of $z_{phot}$.
Our performance is poor both at low-$z$ ($z\lesssim0.2$) and high-$z$ ($z\gtrsim1.5$) ends.
This is expected because our filter set ($grizy$) does not straddle the 4000\AA\ break
at these redshifts.  We are not able to break the degeneracy between $z\sim0$ and
$z\sim2$ solutions (i.e., 4000\AA\ break and Lyman break degeneracy), resulting in poor
performance at low-$z$.  At $z\gtrsim1.5$,
we probe only the featureless UV continuum and it makes it difficult to obtain good photo-$z$'s there.
The Lyman break comes in the $g$-band and some codes show improvements at $z\gtrsim3.5$,.
In the good redshift range ($0.2\lesssim z \lesssim1.5$),
our photo-$z$'s are fair -- the outlier rate is about 15\% and the scatter is
about 0.05.  Note that these numbers are for all galaxies down to $i=25$.  If we use
brighter galaxies with $i<24$, the numbers improve to about 8\% and 0.04.  Our photo-$z$'s should thus
be sufficient to enable many of the science goals in HSC-SSP.
Also, we can clip potential photo-$z$ outliers to further improve the accuracy (see Section \ref{ssec:riskcut}).
Table~\ref{tab:stat_zphot} summarizes the statistical measures for all the codes.

{\scriptsize
  \begin{longtable}{llcccccc}
    \caption{
      Photo-$z$ statistics for all the codes as a function of $z_{phot}$.  The number are for all
      galaxies down to $i=25$.
    }
    \label{tab:stat_zphot}
    \hline
    Code   &  $z_{phot}$  &  bias  &  $\sigma_{conv}$ & $f_{outlier,conv}$ & $\sigma$ & $f_{outlier}$ & $<L(\Delta z))>$\\
    \endfirsthead
    \endhead
    \hline
    \endfoot
    \hline
    \endlastfoot
    \hline
     & $0.00 - 0.10$ & $+0.014$ & $0.049$ & $0.213$ & $0.031$ & $0.260$ & $0.221$\\
 & $0.10 - 0.20$ & $+0.004$ & $0.044$ & $0.112$ & $0.042$ & $0.168$ & $0.157$\\
 & $0.20 - 0.30$ & $+0.004$ & $0.049$ & $0.192$ & $0.046$ & $0.269$ & $0.219$\\
 & $0.30 - 0.40$ & $+0.002$ & $0.042$ & $0.135$ & $0.042$ & $0.238$ & $0.181$\\
 & $0.40 - 0.50$ & $+0.003$ & $0.057$ & $0.170$ & $0.053$ & $0.220$ & $0.213$\\
 & $0.50 - 0.60$ & $+0.007$ & $0.055$ & $0.200$ & $0.047$ & $0.256$ & $0.231$\\
 & $0.60 - 0.70$ & $+0.004$ & $0.037$ & $0.132$ & $0.033$ & $0.228$ & $0.163$\\
 & $0.70 - 0.80$ & $+0.003$ & $0.041$ & $0.115$ & $0.042$ & $0.210$ & $0.154$\\
 & $0.80 - 0.90$ & $-0.006$ & $0.047$ & $0.100$ & $0.049$ & $0.171$ & $0.149$\\
 & $0.90 - 1.00$ & $-0.000$ & $0.040$ & $0.148$ & $0.038$ & $0.239$ & $0.165$\\
 & $1.00 - 1.10$ & $-0.009$ & $0.067$ & $0.213$ & $0.090$ & $0.162$ & $0.225$\\
 & $1.10 - 1.20$ & $-0.006$ & $0.050$ & $0.141$ & $0.052$ & $0.203$ & $0.171$\\
	DEmP	& $1.20 - 1.30$ & $-0.001$ & $0.054$ & $0.115$ & $0.055$ & $0.164$ & $0.160$\\
 & $1.30 - 1.40$ & $-0.001$ & $0.056$ & $0.123$ & $0.061$ & $0.163$ & $0.172$\\
 & $1.40 - 1.50$ & $-0.006$ & $0.057$ & $0.202$ & $0.062$ & $0.239$ & $0.218$\\
 & $1.50 - 1.60$ & $-0.007$ & $0.070$ & $0.268$ & $0.091$ & $0.231$ & $0.268$\\
 & $1.60 - 1.70$ & $-0.010$ & $0.106$ & $0.301$ & $0.142$ & $0.115$ & $0.316$\\
 & $1.70 - 1.80$ & $+0.000$ & $0.113$ & $0.273$ & $0.132$ & $0.136$ & $0.319$\\
 & $1.80 - 1.90$ & $+0.011$ & $0.138$ & $0.291$ & $0.140$ & $0.119$ & $0.345$\\
 & $1.90 - 2.00$ & $+0.003$ & $0.124$ & $0.280$ & $0.118$ & $0.117$ & $0.323$\\
 & $2.00 - 2.10$ & $+0.017$ & $0.143$ & $0.308$ & $0.130$ & $0.134$ & $0.357$\\
 & $2.10 - 2.20$ & $+0.014$ & $0.154$ & $0.330$ & $0.139$ & $0.119$ & $0.373$\\
 & $2.20 - 2.30$ & $+0.019$ & $0.149$ & $0.342$ & $0.141$ & $0.141$ & $0.380$\\
 & $2.30 - 2.40$ & $+0.020$ & $0.153$ & $0.337$ & $0.140$ & $0.166$ & $0.387$\\
 & $2.40 - 2.50$ & $+0.021$ & $0.149$ & $0.370$ & $0.150$ & $0.186$ & $0.402$\\
 & $2.50 - 2.60$ & $+0.014$ & $0.132$ & $0.342$ & $0.122$ & $0.250$ & $0.388$\\
 & $2.60 - 2.70$ & $-0.002$ & $0.130$ & $0.356$ & $0.101$ & $0.314$ & $0.407$\\
 & $2.70 - 2.80$ & $+0.007$ & $0.120$ & $0.334$ & $0.086$ & $0.307$ & $0.382$\\
 & $2.80 - 2.90$ & $+0.001$ & $0.108$ & $0.327$ & $0.073$ & $0.328$ & $0.386$\\
 & $2.90 - 3.00$ & $-0.001$ & $0.109$ & $0.364$ & $0.067$ & $0.373$ & $0.406$\\
 & $3.00 - 3.10$ & $+0.004$ & $0.114$ & $0.374$ & $0.065$ & $0.382$ & $0.430$\\
 & $3.10 - 3.20$ & $-0.005$ & $0.129$ & $0.434$ & $0.059$ & $0.460$ & $0.482$\\
 & $3.20 - 3.30$ & $-0.002$ & $0.135$ & $0.442$ & $0.059$ & $0.464$ & $0.483$\\
 & $3.30 - 3.40$ & $-0.008$ & $0.117$ & $0.406$ & $0.051$ & $0.435$ & $0.450$\\
 & $3.40 - 3.50$ & $-0.006$ & $0.094$ & $0.337$ & $0.053$ & $0.372$ & $0.391$\\
 & $3.50 - 3.60$ & $+0.002$ & $0.070$ & $0.312$ & $0.044$ & $0.361$ & $0.353$\\
 & $3.60 - 3.70$ & $-0.002$ & $0.073$ & $0.305$ & $0.046$ & $0.331$ & $0.356$\\
 & $3.70 - 3.80$ & $+0.007$ & $0.086$ & $0.321$ & $0.052$ & $0.337$ & $0.374$\\
 & $3.80 - 3.90$ & $+0.016$ & $0.101$ & $0.344$ & $0.057$ & $0.357$ & $0.401$\\
 & $3.90 - 4.00$ & $+0.024$ & $0.099$ & $0.313$ & $0.058$ & $0.322$ & $0.377$\\
\hline
 & $0.00 - 0.10$ & $-0.045$ & $0.092$ & $0.408$ & $0.338$ & $0.003$ & $0.420$\\
 & $0.10 - 0.20$ & $+0.002$ & $0.054$ & $0.155$ & $0.048$ & $0.202$ & $0.207$\\
 & $0.20 - 0.30$ & $+0.002$ & $0.052$ & $0.190$ & $0.049$ & $0.268$ & $0.219$\\
 & $0.30 - 0.40$ & $-0.003$ & $0.042$ & $0.152$ & $0.039$ & $0.253$ & $0.194$\\
 & $0.40 - 0.50$ & $-0.000$ & $0.052$ & $0.181$ & $0.046$ & $0.248$ & $0.216$\\
 & $0.50 - 0.60$ & $+0.005$ & $0.065$ & $0.251$ & $0.050$ & $0.296$ & $0.274$\\
 & $0.60 - 0.70$ & $+0.000$ & $0.040$ & $0.161$ & $0.034$ & $0.254$ & $0.188$\\
 & $0.70 - 0.80$ & $-0.004$ & $0.046$ & $0.157$ & $0.044$ & $0.237$ & $0.188$\\
 & $0.80 - 0.90$ & $-0.012$ & $0.047$ & $0.101$ & $0.046$ & $0.176$ & $0.149$\\
 & $0.90 - 1.00$ & $-0.003$ & $0.045$ & $0.182$ & $0.040$ & $0.262$ & $0.190$\\
 & $1.00 - 1.10$ & $-0.017$ & $0.064$ & $0.215$ & $0.090$ & $0.159$ & $0.223$\\
 & $1.10 - 1.20$ & $-0.016$ & $0.053$ & $0.147$ & $0.052$ & $0.192$ & $0.178$\\
	EPHOR	& $1.20 - 1.30$ & $-0.006$ & $0.049$ & $0.083$ & $0.048$ & $0.139$ & $0.132$\\
 & $1.30 - 1.40$ & $-0.002$ & $0.049$ & $0.094$ & $0.051$ & $0.155$ & $0.143$\\
 & $1.40 - 1.50$ & $-0.004$ & $0.052$ & $0.172$ & $0.051$ & $0.237$ & $0.193$\\
 & $1.50 - 1.60$ & $-0.000$ & $0.065$ & $0.262$ & $0.067$ & $0.280$ & $0.262$\\
 & $1.60 - 1.70$ & $-0.007$ & $0.109$ & $0.316$ & $0.152$ & $0.111$ & $0.327$\\
 & $1.70 - 1.80$ & $-0.009$ & $0.127$ & $0.313$ & $0.147$ & $0.114$ & $0.341$\\
 & $1.80 - 1.90$ & $+0.002$ & $0.119$ & $0.249$ & $0.122$ & $0.112$ & $0.304$\\
 & $1.90 - 2.00$ & $+0.002$ & $0.133$ & $0.274$ & $0.124$ & $0.098$ & $0.322$\\
 & $2.00 - 2.10$ & $+0.035$ & $0.142$ & $0.294$ & $0.133$ & $0.100$ & $0.344$\\
 & $2.10 - 2.20$ & $+0.032$ & $0.137$ & $0.343$ & $0.140$ & $0.110$ & $0.354$\\
 & $2.20 - 2.30$ & $+0.034$ & $0.155$ & $0.338$ & $0.148$ & $0.120$ & $0.385$\\
 & $2.30 - 2.40$ & $+0.007$ & $0.134$ & $0.312$ & $0.128$ & $0.204$ & $0.368$\\
 & $2.40 - 2.50$ & $+0.021$ & $0.142$ & $0.349$ & $0.138$ & $0.201$ & $0.388$\\
 & $2.50 - 2.60$ & $+0.033$ & $0.132$ & $0.370$ & $0.143$ & $0.212$ & $0.389$\\
 & $2.60 - 2.70$ & $+0.039$ & $0.121$ & $0.338$ & $0.127$ & $0.221$ & $0.377$\\
 & $2.70 - 2.80$ & $+0.020$ & $0.118$ & $0.275$ & $0.092$ & $0.221$ & $0.344$\\
 & $2.80 - 2.90$ & $+0.010$ & $0.095$ & $0.281$ & $0.076$ & $0.270$ & $0.343$\\
 & $2.90 - 3.00$ & $+0.007$ & $0.090$ & $0.312$ & $0.069$ & $0.315$ & $0.353$\\
 & $3.00 - 3.10$ & $+0.011$ & $0.126$ & $0.368$ & $0.080$ & $0.354$ & $0.423$\\
 & $3.10 - 3.20$ & $+0.004$ & $0.113$ & $0.364$ & $0.070$ & $0.370$ & $0.419$\\
 & $3.20 - 3.30$ & $+0.005$ & $0.125$ & $0.406$ & $0.069$ & $0.409$ & $0.453$\\
 & $3.30 - 3.40$ & $-0.001$ & $0.104$ & $0.354$ & $0.062$ & $0.379$ & $0.412$\\
 & $3.40 - 3.50$ & $+0.004$ & $0.099$ & $0.364$ & $0.054$ & $0.405$ & $0.410$\\
 & $3.50 - 3.60$ & $+0.007$ & $0.062$ & $0.262$ & $0.042$ & $0.316$ & $0.304$\\
 & $3.60 - 3.70$ & $+0.004$ & $0.072$ & $0.204$ & $0.053$ & $0.230$ & $0.276$\\
 & $3.70 - 3.80$ & $+0.000$ & $0.081$ & $0.281$ & $0.057$ & $0.299$ & $0.340$\\
 & $3.80 - 3.90$ & $+0.016$ & $0.091$ & $0.274$ & $0.058$ & $0.289$ & $0.342$\\
 & $3.90 - 4.00$ & $-0.016$ & $0.087$ & $0.380$ & $0.039$ & $0.441$ & $0.419$\\
\hline
 & $0.00 - 0.10$ & $+0.001$ & $0.068$ & $0.218$ & $0.054$ & $0.248$ & $0.254$\\
 & $0.10 - 0.20$ & $-0.005$ & $0.044$ & $0.125$ & $0.040$ & $0.185$ & $0.168$\\
 & $0.20 - 0.30$ & $-0.004$ & $0.051$ & $0.190$ & $0.048$ & $0.266$ & $0.221$\\
 & $0.30 - 0.40$ & $-0.002$ & $0.034$ & $0.106$ & $0.032$ & $0.206$ & $0.140$\\
 & $0.40 - 0.50$ & $+0.004$ & $0.058$ & $0.187$ & $0.052$ & $0.232$ & $0.226$\\
 & $0.50 - 0.60$ & $+0.001$ & $0.057$ & $0.198$ & $0.048$ & $0.254$ & $0.229$\\
 & $0.60 - 0.70$ & $-0.003$ & $0.036$ & $0.105$ & $0.033$ & $0.202$ & $0.139$\\
 & $0.70 - 0.80$ & $+0.001$ & $0.038$ & $0.076$ & $0.041$ & $0.166$ & $0.124$\\
 & $0.80 - 0.90$ & $-0.005$ & $0.045$ & $0.080$ & $0.044$ & $0.152$ & $0.129$\\
 & $0.90 - 1.00$ & $-0.003$ & $0.043$ & $0.144$ & $0.041$ & $0.225$ & $0.166$\\
 & $1.00 - 1.10$ & $-0.019$ & $0.075$ & $0.237$ & $0.106$ & $0.108$ & $0.242$\\
 & $1.10 - 1.20$ & $-0.006$ & $0.051$ & $0.135$ & $0.053$ & $0.194$ & $0.167$\\
	EPHOR\_AB	& $1.20 - 1.30$ & $-0.004$ & $0.055$ & $0.101$ & $0.053$ & $0.146$ & $0.149$\\
 & $1.30 - 1.40$ & $-0.007$ & $0.056$ & $0.115$ & $0.058$ & $0.159$ & $0.164$\\
 & $1.40 - 1.50$ & $-0.012$ & $0.057$ & $0.190$ & $0.054$ & $0.255$ & $0.209$\\
 & $1.50 - 1.60$ & $-0.005$ & $0.064$ & $0.225$ & $0.078$ & $0.214$ & $0.236$\\
 & $1.60 - 1.70$ & $-0.010$ & $0.098$ & $0.266$ & $0.128$ & $0.132$ & $0.288$\\
 & $1.70 - 1.80$ & $-0.013$ & $0.111$ & $0.250$ & $0.126$ & $0.104$ & $0.297$\\
 & $1.80 - 1.90$ & $-0.011$ & $0.118$ & $0.232$ & $0.118$ & $0.091$ & $0.293$\\
 & $1.90 - 2.00$ & $-0.009$ & $0.118$ & $0.250$ & $0.112$ & $0.092$ & $0.296$\\
 & $2.00 - 2.10$ & $-0.011$ & $0.118$ & $0.211$ & $0.110$ & $0.068$ & $0.281$\\
 & $2.10 - 2.20$ & $-0.005$ & $0.114$ & $0.208$ & $0.106$ & $0.095$ & $0.282$\\
 & $2.20 - 2.30$ & $+0.002$ & $0.129$ & $0.256$ & $0.124$ & $0.119$ & $0.319$\\
 & $2.30 - 2.40$ & $+0.008$ & $0.124$ & $0.270$ & $0.134$ & $0.110$ & $0.329$\\
 & $2.40 - 2.50$ & $+0.014$ & $0.124$ & $0.290$ & $0.134$ & $0.176$ & $0.348$\\
 & $2.50 - 2.60$ & $-0.013$ & $0.113$ & $0.279$ & $0.099$ & $0.252$ & $0.344$\\
 & $2.60 - 2.70$ & $-0.003$ & $0.114$ & $0.308$ & $0.088$ & $0.287$ & $0.370$\\
 & $2.70 - 2.80$ & $-0.006$ & $0.111$ & $0.329$ & $0.072$ & $0.336$ & $0.384$\\
 & $2.80 - 2.90$ & $-0.009$ & $0.111$ & $0.363$ & $0.067$ & $0.390$ & $0.414$\\
 & $2.90 - 3.00$ & $-0.010$ & $0.109$ & $0.369$ & $0.064$ & $0.383$ & $0.416$\\
 & $3.00 - 3.10$ & $-0.007$ & $0.103$ & $0.359$ & $0.058$ & $0.384$ & $0.418$\\
 & $3.10 - 3.20$ & $-0.001$ & $0.109$ & $0.405$ & $0.057$ & $0.431$ & $0.448$\\
 & $3.20 - 3.30$ & $-0.008$ & $0.100$ & $0.351$ & $0.058$ & $0.385$ & $0.406$\\
 & $3.30 - 3.40$ & $-0.005$ & $0.078$ & $0.312$ & $0.050$ & $0.360$ & $0.364$\\
 & $3.40 - 3.50$ & $-0.003$ & $0.066$ & $0.274$ & $0.046$ & $0.308$ & $0.322$\\
 & $3.50 - 3.60$ & $-0.005$ & $0.057$ & $0.230$ & $0.043$ & $0.280$ & $0.278$\\
 & $3.60 - 3.70$ & $-0.010$ & $0.074$ & $0.232$ & $0.050$ & $0.255$ & $0.293$\\
 & $3.70 - 3.80$ & $-0.003$ & $0.080$ & $0.291$ & $0.053$ & $0.318$ & $0.339$\\
 & $3.80 - 3.90$ & $-0.005$ & $0.095$ & $0.303$ & $0.058$ & $0.320$ & $0.368$\\
 & $3.90 - 4.00$ & $-0.008$ & $0.080$ & $0.209$ & $0.054$ & $0.215$ & $0.283$\\
\hline
 & $0.00 - 0.10$ & $+0.011$ & $0.044$ & $0.281$ & $0.021$ & $0.402$ & $0.282$\\
 & $0.10 - 0.20$ & $-0.006$ & $0.045$ & $0.144$ & $0.039$ & $0.207$ & $0.182$\\
 & $0.20 - 0.30$ & $-0.003$ & $0.044$ & $0.198$ & $0.036$ & $0.303$ & $0.225$\\
 & $0.30 - 0.40$ & $-0.002$ & $0.035$ & $0.119$ & $0.033$ & $0.233$ & $0.157$\\
 & $0.40 - 0.50$ & $-0.001$ & $0.053$ & $0.162$ & $0.048$ & $0.230$ & $0.204$\\
 & $0.50 - 0.60$ & $+0.002$ & $0.052$ & $0.193$ & $0.045$ & $0.261$ & $0.223$\\
 & $0.60 - 0.70$ & $-0.001$ & $0.036$ & $0.125$ & $0.033$ & $0.225$ & $0.156$\\
 & $0.70 - 0.80$ & $-0.002$ & $0.039$ & $0.101$ & $0.040$ & $0.196$ & $0.140$\\
 & $0.80 - 0.90$ & $-0.008$ & $0.044$ & $0.086$ & $0.044$ & $0.169$ & $0.135$\\
 & $0.90 - 1.00$ & $-0.002$ & $0.040$ & $0.147$ & $0.037$ & $0.236$ & $0.162$\\
 & $1.00 - 1.10$ & $-0.008$ & $0.060$ & $0.198$ & $0.074$ & $0.195$ & $0.211$\\
 & $1.10 - 1.20$ & $-0.007$ & $0.045$ & $0.119$ & $0.045$ & $0.198$ & $0.153$\\
	Franken-Z	& $1.20 - 1.30$ & $-0.003$ & $0.049$ & $0.091$ & $0.049$ & $0.151$ & $0.139$\\
 & $1.30 - 1.40$ & $-0.004$ & $0.050$ & $0.109$ & $0.052$ & $0.175$ & $0.153$\\
 & $1.40 - 1.50$ & $-0.006$ & $0.051$ & $0.191$ & $0.050$ & $0.262$ & $0.206$\\
 & $1.50 - 1.60$ & $-0.005$ & $0.062$ & $0.245$ & $0.062$ & $0.275$ & $0.251$\\
 & $1.60 - 1.70$ & $-0.012$ & $0.095$ & $0.268$ & $0.122$ & $0.145$ & $0.288$\\
 & $1.70 - 1.80$ & $+0.008$ & $0.105$ & $0.237$ & $0.120$ & $0.139$ & $0.295$\\
 & $1.80 - 1.90$ & $+0.000$ & $0.116$ & $0.249$ & $0.122$ & $0.116$ & $0.302$\\
 & $1.90 - 2.00$ & $-0.005$ & $0.117$ & $0.257$ & $0.117$ & $0.101$ & $0.301$\\
 & $2.00 - 2.10$ & $+0.002$ & $0.130$ & $0.276$ & $0.120$ & $0.115$ & $0.321$\\
 & $2.10 - 2.20$ & $+0.014$ & $0.129$ & $0.278$ & $0.121$ & $0.115$ & $0.329$\\
 & $2.20 - 2.30$ & $+0.018$ & $0.147$ & $0.291$ & $0.138$ & $0.119$ & $0.353$\\
 & $2.30 - 2.40$ & $+0.024$ & $0.147$ & $0.326$ & $0.146$ & $0.144$ & $0.378$\\
 & $2.40 - 2.50$ & $+0.019$ & $0.121$ & $0.334$ & $0.132$ & $0.203$ & $0.356$\\
 & $2.50 - 2.60$ & $+0.016$ & $0.128$ & $0.332$ & $0.132$ & $0.239$ & $0.380$\\
 & $2.60 - 2.70$ & $+0.005$ & $0.111$ & $0.297$ & $0.093$ & $0.270$ & $0.354$\\
 & $2.70 - 2.80$ & $-0.001$ & $0.098$ & $0.310$ & $0.075$ & $0.310$ & $0.356$\\
 & $2.80 - 2.90$ & $+0.001$ & $0.105$ & $0.328$ & $0.070$ & $0.338$ & $0.377$\\
 & $2.90 - 3.00$ & $+0.007$ & $0.105$ & $0.324$ & $0.072$ & $0.324$ & $0.378$\\
 & $3.00 - 3.10$ & $+0.004$ & $0.096$ & $0.340$ & $0.059$ & $0.354$ & $0.393$\\
 & $3.10 - 3.20$ & $+0.007$ & $0.098$ & $0.364$ & $0.055$ & $0.384$ & $0.413$\\
 & $3.20 - 3.30$ & $+0.006$ & $0.116$ & $0.421$ & $0.053$ & $0.444$ & $0.465$\\
 & $3.30 - 3.40$ & $+0.001$ & $0.092$ & $0.387$ & $0.047$ & $0.430$ & $0.421$\\
 & $3.40 - 3.50$ & $-0.001$ & $0.078$ & $0.299$ & $0.049$ & $0.348$ & $0.350$\\
 & $3.50 - 3.60$ & $+0.004$ & $0.064$ & $0.277$ & $0.042$ & $0.324$ & $0.319$\\
 & $3.60 - 3.70$ & $+0.007$ & $0.065$ & $0.237$ & $0.046$ & $0.273$ & $0.293$\\
 & $3.70 - 3.80$ & $+0.003$ & $0.079$ & $0.272$ & $0.052$ & $0.295$ & $0.323$\\
 & $3.80 - 3.90$ & $+0.013$ & $0.092$ & $0.334$ & $0.050$ & $0.344$ & $0.376$\\
 & $3.90 - 4.00$ & $+0.010$ & $0.084$ & $0.294$ & $0.054$ & $0.313$ & $0.348$\\
\hline
 & $0.00 - 0.10$ & $-0.272$ & $0.286$ & $0.550$ & $0.330$ & $0.000$ & $0.522$\\
 & $0.10 - 0.20$ & $+0.001$ & $0.070$ & $0.283$ & $0.048$ & $0.317$ & $0.301$\\
 & $0.20 - 0.30$ & $+0.007$ & $0.059$ & $0.193$ & $0.055$ & $0.257$ & $0.231$\\
 & $0.30 - 0.40$ & $+0.005$ & $0.042$ & $0.132$ & $0.039$ & $0.224$ & $0.173$\\
 & $0.40 - 0.50$ & $+0.013$ & $0.074$ & $0.244$ & $0.061$ & $0.270$ & $0.287$\\
 & $0.50 - 0.60$ & $+0.016$ & $0.081$ & $0.258$ & $0.073$ & $0.246$ & $0.290$\\
 & $0.60 - 0.70$ & $-0.001$ & $0.047$ & $0.137$ & $0.044$ & $0.218$ & $0.177$\\
 & $0.70 - 0.80$ & $-0.010$ & $0.053$ & $0.088$ & $0.054$ & $0.138$ & $0.152$\\
 & $0.80 - 0.90$ & $-0.012$ & $0.053$ & $0.095$ & $0.050$ & $0.161$ & $0.151$\\
 & $0.90 - 1.00$ & $-0.002$ & $0.055$ & $0.144$ & $0.055$ & $0.192$ & $0.184$\\
 & $1.00 - 1.10$ & $-0.028$ & $0.088$ & $0.239$ & $0.111$ & $0.091$ & $0.259$\\
 & $1.10 - 1.20$ & $-0.019$ & $0.064$ & $0.146$ & $0.076$ & $0.134$ & $0.195$\\
	Mizuki	& $1.20 - 1.30$ & $-0.003$ & $0.061$ & $0.101$ & $0.062$ & $0.137$ & $0.162$\\
 & $1.30 - 1.40$ & $-0.003$ & $0.064$ & $0.156$ & $0.067$ & $0.179$ & $0.204$\\
 & $1.40 - 1.50$ & $-0.006$ & $0.064$ & $0.228$ & $0.069$ & $0.244$ & $0.242$\\
 & $1.50 - 1.60$ & $+0.000$ & $0.074$ & $0.304$ & $0.107$ & $0.243$ & $0.298$\\
 & $1.60 - 1.70$ & $+0.012$ & $0.113$ & $0.329$ & $0.153$ & $0.101$ & $0.332$\\
 & $1.70 - 1.80$ & $+0.043$ & $0.161$ & $0.369$ & $0.175$ & $0.077$ & $0.394$\\
 & $1.80 - 1.90$ & $+0.033$ & $0.169$ & $0.385$ & $0.172$ & $0.080$ & $0.398$\\
 & $1.90 - 2.00$ & $+0.017$ & $0.160$ & $0.361$ & $0.160$ & $0.117$ & $0.383$\\
 & $2.00 - 2.10$ & $+0.019$ & $0.161$ & $0.341$ & $0.149$ & $0.102$ & $0.376$\\
 & $2.10 - 2.20$ & $+0.009$ & $0.148$ & $0.297$ & $0.139$ & $0.108$ & $0.361$\\
 & $2.20 - 2.30$ & $+0.024$ & $0.166$ & $0.347$ & $0.153$ & $0.128$ & $0.404$\\
 & $2.30 - 2.40$ & $+0.028$ & $0.166$ & $0.369$ & $0.161$ & $0.157$ & $0.420$\\
 & $2.40 - 2.50$ & $+0.011$ & $0.146$ & $0.362$ & $0.148$ & $0.199$ & $0.405$\\
 & $2.50 - 2.60$ & $+0.029$ & $0.143$ & $0.389$ & $0.150$ & $0.213$ & $0.413$\\
 & $2.60 - 2.70$ & $+0.019$ & $0.142$ & $0.385$ & $0.132$ & $0.285$ & $0.426$\\
 & $2.70 - 2.80$ & $+0.000$ & $0.141$ & $0.390$ & $0.101$ & $0.355$ & $0.433$\\
 & $2.80 - 2.90$ & $-0.004$ & $0.165$ & $0.445$ & $0.081$ & $0.439$ & $0.491$\\
 & $2.90 - 3.00$ & $+0.006$ & $0.181$ & $0.483$ & $0.076$ & $0.479$ & $0.508$\\
 & $3.00 - 3.10$ & $+0.006$ & $0.130$ & $0.435$ & $0.062$ & $0.444$ & $0.476$\\
 & $3.10 - 3.20$ & $+0.790$ & $0.834$ & $0.532$ & $1.229$ & $0.000$ & $0.566$\\
 & $3.20 - 3.30$ & $+0.082$ & $0.275$ & $0.504$ & $1.196$ & $0.083$ & $0.551$\\
 & $3.30 - 3.40$ & $+0.116$ & $0.327$ & $0.518$ & $1.787$ & $0.000$ & $0.558$\\
 & $3.40 - 3.50$ & $+0.006$ & $0.151$ & $0.465$ & $0.050$ & $0.503$ & $0.504$\\
 & $3.50 - 3.60$ & $+0.005$ & $0.079$ & $0.337$ & $0.047$ & $0.381$ & $0.382$\\
 & $3.60 - 3.70$ & $+0.006$ & $0.077$ & $0.345$ & $0.048$ & $0.375$ & $0.390$\\
 & $3.70 - 3.80$ & $+0.006$ & $0.119$ & $0.444$ & $0.055$ & $0.464$ & $0.485$\\
 & $3.80 - 3.90$ & $+0.026$ & $0.103$ & $0.316$ & $0.061$ & $0.321$ & $0.380$\\
 & $3.90 - 4.00$ & $-0.003$ & $0.160$ & $0.491$ & $0.054$ & $0.505$ & $0.525$\\
\hline
 & $0.00 - 0.10$ & $-99.000$ & $-99.000$ & $-99.000$ & $-99.000$ & $-99.000$ & $-99.000$\\
 & $0.10 - 0.20$ & $+0.016$ & $0.068$ & $0.170$ & $0.054$ & $0.207$ & $0.234$\\
 & $0.20 - 0.30$ & $+0.013$ & $0.072$ & $0.256$ & $0.055$ & $0.297$ & $0.287$\\
 & $0.30 - 0.40$ & $+0.002$ & $0.053$ & $0.182$ & $0.050$ & $0.251$ & $0.226$\\
 & $0.40 - 0.50$ & $+0.011$ & $0.070$ & $0.194$ & $0.064$ & $0.210$ & $0.241$\\
 & $0.50 - 0.60$ & $-0.001$ & $0.065$ & $0.217$ & $0.054$ & $0.258$ & $0.249$\\
 & $0.60 - 0.70$ & $+0.000$ & $0.038$ & $0.117$ & $0.035$ & $0.207$ & $0.148$\\
 & $0.70 - 0.80$ & $-0.005$ & $0.050$ & $0.132$ & $0.054$ & $0.184$ & $0.171$\\
 & $0.80 - 0.90$ & $-0.009$ & $0.053$ & $0.089$ & $0.054$ & $0.142$ & $0.148$\\
 & $0.90 - 1.00$ & $-0.001$ & $0.048$ & $0.148$ & $0.046$ & $0.221$ & $0.174$\\
 & $1.00 - 1.10$ & $-0.019$ & $0.075$ & $0.205$ & $0.095$ & $0.135$ & $0.229$\\
 & $1.10 - 1.20$ & $-0.011$ & $0.054$ & $0.124$ & $0.056$ & $0.164$ & $0.167$\\
	MLZ	& $1.20 - 1.30$ & $-0.010$ & $0.060$ & $0.093$ & $0.058$ & $0.128$ & $0.152$\\
 & $1.30 - 1.40$ & $-0.009$ & $0.063$ & $0.140$ & $0.068$ & $0.159$ & $0.189$\\
 & $1.40 - 1.50$ & $-0.000$ & $0.082$ & $0.319$ & $0.128$ & $0.177$ & $0.305$\\
 & $1.50 - 1.60$ & $-0.018$ & $0.072$ & $0.291$ & $0.106$ & $0.215$ & $0.285$\\
 & $1.60 - 1.70$ & $-0.032$ & $0.115$ & $0.335$ & $0.152$ & $0.087$ & $0.338$\\
 & $1.70 - 1.80$ & $+0.006$ & $0.134$ & $0.320$ & $0.159$ & $0.076$ & $0.355$\\
 & $1.80 - 1.90$ & $+0.024$ & $0.143$ & $0.315$ & $0.151$ & $0.106$ & $0.357$\\
 & $1.90 - 2.00$ & $+0.032$ & $0.152$ & $0.319$ & $0.145$ & $0.102$ & $0.362$\\
 & $2.00 - 2.10$ & $+0.013$ & $0.139$ & $0.275$ & $0.129$ & $0.097$ & $0.334$\\
 & $2.10 - 2.20$ & $+0.023$ & $0.137$ & $0.285$ & $0.133$ & $0.081$ & $0.338$\\
 & $2.20 - 2.30$ & $+0.034$ & $0.155$ & $0.316$ & $0.144$ & $0.090$ & $0.371$\\
 & $2.30 - 2.40$ & $+0.044$ & $0.168$ & $0.369$ & $0.149$ & $0.102$ & $0.403$\\
 & $2.40 - 2.50$ & $+0.073$ & $0.174$ & $0.433$ & $0.165$ & $0.131$ & $0.430$\\
 & $2.50 - 2.60$ & $+0.059$ & $0.154$ & $0.391$ & $0.160$ & $0.167$ & $0.418$\\
 & $2.60 - 2.70$ & $+0.031$ & $0.134$ & $0.371$ & $0.145$ & $0.226$ & $0.402$\\
 & $2.70 - 2.80$ & $+0.030$ & $0.157$ & $0.419$ & $0.149$ & $0.298$ & $0.445$\\
 & $2.80 - 2.90$ & $+0.019$ & $0.155$ & $0.429$ & $0.110$ & $0.368$ & $0.456$\\
 & $2.90 - 3.00$ & $+0.004$ & $0.150$ & $0.441$ & $0.080$ & $0.431$ & $0.472$\\
 & $3.00 - 3.10$ & $-0.013$ & $0.144$ & $0.450$ & $0.064$ & $0.460$ & $0.494$\\
 & $3.10 - 3.20$ & $-0.013$ & $0.171$ & $0.479$ & $0.066$ & $0.487$ & $0.526$\\
 & $3.20 - 3.30$ & $-0.003$ & $0.145$ & $0.449$ & $0.061$ & $0.465$ & $0.488$\\
 & $3.30 - 3.40$ & $-0.011$ & $0.115$ & $0.407$ & $0.057$ & $0.440$ & $0.454$\\
 & $3.40 - 3.50$ & $-0.008$ & $0.092$ & $0.355$ & $0.048$ & $0.389$ & $0.400$\\
 & $3.50 - 3.60$ & $-0.002$ & $0.077$ & $0.318$ & $0.050$ & $0.344$ & $0.368$\\
 & $3.60 - 3.70$ & $-0.001$ & $0.080$ & $0.326$ & $0.048$ & $0.360$ & $0.371$\\
 & $3.70 - 3.80$ & $-0.007$ & $0.081$ & $0.266$ & $0.051$ & $0.282$ & $0.327$\\
 & $3.80 - 3.90$ & $+0.015$ & $0.101$ & $0.293$ & $0.064$ & $0.309$ & $0.355$\\
 & $3.90 - 4.00$ & $-0.000$ & $0.076$ & $0.290$ & $0.050$ & $0.303$ & $0.343$\\
\hline
 & $0.00 - 0.10$ & $+0.009$ & $0.042$ & $0.067$ & $0.037$ & $0.113$ & $0.109$\\
 & $0.10 - 0.20$ & $-0.001$ & $0.054$ & $0.147$ & $0.047$ & $0.197$ & $0.196$\\
 & $0.20 - 0.30$ & $-0.000$ & $0.061$ & $0.207$ & $0.060$ & $0.252$ & $0.242$\\
 & $0.30 - 0.40$ & $-0.004$ & $0.045$ & $0.153$ & $0.043$ & $0.243$ & $0.198$\\
 & $0.40 - 0.50$ & $-0.002$ & $0.061$ & $0.199$ & $0.053$ & $0.245$ & $0.237$\\
 & $0.50 - 0.60$ & $+0.001$ & $0.067$ & $0.230$ & $0.054$ & $0.268$ & $0.262$\\
 & $0.60 - 0.70$ & $+0.001$ & $0.042$ & $0.169$ & $0.036$ & $0.257$ & $0.196$\\
 & $0.70 - 0.80$ & $+0.000$ & $0.047$ & $0.165$ & $0.045$ & $0.243$ & $0.195$\\
 & $0.80 - 0.90$ & $-0.010$ & $0.052$ & $0.123$ & $0.054$ & $0.180$ & $0.170$\\
 & $0.90 - 1.00$ & $-0.004$ & $0.047$ & $0.192$ & $0.041$ & $0.271$ & $0.198$\\
 & $1.00 - 1.10$ & $-0.021$ & $0.069$ & $0.222$ & $0.097$ & $0.137$ & $0.230$\\
 & $1.10 - 1.20$ & $-0.012$ & $0.051$ & $0.139$ & $0.051$ & $0.192$ & $0.170$\\
	NNPZ	& $1.20 - 1.30$ & $-0.004$ & $0.050$ & $0.092$ & $0.050$ & $0.146$ & $0.141$\\
 & $1.30 - 1.40$ & $-0.002$ & $0.055$ & $0.121$ & $0.057$ & $0.164$ & $0.164$\\
 & $1.40 - 1.50$ & $-0.004$ & $0.057$ & $0.198$ & $0.059$ & $0.243$ & $0.216$\\
 & $1.50 - 1.60$ & $-0.005$ & $0.067$ & $0.268$ & $0.069$ & $0.277$ & $0.269$\\
 & $1.60 - 1.70$ & $-0.018$ & $0.107$ & $0.302$ & $0.142$ & $0.124$ & $0.317$\\
 & $1.70 - 1.80$ & $-0.001$ & $0.132$ & $0.333$ & $0.158$ & $0.089$ & $0.353$\\
 & $1.80 - 1.90$ & $+0.008$ & $0.137$ & $0.311$ & $0.145$ & $0.125$ & $0.350$\\
 & $1.90 - 2.00$ & $+0.009$ & $0.131$ & $0.267$ & $0.124$ & $0.096$ & $0.322$\\
 & $2.00 - 2.10$ & $+0.013$ & $0.124$ & $0.251$ & $0.118$ & $0.110$ & $0.312$\\
 & $2.10 - 2.20$ & $+0.026$ & $0.140$ & $0.295$ & $0.132$ & $0.108$ & $0.353$\\
 & $2.20 - 2.30$ & $+0.016$ & $0.146$ & $0.321$ & $0.138$ & $0.130$ & $0.362$\\
 & $2.30 - 2.40$ & $+0.022$ & $0.138$ & $0.299$ & $0.136$ & $0.138$ & $0.355$\\
 & $2.40 - 2.50$ & $+0.036$ & $0.141$ & $0.382$ & $0.149$ & $0.163$ & $0.393$\\
 & $2.50 - 2.60$ & $+0.026$ & $0.136$ & $0.348$ & $0.134$ & $0.214$ & $0.389$\\
 & $2.60 - 2.70$ & $+0.019$ & $0.119$ & $0.347$ & $0.118$ & $0.254$ & $0.389$\\
 & $2.70 - 2.80$ & $-0.002$ & $0.106$ & $0.310$ & $0.081$ & $0.300$ & $0.357$\\
 & $2.80 - 2.90$ & $+0.001$ & $0.104$ & $0.277$ & $0.077$ & $0.272$ & $0.348$\\
 & $2.90 - 3.00$ & $+0.006$ & $0.088$ & $0.291$ & $0.071$ & $0.293$ & $0.344$\\
 & $3.00 - 3.10$ & $+0.003$ & $0.101$ & $0.327$ & $0.066$ & $0.336$ & $0.381$\\
 & $3.10 - 3.20$ & $-0.004$ & $0.100$ & $0.317$ & $0.068$ & $0.341$ & $0.378$\\
 & $3.20 - 3.30$ & $-0.006$ & $0.086$ & $0.326$ & $0.055$ & $0.355$ & $0.377$\\
 & $3.30 - 3.40$ & $-0.009$ & $0.102$ & $0.375$ & $0.055$ & $0.412$ & $0.422$\\
 & $3.40 - 3.50$ & $-0.005$ & $0.068$ & $0.322$ & $0.041$ & $0.380$ & $0.365$\\
 & $3.50 - 3.60$ & $+0.003$ & $0.073$ & $0.301$ & $0.048$ & $0.339$ & $0.349$\\
 & $3.60 - 3.70$ & $+0.007$ & $0.070$ & $0.253$ & $0.050$ & $0.289$ & $0.309$\\
 & $3.70 - 3.80$ & $+0.011$ & $0.081$ & $0.259$ & $0.055$ & $0.274$ & $0.318$\\
 & $3.80 - 3.90$ & $+0.013$ & $0.070$ & $0.230$ & $0.050$ & $0.246$ & $0.289$\\
 & $3.90 - 4.00$ & $+0.001$ & $0.088$ & $0.277$ & $0.061$ & $0.298$ & $0.345$\\
\hline

    \hline
  \end{longtable}
}

\subsection{Code-code comparisons}
\label{ssec:code_code_comparisons}

\begin{figure}
  \begin{center}
    \includegraphics[width=8cm]{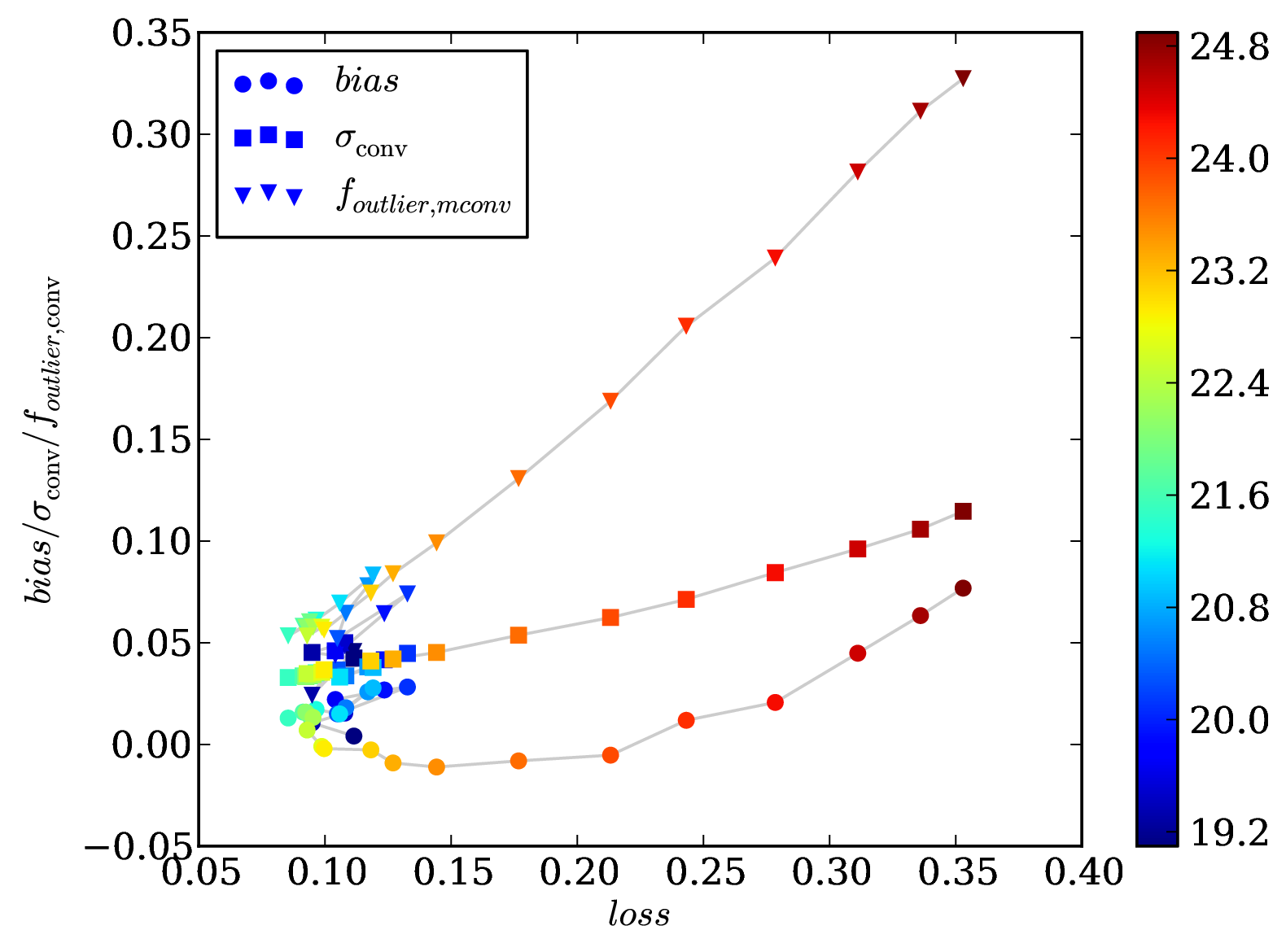}
  \end{center}
  \caption{
    Relationship between loss and other metrics.  The symbols are color-coded according
    to the $i$-band magnitude cut applied.  This is for MLZ using the Wide-depth
    median seeing catalog, but the other codes show similar trends.
 }
 \label{fig:loss_property}
\end{figure}

In the previous subsection, we plot the three metrics (bias, dispersion, and
outlier rate) separately for each code.  But, it is useful to use a single metric
to compare the performance between the codes.  For this, we use the loss parameter,
$L(\Delta z)$, introduced earlier.  Because this is not a popular statistic used
in the literature, we first show its relationship between the other statistical
measures in Fig.~\ref{fig:loss_property}.  The figure is for \texttt{MLZ}, but
the other codes behave similarly.  While all of the bias, scatter and outlier
rate are correlated (all of them get worse at fainter magnitudes), it is clear
that the mean loss most strongly correlates with the outlier rate.
Loss should also change with bias and scatter at fixed outlier rate by definition,
but it is the outlier rate that increases drastically at faint mags and the mean loss
most strongly correlates with that parameter.

\begin{figure}
  \begin{center}
    \includegraphics[width=8cm]{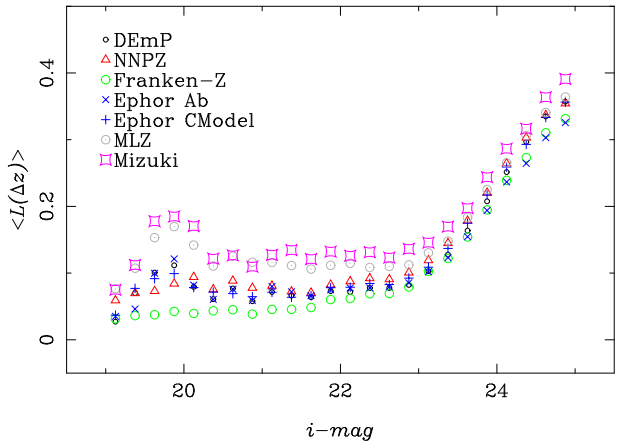}
  \end{center}
  \caption{
    Mean loss as a function of $i$-band magnitude for all the codes.  The symbols are
    explained in the figure.
 }
 \label{fig:loss_mag}
\end{figure}

\begin{figure}
  \begin{center}
    \includegraphics[width=8cm]{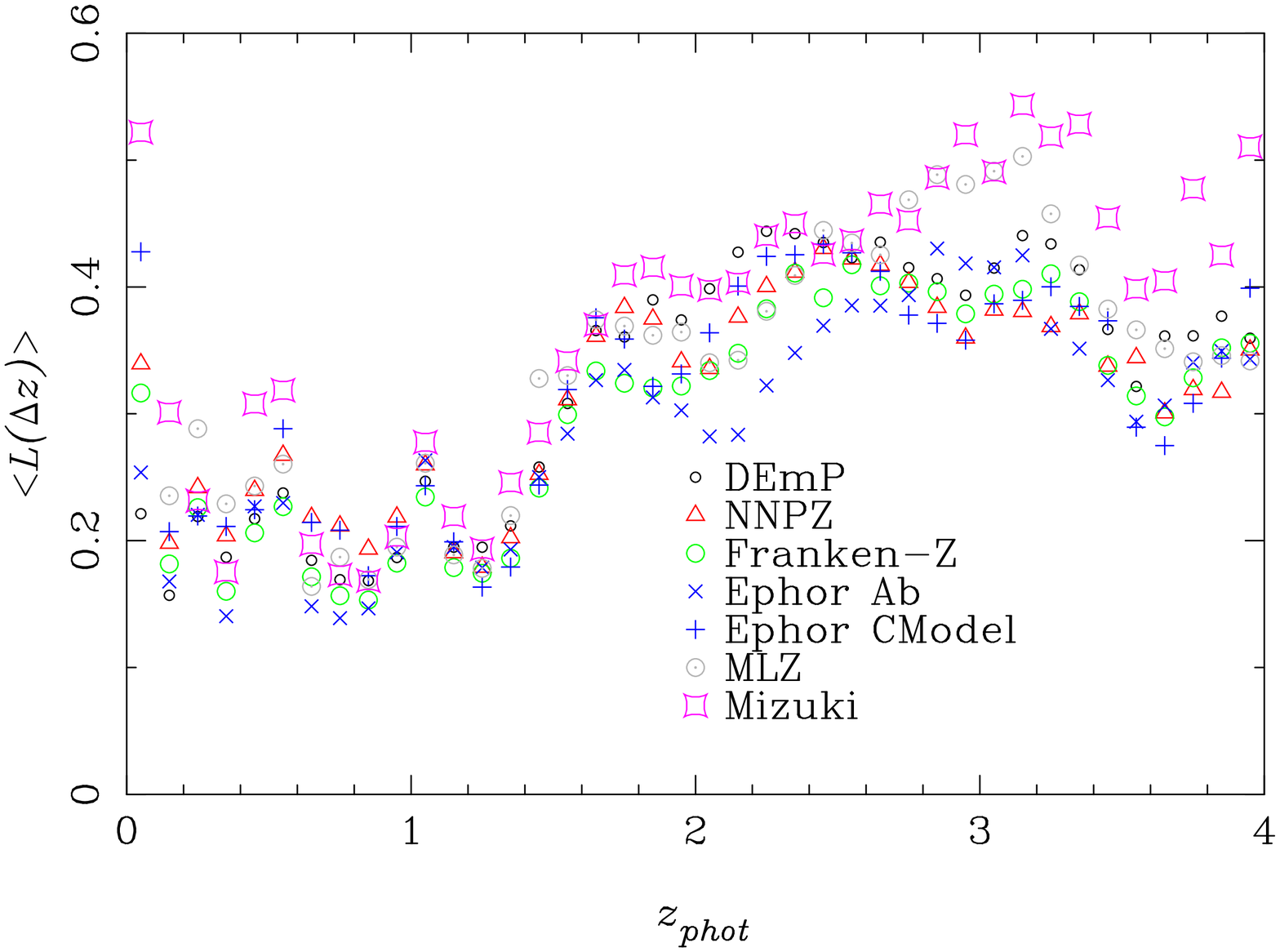}
  \end{center}
  \caption{
    Same as Fig.~\ref{fig:loss_mag} but as a function of $z_{phot}$.
 }
 \label{fig:loss_zphot}
\end{figure}

Figs.~\ref{fig:loss_mag} and \ref{fig:loss_zphot} shows the mean loss as
a function of magnitude and $z_{phot}$, respectively.  All the codes show a similar behavior in these
figures; the accuracy starts to get worse around $i=23$ and the redshift range of
$0.2\lesssim z \lesssim1.5$ shows the best performance.  Mizuki tends to perform worse
than the other codes.  Although it was trained on an earlier version of the training
sample with sub-optimal weights (see Section \ref{ssec:mizuki}),
it is the only
classical template-fitting code and it might suggest that machine-learning codes
outperform template fitting.  There are advantages and disadvantages in both
techniques and we will discuss them further in Section \ref{sec:summary}.
Again, note that the metrics for \texttt{FRANKEN-Z} are likely overly optimistic
given some degree of over-fitting.


\subsection{Seeing and depth dependence}
\label{ssec:seeing}

The photometric accuracy is not only a function of integration time and sky transparency,
but also seeing.  As described in Section~\ref{ssec:test_samples}, we have generated
the COSMOS wide-depth stacks for three different seeing FWHMs.  We use them to evaluate
the seeing dependence of our photo-$z$ accuracy at the Wide-depth.

\begin{figure}
  \begin{center}
    \includegraphics[width=8cm]{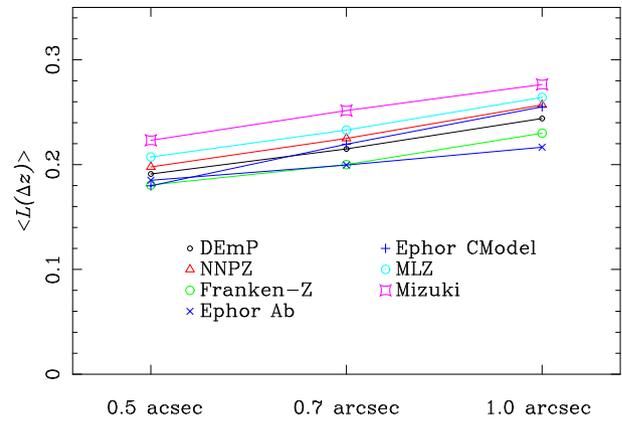}
  \end{center}
  \caption{
    Mean loss as a function of seeing.
 }
 \label{fig:loss_seeing}
\end{figure}

Fig.~\ref{fig:loss_seeing} shows $<L(\Delta z)>$ as a function of seeing. Loss is larger
at worse seeing as expected and we find $\Delta <L(\Delta z)>\sim0.05$ between the two extremes.  Most of
the HSC data are taken under $0.5-1.0$ arcsec seeing \citep{aihara17}, and Fig.~\ref{fig:loss_seeing}
gives the peak-peak variation of our photo-$z$ performance across the Wide survey.
\texttt{EPHOR} delivers photo-$z$'s computed with CModel and PSF-matched aperture photometry
(\texttt{EPHOR} and \texttt{EPHOR\_AB}, respectively).  A comparison between them show how strongly each
photometry technique suffers from the seeing variation.  The PSF-matched photometry turns
out to be less strongly affected by seeing than CModel; $\Delta <L(\Delta z)>\sim0.03$ and $0.06$
for PSF-matched and CModel photometry, respectively.  The weaker seeing dependence of the PSF-matched photometry
is not surprising because the images are smoothed to 1.1 arcsec FWHM, regardless of the native seeing.
It is, however, rather surprising that the measurements under the native seeing deliver poorer photo-$z$
accuracy.  But, we note that
the current CModel has issues with a prior, which affects the resultant photometry
\citep{bosch17, huang17}.  It is unlikely that the color measurements are severely
affected, but fluxes are undoubtedly affected.  Also, the deblending algorithm tends to fail in
dense regions such as cluster cores \citep{aihara17}, which also affects CModel
measurements.  The PSF-matched photometry suffers less from the deblending issue because
it is performed without deblending.
Future improvements in the measurement algorithms will make CModel work better.

\begin{figure}
  \begin{center}
    \includegraphics[width=8cm]{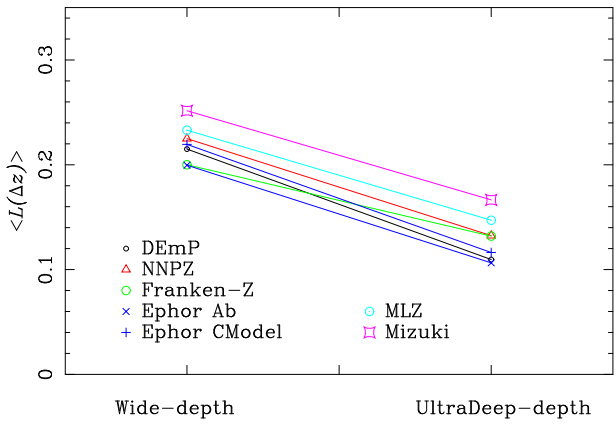}
  \end{center}
  \caption{
    Same as Fig. \ref{fig:loss_seeing} but as a function of depth.
 }
 \label{fig:loss_depth}
\end{figure}

The depth dependence is shown in Fig.~\ref{fig:loss_depth}.  Again, all the codes behave
similarly and the mean loss is smaller by $\sim0.1$ at the UltraDeep depth.  Although
not shown in the figure, the improvement is not limited to $0.2<z<1.5$ but is observed
at all redshifts.  This implies that obtaining photometry in more filters is not the only
way to improve photo-$z$'s.  Going deeper can be a useful alternative.

\subsection{Cut on the risk parameter}
\label{ssec:riskcut}

We have characterized our photo-$z$ performance using all galaxies
down to $i=25$ without any clipping of potential outliers.
We can achieve reasonably good photo-$z$ accuracy
at a somewhat limited redshift range due to the filter set as discussed above.
Also, our photo-$z$'s are of course not perfect and there are always outliers even
within the good redshift range.
There are a few quantities that can be used to indicate a reliability
of photo-$z$  such as $C(z)$ and \texttt{odds} \citep{benitez2000} that allow us to remove potential outliers.
We have introduced a new parameter, $R(z)$, in Section~\ref{ssec:zp_best}
and here we compare this new parameter with the commonly used $C(z)$.

\begin{figure}
  \begin{center}
    \includegraphics[width=8cm]{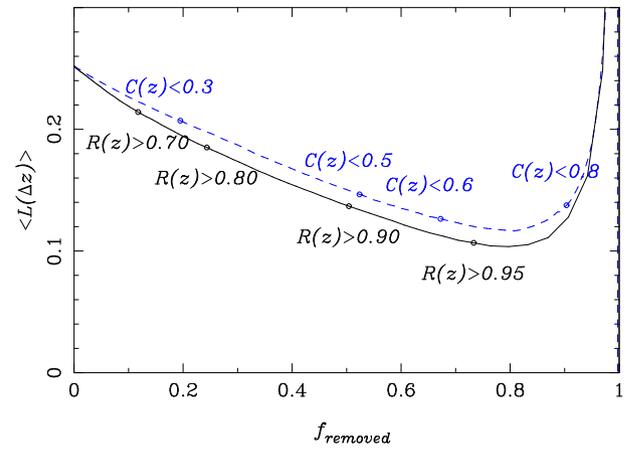}
  \end{center}
  \caption{
    Loss plotted against the fraction of objects removed by applying a cut on
    $C(z_{phot})$ and $R(z_{phot})$.  $z_{phot}$ is denoted as $z$ in the figure for simplicity.
    The dashed and solid curves are for $C(z_{phot})$
    and $R(z_{phot})$ and threshold applied for each of them are shown in the figure.
    This is for \texttt{Mizuki}, but the other codes show a similar trend.
 }
 \label{fig:zconf_vs_zrisk}
\end{figure}

Fig.~\ref{fig:zconf_vs_zrisk} compares $C(z_{phot})$ and $R(z_{phot})$.
As defined earlier, $z_{phot}$ is the best point estimate.
We remove
objects with $C(z_{phot})$/$R(z_{phot})$ smaller/larger than a threshold value and
plot the resultant $<L(\Delta z)>$ as a function of the fraction of objects removed.  At a given
fraction of removed objects, $<L(\Delta z)>$ is always smaller for $R(z_{phot})$ than for $C(z_{phot})$.
For instance, at $f_{removed}=0.5$ (i.e., we remove a half of the objects),
which roughly corresponds to $C(z_{phot})<0.5$ and $R(z_{phot})>0.9$ cuts, loss
is smaller for $R(z_{phot})$ by about 0.02.  $R(z_{phot})$ is designed to minimize
loss and thus this may not be a fair comparison, but we observe the same
trend if we plot other quantities such as the outlier rate.  This demonstrates
that $R(z)$ works better at identifying outliers than the commonly used $C(z)$.

\section{Accuracy of PDF}
\label{sec:pdf}

We have focused on the point statistics in the previous section.  We now move
on to discuss the accuracy of the full PDF.  We first focus on the $N(z)$
distribution of galaxies and then turn our attention to Probability Integral
Transform (PIT) and Continuous Ranked Probability Score (CRPS) to evaluate the PDF accuracy.

\subsection{N(z) distribution}
\label{ssec:dndz}

In various scientific uses, we often consider not only the redshift
for single galaxy but also the global properties averaged over a
number of objects.  In this section, we show redshift distributions of
photometric sample from the S16A internal release and compare them among the seven different photo-$z$ codes.

\subsubsection{Internal comparisons}
              
\begin{figure}
  \includegraphics[width=\linewidth]{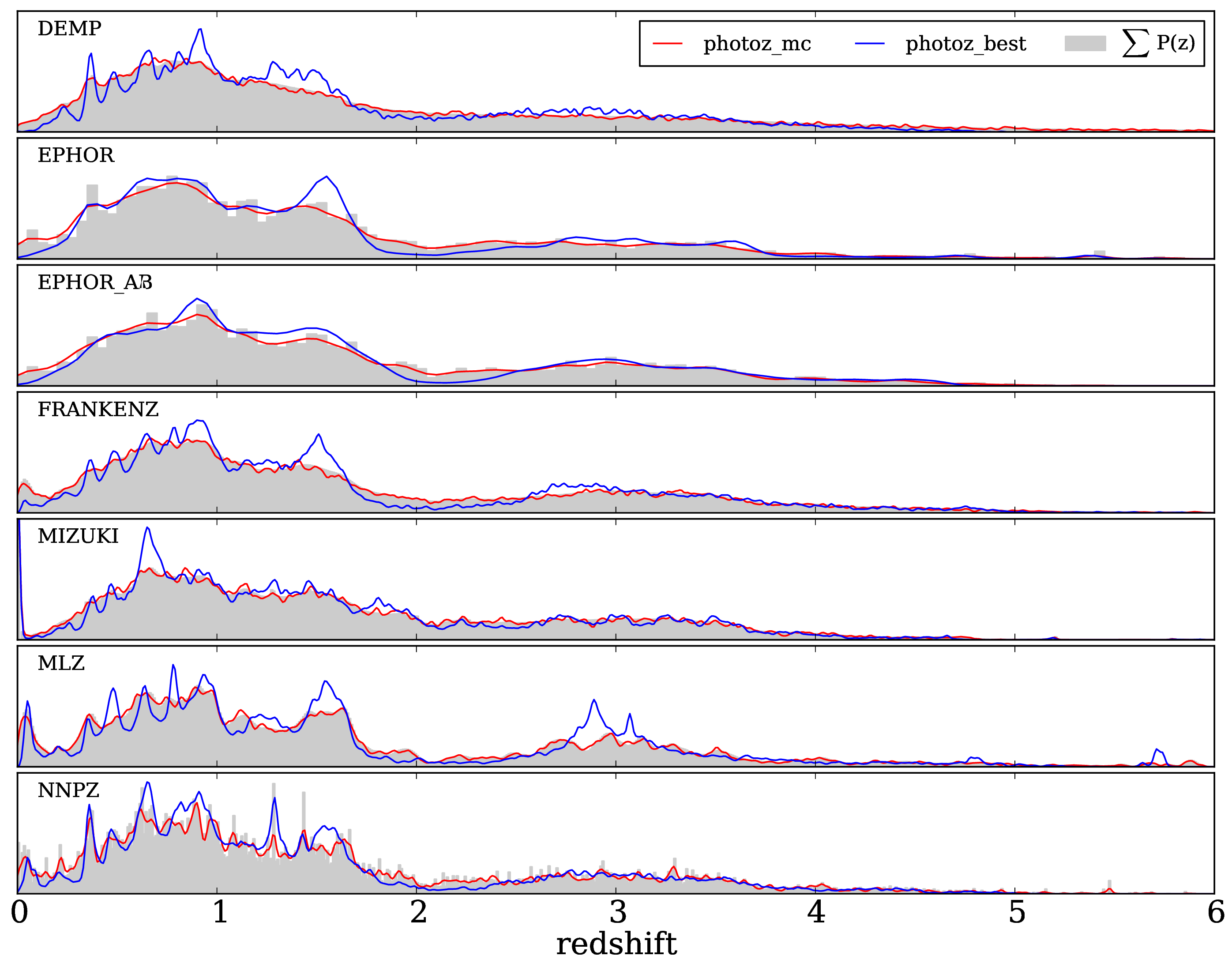}
  \caption{
    $N(z)$ distributions for all galaxies in the Wide layer
    inferred using a few different estimators;
    sum of full PDF (gray
    histogram), Gaussian KDE for $z_{\rm MC}$ (red line)
    and $z_{\rm best}$ (blue line). Sum of full PDF and
    $N(z_{\rm MC})$ agrees very well, while $N(z_{\rm best})$
    estimates show sharp redshift spikes.  This is likely due
    to the spikes present in the training data from COSMOS.
    \label{fig:dndz_wide}}
\end{figure}

As we will discuss in Section \ref{sec:products}, we randomly draw a redshift
from $P(z)$ for each object ($z_{MC}$).  We first demonstrate that this Monte-Carlo draw
from the PDF well reproduces the original PDF and is a very useful point estimate
for a statistical sample.  In Fig. \ref{fig:dndz_wide}, we
compare the stacked PDF and the sum of $z_{MC}$ using Gaussian Kernel Density Estimator (KDE),

\begin{equation}
  N^{\rm P}(z)
  =
      \frac{1}{n} \sum_i^{n} P_i(z)
\end{equation}

\begin{equation}
  \label{eq:kde}
  N^{\rm MC}(z)
  =
  \frac{1}{\sqrt{2\pi}nh} \sum_i^n \exp\left[ \frac{(z-z_{{\rm MC}, i})^2}{2h^2} \right],
\end{equation}

\noindent
where the kernel width $h$ is set to the PDF resolution, 0.05 for
\texttt{EPHOR} and \texttt{EPHOR\_AB} and 0.01 for all the other
codes. The estimator reduces the discreteness of the sample, but we
found that given the large number of objects, we do not see any
major differences between the classical count-up histogram and KDE.

As shown in the figure, we see a good agreement between $N^{\rm MC}$ and
$N^{\rm P}$ for most codes, although $N^{\rm MC}$ fails to trace small scale spiky
features in \texttt{NNPZ}, \texttt{EPHOR}, \texttt{EPHOR\_AB} seen in
$N^{\rm P}$.
In the same figure, we also plot the $N(z)$ distribution from $z_{best}$ using Eq. \ref{eq:kde}.
Although the $z_{\rm best}$ is the optimal
point estimate in terms of minimizing the risk function (see section
\ref{ssec:zp_best}), $N^{\rm best}$ amplifies the wiggle feature of
the $N(z)$ distribution. This might imply that the point estimates
are affected by inhomogeneities in
the training sample.  For instance, the local
peaks around $z\sim 1.5$ are a consequence of the bumpy structure in the
COSMOS 30-band photo-$z$, on which we highly rely to calibrate our
photo-$z$'s (see also Fig.~\ref{fig:dndz_cosmos_wide}).  $N^{\rm
  best}$ for \texttt{Mizuki} is least affected since the template fitting does
not rely on the training sample very much, while machine learning codes do.
In the following, we use $N^{\rm MC}$ as the representative of the redshift
distribution instead of $N^{\rm P}$ since summing up the full PDF is
computationally much more expensive.

\begin{figure}
  \includegraphics[width=\linewidth]{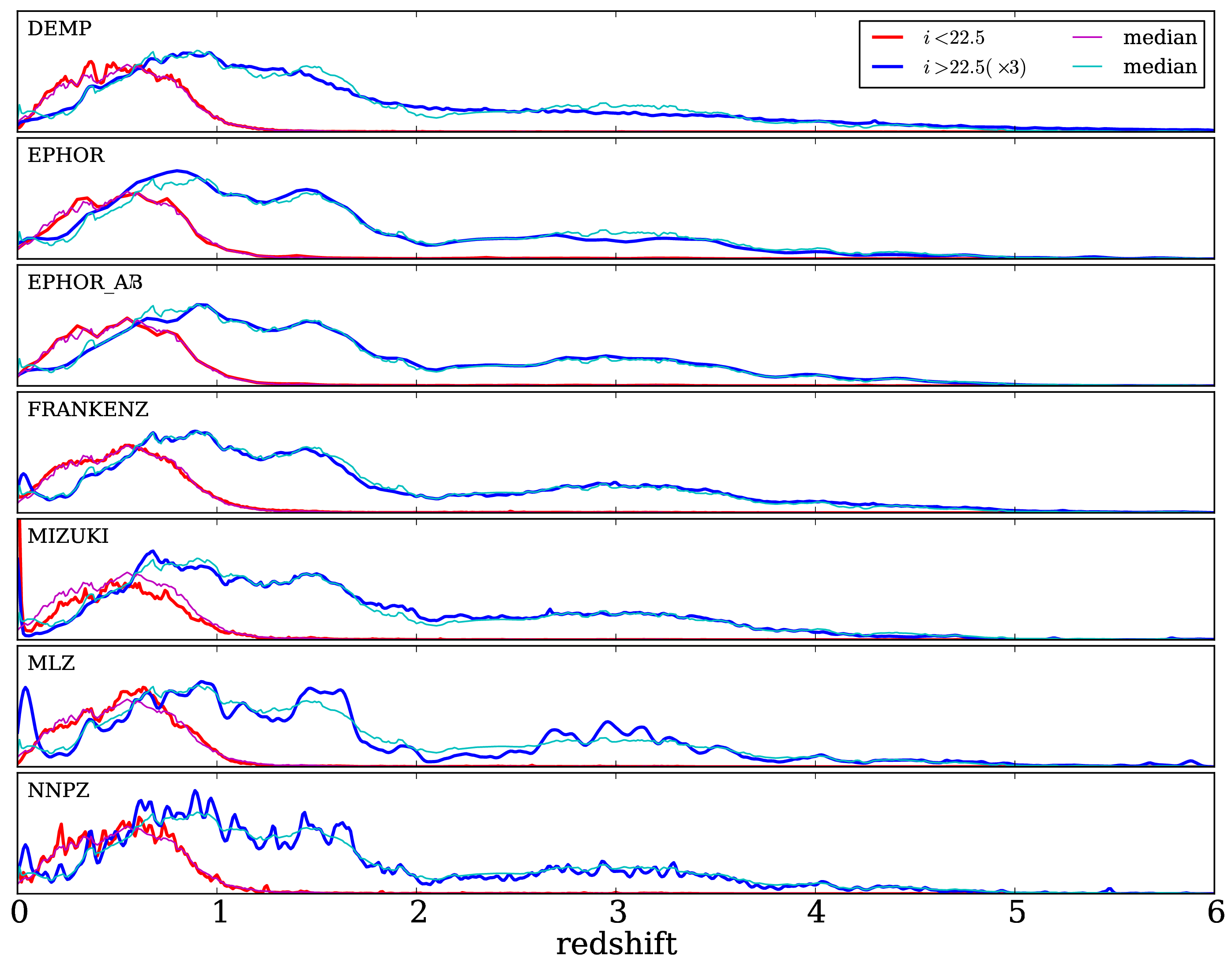}
  \caption{$N(z)$ distribution for bright (red) and faint (blue)
    samples. The median distribution of all the photo-$z$ codes
    are also shown for bright (magenta thin line) and faint (cyan thin
    line) samples. 
    \texttt{EPHOR} looks smoother than the others due to a lager redshift bin size.
    \label{fig:dndz_brightfaint}}
\end{figure}

Figure \ref{fig:dndz_brightfaint} shows the $N(z)$ distribution from
$z_{\rm MC}$ for bright ($i<22.5$) and faint ( $i>22.5$) sample.
Sharp drop of bright sample at $z\sim 1$ reflects that we have
few bright objects at $z>1$.
On the other hand, we have galaxies out to $z\sim6$ in the fainter sample.
Although there are subtle differences between the codes,
the overall redshift distributions are similar for all of the codes, which
is encouraging.

\subsubsection{External comparisons}

\begin{figure}
  \includegraphics[width=\linewidth]{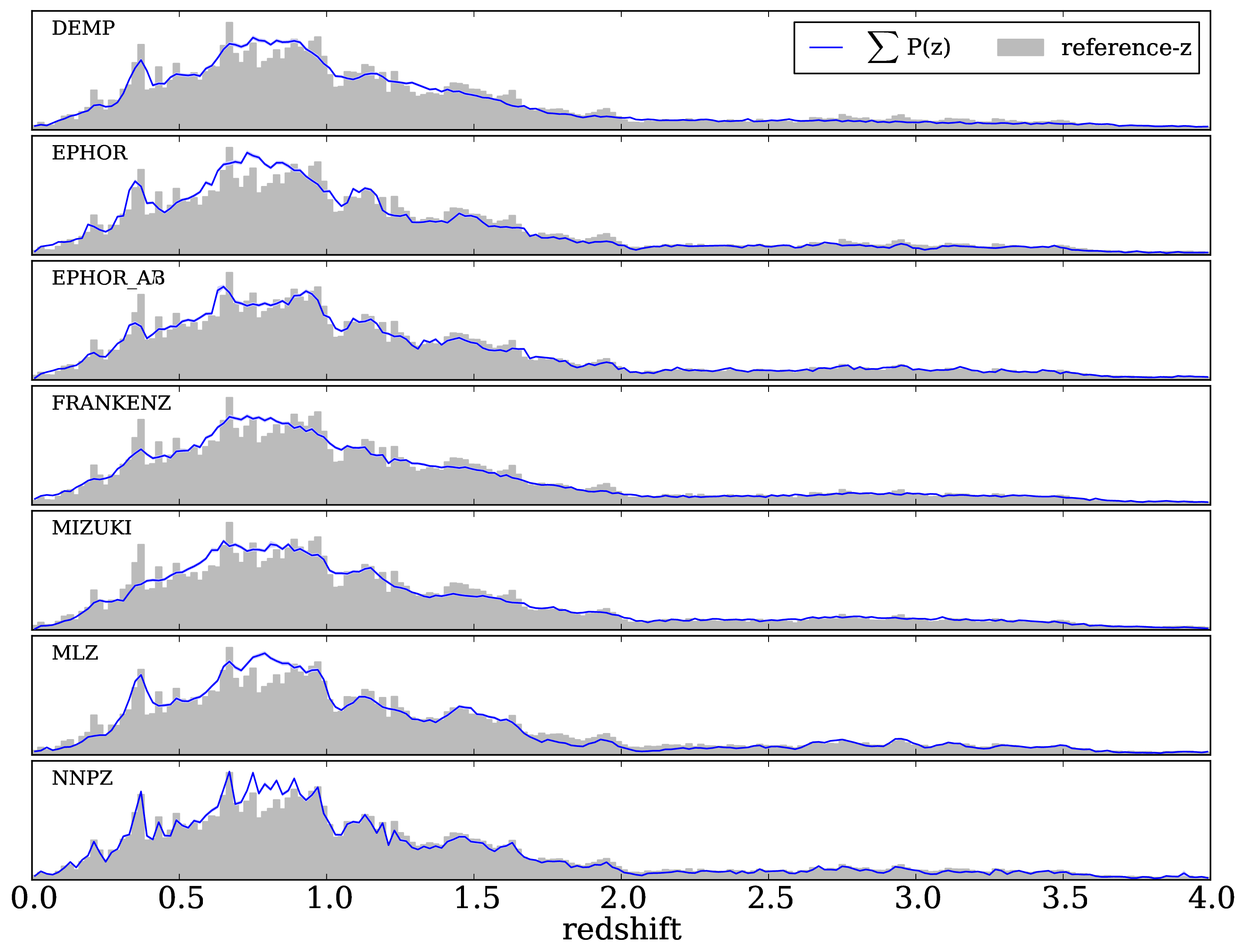}
  \caption{$N(z)$ distributions from COSMOS wide depth stacked image
    with median seeing (blue lines) and reference redshifts (gray
    shaded histogram).
    \label{fig:dndz_cosmos_wide}}
\end{figure}

We have compared the internal consistency in the previous section.
We now turn our attention to external comparisons using the reference redshifts
in the COSMOS Wide-depth median stack.
Fig. \ref{fig:dndz_cosmos_wide} 
shows the comparison between $N^{\rm MC}$ and $N(z)$ based on the reference redshifts.
Assuming that the reference redshifts are correct, the deviations from
their $N(z)$ is an indication of incorrect PDF.  While all the codes
reproduce the overall $N(z)$ reasonably well, \texttt{NNPZ} reproduces
the $N(z)$ most accurately.  \texttt{Mizuki} misses a peak at $z\sim0.35$,
and \texttt{EPHOR}, \texttt{MLZ}, and \texttt{NNPZ} tend to overestimate
at $z\sim0.7$.
This has implications for weak-lensing science, which often relies on $N(z)$ from
photo-$z$.  However, detailed discussions of the over/underestimated $N(z)$ for
weak-lensing are beyond the scope of the paper and can be found elsewhere (More et al. in prep.).

We have trained our codes using galaxies that are primarily from COSMOS especially at faint mags,
and we have compared our $N(z)$ against COSMOS.  Re-weighting the training galaxies
to reproduce the HSC Wide sample largely eliminates the circularity here.
However, it will certainly be useful to have a separate field with different $N(z)$ for
more comparisons.  Such a COSMOS-like field with accurate photo-$z$'s down to faint mags
is currently not available, which is a one of the major limitations of our photo-$z$ tests.
We will discuss our future directions in Section \ref{sec:summary}.


\subsection{Tests on PDF}
\label{ssec:test_on_pdf}

As a further test of the accuracy of PDF, we apply two techniques;
Probability Integral Transform (PIT) and Continuous Ranked Probability Score (CRPS).
They are summarized in \citet{polsterer16}, but a brief description is given here.

\begin{figure*}
  \begin{center}
    \includegraphics[width=6cm]{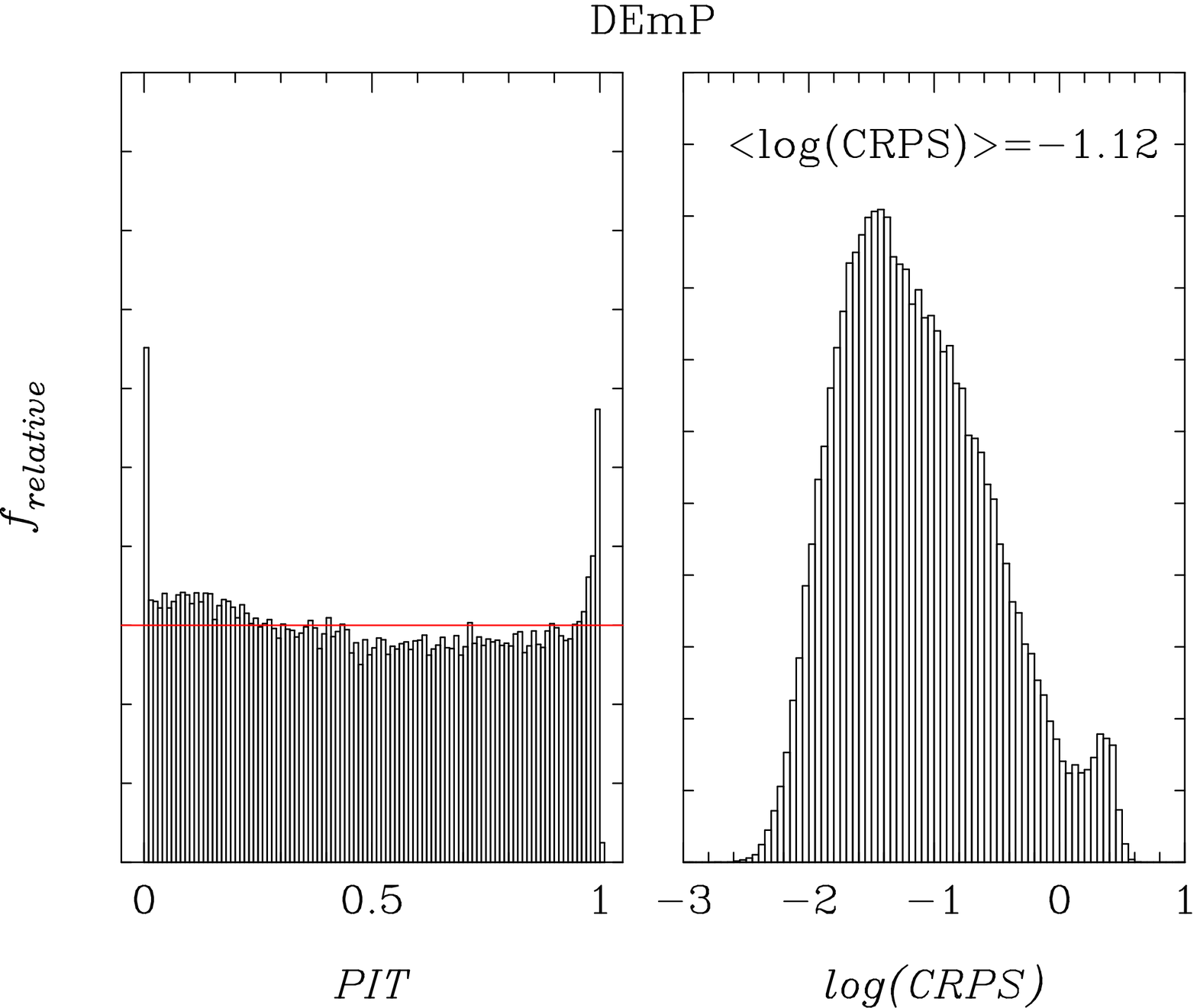}\hspace{0.5cm}
    \includegraphics[width=6cm]{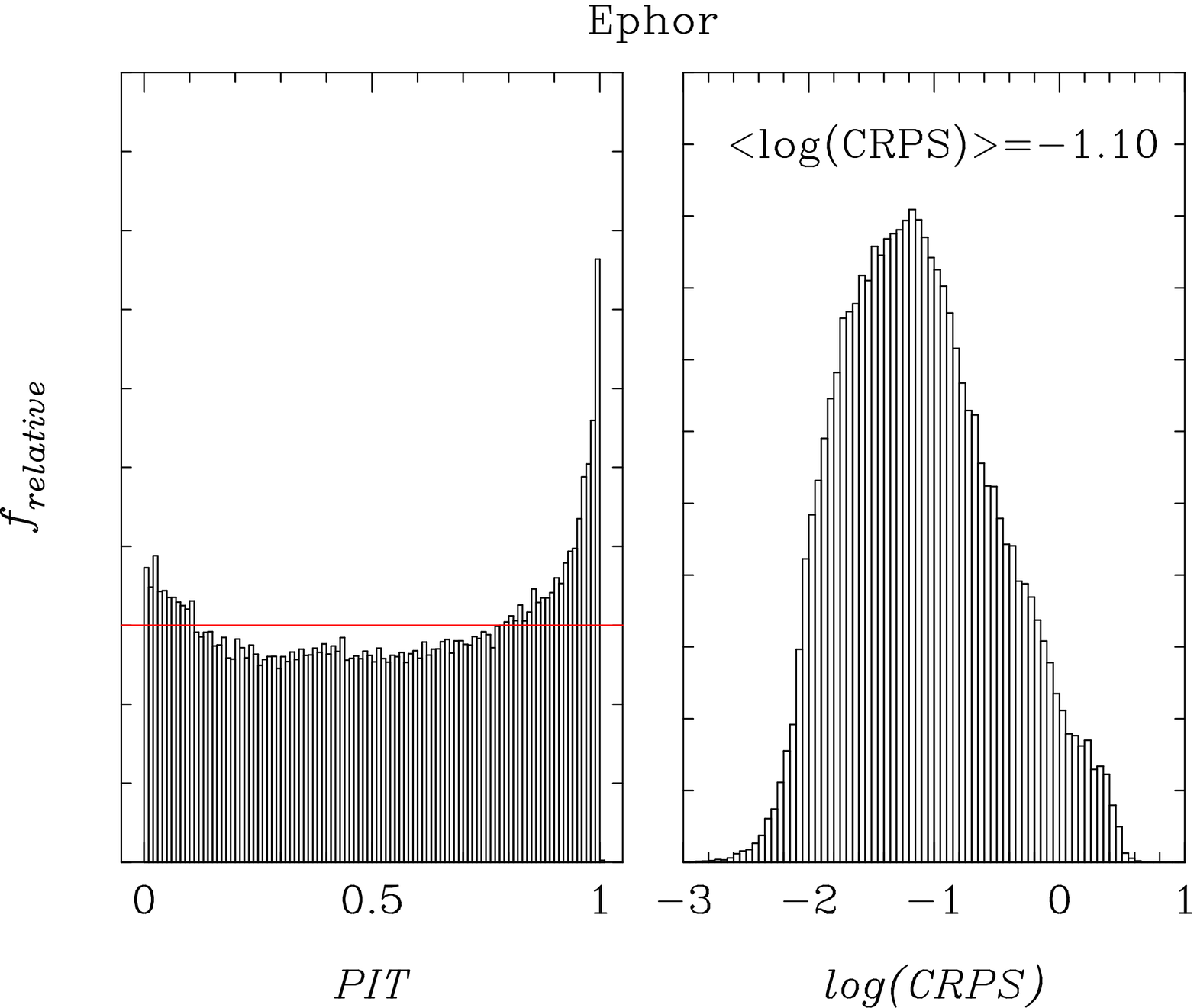}\\\vspace{0.5cm}
    \includegraphics[width=6cm]{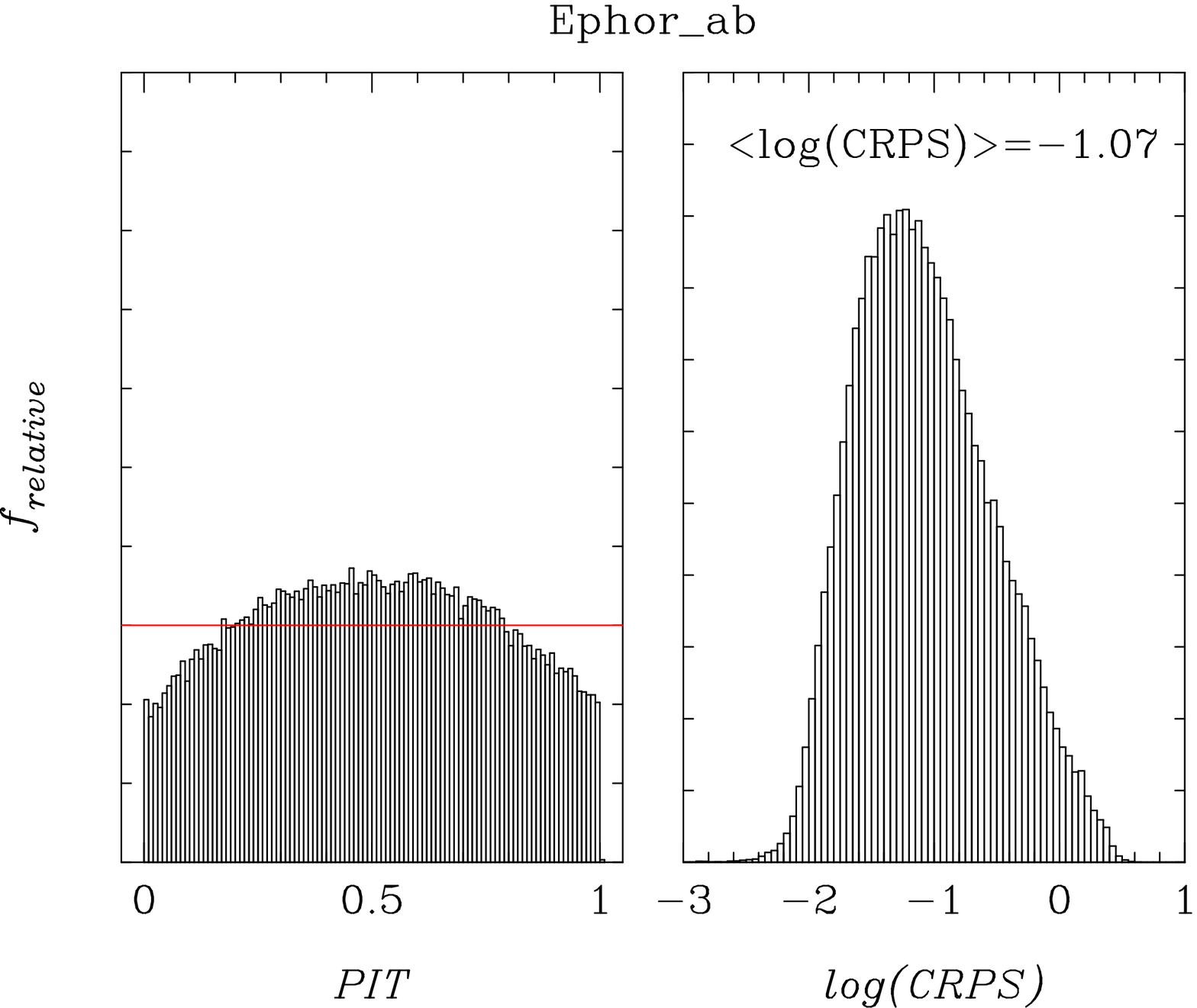}\hspace{0.5cm}
    \includegraphics[width=6cm]{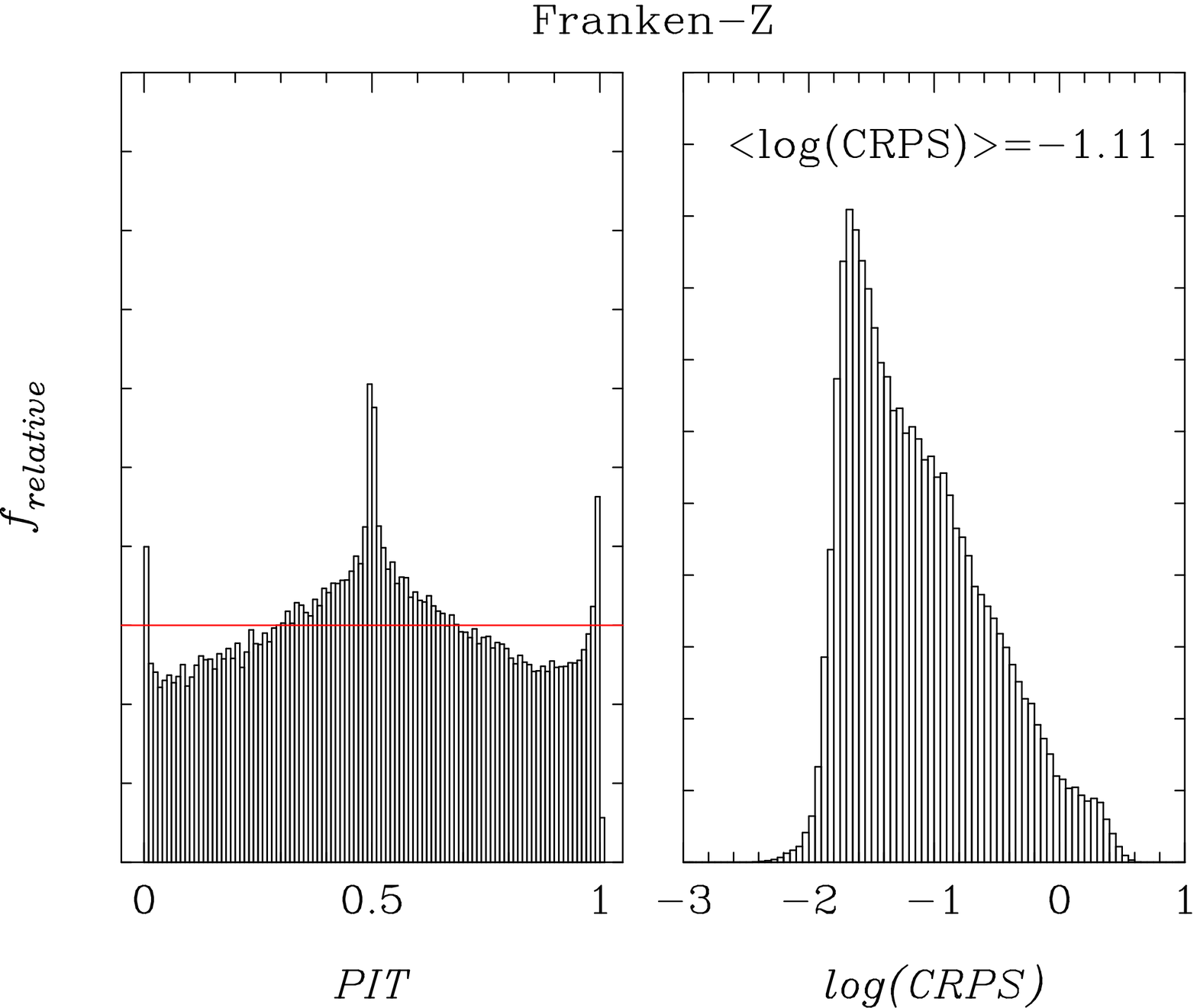}\\\vspace{0.5cm}
    \includegraphics[width=6cm]{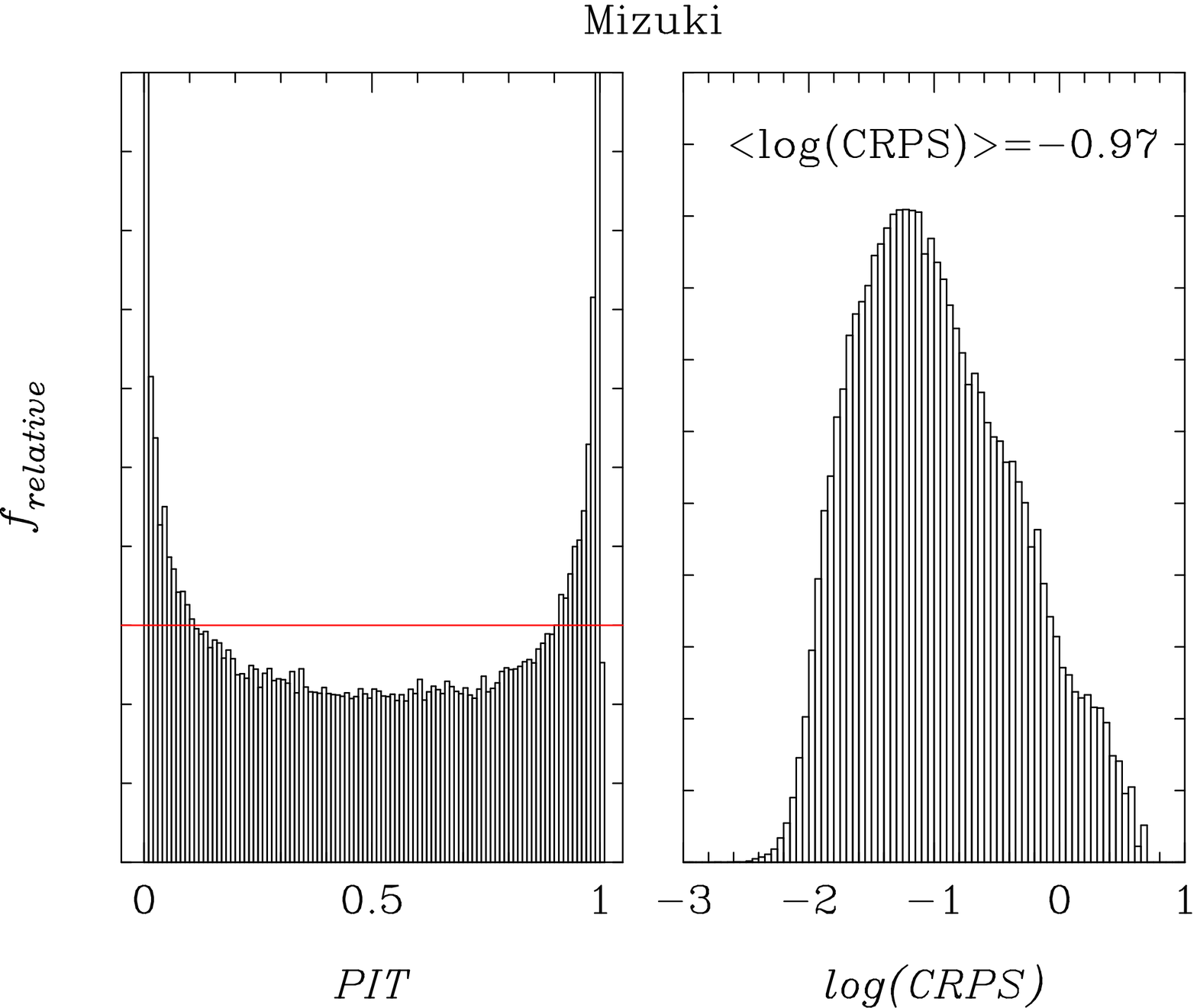}\hspace{0.5cm}
    \includegraphics[width=6cm]{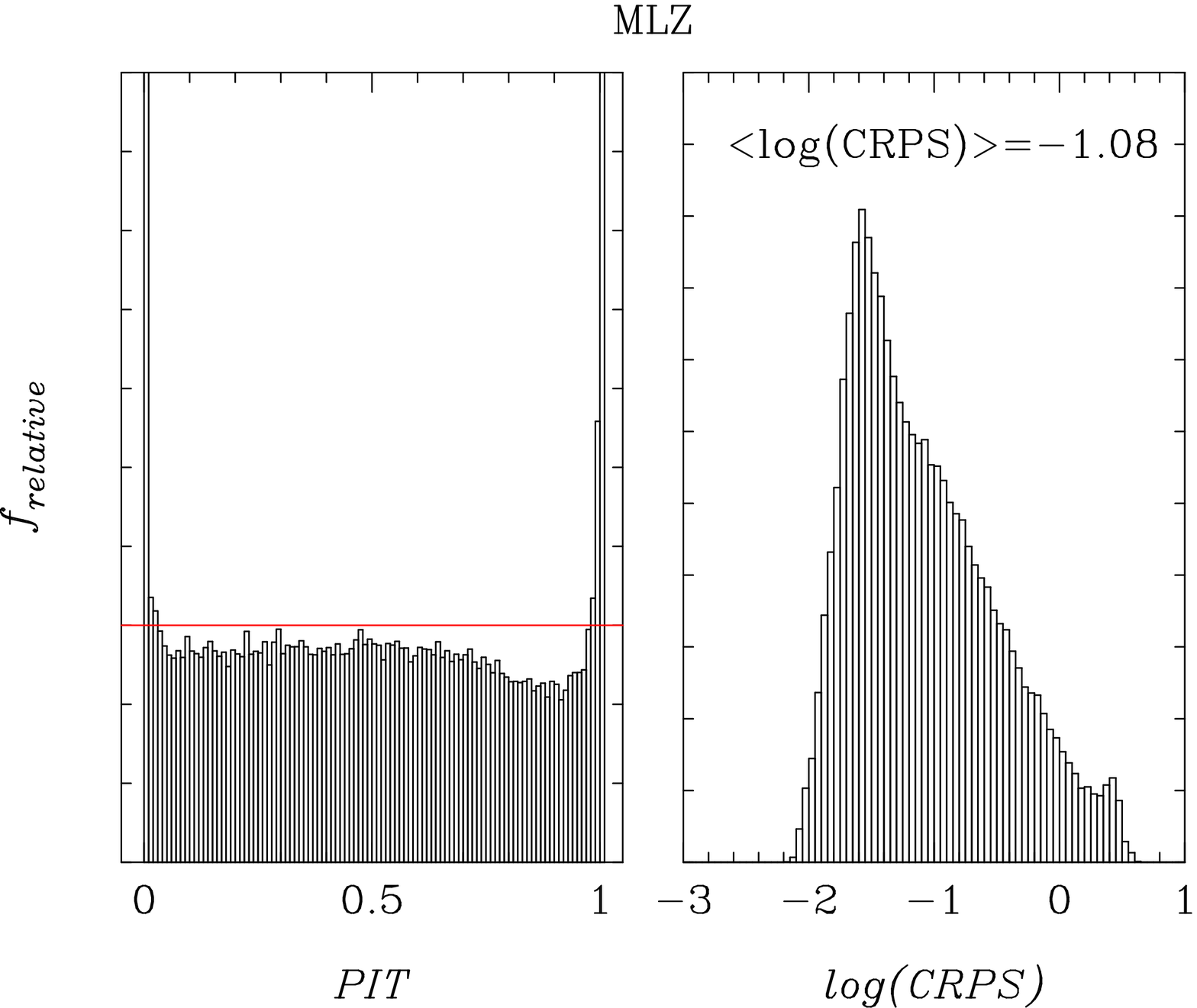}\\\vspace{0.5cm}
    \includegraphics[width=6cm]{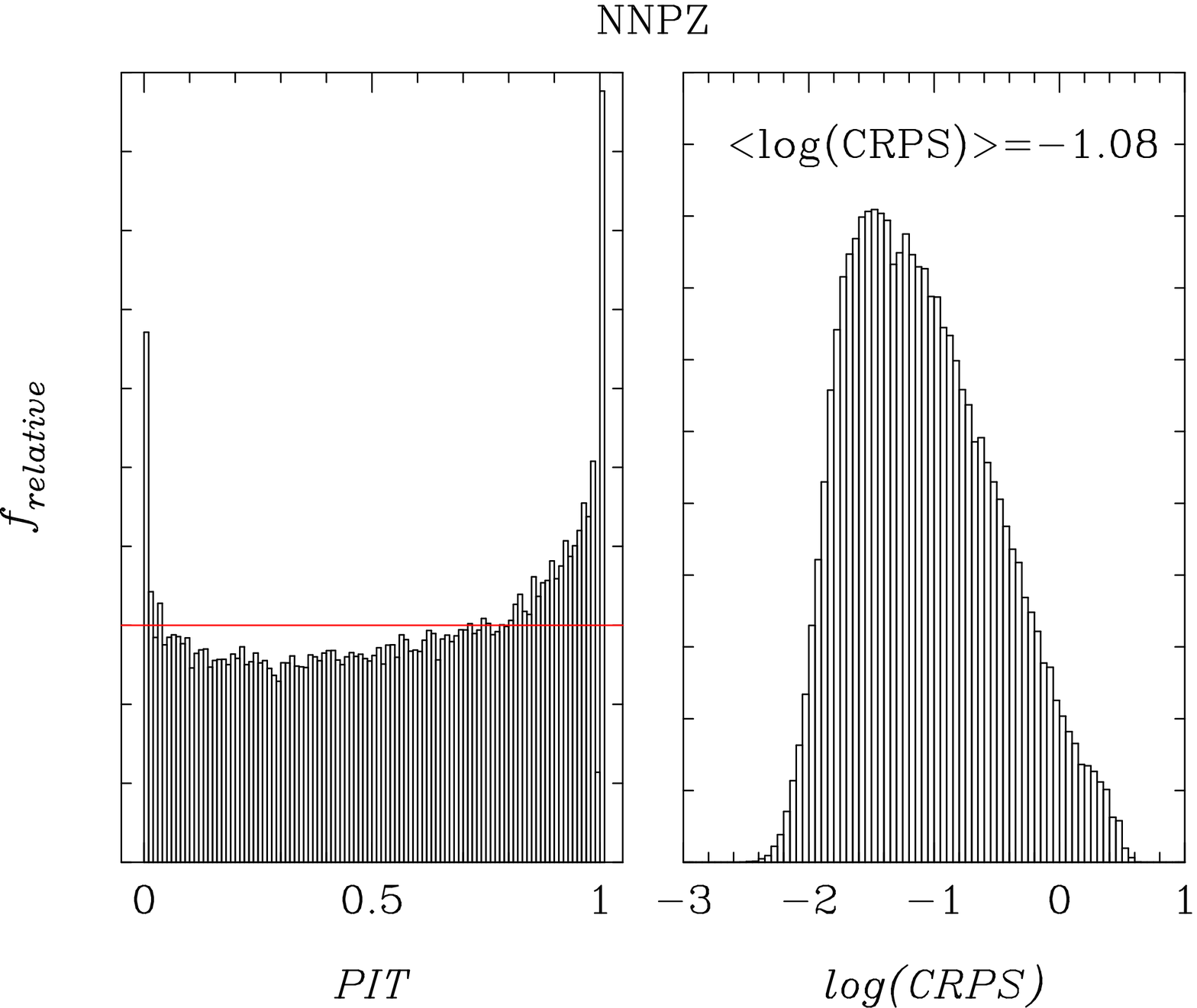}
  \end{center}
  \caption{
    PIT (left) and CRPS (right) for all the codes.  The red horizontal line in the left panel is
    just to guide the eye.
 }
 \label{fig:pdf_stat}
\end{figure*}

PIT was proposed as a visual diagnostic tool to check the calibration of PDF.
It is a very simple diagnostic and one only needs to draw a histogram of the following
integrated probability,

\begin{equation}
  PIT(z_{ref})=\int_0^{z_{ref}} P(z) dz,
\end{equation}

\noindent
for all objects in the test sample.
The left panels in
Fig.~\ref{fig:pdf_stat} show the PIT histograms for all the codes.
If the PDF is calibrated well, we expect to observe a flat PIT distribution.
Deviations from the flat distribution is an indication of incorrect PDF and
this formed a basis of the empirical PDF re-calibration by \citet{bordoloi10}.
\texttt{EPHOR\_AB} shows a convex shape, which is a clear indication of 
overdispersed PDF, i.e., PDF is too wide.  On the other hand, \texttt{Mizuki}
has a concave shape and it indicates that the PDFs are underdispersed, i.e.,
PDF is too narrow.  Most of the other codes show a relatively flat distribution,
except at the two extremes of the distribution, where many codes show a spike.
These spikes are caused by outliers and the figures suggest that the outliers are not
properly captured in the PDFs.

\texttt{FRANKEN-Z} shows an interesting PIT distribution with a peak at the center.
The peak indicates that a larger-than-expected fraction of objects have the median redshift almost exactly
at $z_{ref}$, which suggests that PDF is too accurate. We do not expect to
see such a feature in the presence of random uncertainties.
While we have not fully understood the origin of the peak, we tentatively
interpret it as a sign of over-fitting. Most likely, this peak is due to 
\texttt{FRANKEN-Z}s inclusion of both the training and target errors when deriving likelihoods.
Unlike other nearest neighbor methods such as \texttt{NNPZ} which select neighbors and derive weights
using Monte Carlo procedures based on (modifications to the) Euclidean norm, \texttt{FRANKEN-Z} 
computes the intrinsic likelihood expected if training/testing objects were Monte Carlo realizations of the same
underlying galaxy (Speagle et al. in prep.). Objects whose photometry between the 
Wide-depth stacks and Deep/UltraDeep observations are not fully independent can thus sometimes 
deviate much less than expected, leading to large contributions to the posterior and
subsequent signs of over-fitting. We note that numerous cross-validation and hold-out tests
have not found evidence of such behavior in the native training sample.

All of the codes have some degree of deviations from the flat PIT distribution.
This motivates us to use the PIT distribution to empirically re-calibrate
our $P(z)$ \citep{bordoloi10} in our future releases as it will likely improve
our overall performance.

We turn our attention to the other technique, CRPS.
CRPS is a measure of a 'distance' between PDF and $z_{ref}$ and is defined as

\begin{equation}
  CRPS=\int_{-\infty}^{+\infty} \{PIT(z) - H(z-z_{ref})\}^2 dz,
\end{equation}

\noindent
where $H(z-z_{ref})$ the Heaviside step-function;

\begin{equation}
  H(x)=\left\{ \begin{array}{l}
      0\ \mathrm{if}\ x<0\\
      1\ \mathrm{if}\ x\geq0.
  \end{array}\right.
\end{equation}

\noindent
The right panels of Fig.~\ref{fig:pdf_stat} show CRPS for all the codes.
When PDFs are calibrated well, the mean CRPS is small.  A large CRPS is
an indication of incorrect PDF.
To the first order, all the codes perform similarly well; $<\log CRPS>\sim-1$.
However, there are small differences in $CRPS$ between the codes and
machine-learning codes once again tend to perform better than the classical
template-fitting code.

It is interesting to note that a code with good performance with point estimates
does not
necessarily give a small CRPS.  For instance, \texttt{EPHOR\_AB} has a smaller
loss than \texttt{EPHOR} as shown in Fig.~\ref{fig:loss_seeing}.
However, CRPS in Fig. \ref{fig:pdf_stat} is larger, suggesting that PDF is less accurate.
The PIT distribution indicates that \texttt{EPHOR} has over-dispersed PDFs,
and this over-dispersed PDFs are likely driving the slightly larger CRPS.
The analysis here suggests that accurate point estimates do not necessarily mean
that PDFs are accurate.  They are obviously closely related to each other but
not exactly the same.  Thus, in order to evaluate the photo-$z$ performance,
one needs to look both at the point estimates and PDFs.

\section{Data Products}
\label{sec:products}
%
%
We make our photo-$z$ products available to the community.  This section summarizes our target selection
criteria, 'common' outputs that are available for all the codes, as well as
code-specific outputs.

HSC-SSP Public Data Release 1 (PDR1) includes our photo-$z$'s for the Deep and UltraDeep layers, covering
over 30 square degrees in total.  Due to a technical issue during the photo-$z$
production run, we were unable to include our photo-$z$'s for the Wide layer
in PDR1, but
they were made public as part of the first incremental data release
occurred in June 2017.
It is important to note that each code
applies various cuts to select objects for photo-$z$ production.
That is, each code is applied to a different set of objects (but with a significant
overlap) due to features of the code.
Table \ref{tab:photoz_depend} summarizes the target selection by codes.
The table also indicates whether there are additional outputs from the code,
which we will elaborate below.  Most codes imposed \texttt{detect\_is\_primary}
to select primary objects, except for \texttt{EPHOR}.  \texttt{DEmP} and
\texttt{MLZ} compute photo-$z$'s for all the primary
objects, but \texttt{FRANKEN-Z} and \texttt{NNPZ} requires good photometry in all the bands
in addition to the primary flag.
\texttt{Mizuki} computes photo-$z$'s for primary objects with good CModel
photometry in at least 3 bands (inclusive).

All the codes generate a PDF for each object.  We run a common script to
compute various point estimates, confidence intervals and other useful
statistics.  The common outputs are summarized in Table \ref{tab:photoz_common}.
In addition to these common outputs, there are code-specific outputs as follows.

\noindent
\textbf{FRANKEN-Z}
\begin{itemize}
  \item{\texttt{model\_llmin}:} $-2\ln (\max({\cal L}_i))=\min
    (\chi^2_n(i)-n(i))$, where $n(i)=5$ is the number of bands used in the fit.
  \item{\texttt{model\_levidence}:}
    $-2\ln({\rm evidence})=−2\ln(\sum_i {\cal L}_i)$, where
    ${\cal L}_i = \exp\{-0.5[\chi_n^2(i)-n(i)]\}$
    and the sum over $i$ is taken over all unique neighbors.
  \item{\texttt{model\_ntype}:} Number of unique neighbors used in the
    fit grouped by redshift type (spec, g/prism, and many-band photo-$z$).
  \item{\texttt{model\_ptype}:} Fraction of normalized likelihood
    contributed by each redshift type.
  \item{\texttt{model\_nsurvey}:} As above, but grouped by parent survey (SDSS, etc.).
  \item{\texttt{model\_psurvey}:} As above, but contributed by by each parent survey.
\end{itemize}

\noindent
\textbf{Mizuki}
\begin{itemize}
  \item{\texttt{reduced\_chisq}, $\chi^2_\nu$:} Reduced chi-squares of the best-fit model.
    It is recommended to remove objects having $\chi^2_\nu>5$ for
    scientific use.
  \item{\texttt{stellar\_mass}:} Median stellar mass derived from $P(M*)$, which is
    stellar mass PDF marginalized over all the other parameters.
    The 68\% confidence intervals are also available. All the
    uncertainties on physical parameters include uncertainties from
    photo-$z$'s.
  \item{\texttt{sfr}:} Median star formation rate with 68\% intervals.
  \item{\texttt{tauv}, $\tau_V$:} Median dust attenuation in the
    V-band with 68\% intervals. Note that $A_V=1.09\tau_V$.
  \item{\texttt{prob\_x}: } \texttt{x} is either \texttt{gal},
    \texttt{qso} or \texttt{star}, which denote the relative
    probability that an object is galaxy, QSO and star.
  \item{rest-frame magnitudes:} Rest-frame magnitudes in the GALEX,
    SDSS, HSC, and WFCAM filters. Only the magnitudes from the best-fit
    template at the median redshifts are computed and no uncertainties
    are currently available.
\end{itemize}

\textbf{MLZ}
\begin{itemize}
  \item{\texttt{flux\_binary\_flag}: } Binary flag to show how many
    CMmodel fluxes at different filters are available,
    \begin{equation}
      {\rm f} = \sum_{i=0}^4 \left\{2^{9-i}PF_i + 2^{4-i} NF_i \right\},
    \end{equation}
    where $PF_i=1$ if flux$_i > \sigma$flux$_i$, and $NF_i=1$ if
    $|{\rm flux}_i|>\sigma{\rm flux}_i$, and 0 otherwise. Index $i$
    denotes filters with $0$ being g-band and $4$ being y-band.
    If the object is well measured in all five bands, the flag have
    value 1023.
\end{itemize}

All of the catalog products such as photo-$z$ point estimates are available in the database.
The full PDFs are stored
in the fits format and are available from the photo-$z$ page of the PDR1 site.

\begin{table*}
  \begin{center}
    \begin{tabular*}{\textwidth}{l @{\extracolsep{\fill}} ll}\hline\hline
      \texttt{key}                & description \\ \hline \hline
      \texttt{object\_id}         & unique object id to be used to join with the photometry tables \\
      \texttt{photoz\_X}          & Photo-$z$ point estimate where X is
                                    either \texttt{mean}, \texttt{mode}, \texttt{median}, or \texttt{best}. \\
      \texttt{photoz\_mc}         & Monte Carlo draw from the full PDF \\
      \texttt{photoz\_conf\_X}    & Photo-$z$ confidence value defined
                                    by equation \ref{eq:photoz_conf} at \texttt{photoz\_X}. \\
      \texttt{photoz\_risk\_X}    & Risk parameter defined by equation
                                    \ref{eq:risk_function} at \texttt{photoz\_X}. \\
      \texttt{photoz\_std\_X}     &  Second order moment around a point estimate (\texttt{photoz\_X}) derived from full PDF.\\
      \texttt{photoz\_err68\_min} & 16 \% percentile in the PDF\\
      \texttt{photoz\_err68\_max} & 84 \% percentile in the PDF\\
      \texttt{photoz\_err95\_min} & 2.5 \% percentile in the PDF\\
      \texttt{photoz\_err95\_max} & 97.5 \% percentile in the PDF\\ \hline \hline
    \end{tabular*}
  \end{center}
  \caption{
    Common photo-$z$ parameters available for all the codes.\\
    \label{tab:photoz_common}
  }
\end{table*}

\begin{table*}
  \begin{center}
    \begin{tabular*}{\textwidth}{l @{\extracolsep{\fill}} llll}\hline\hline
      \texttt{CODE} &
                 target selection &
                 number of objects &
                 other quantities \\ \hline \hline
      \texttt{DEmP} &
                 \texttt{detect\_is\_primary} is \texttt{True} &
                 171,721,095  &
                 None    \\ \hline
      \texttt{EPHOR} &
                 objects with CModel fluxes  in all five bands &
                 197,227,501 &
                 None \\ \hline
      \texttt{EPHOR\_AB} &
                 objects with afterburner fluxes in all five bands&
                 221,617,662 &
                 None \\ \hline
      \texttt{FRANKEN-Z} &
                 \texttt{detect\_is\_primary} is \texttt{True} &
                 135,966,862 &
                 many\\
                     &
                 objects with afterburner fluxes in all five bands&
                     &
                     \\ \hline
      \texttt{Mizuki} &
                 \texttt{detect\_is\_primary} is \texttt{True} &
                 144,107,354 &
                 many \\
                       &
                 objects with CModel fluxes  in at least three bands&
                       \\ \hline
      \texttt{MLZ} &
                 \texttt{detect\_is\_primary} is \texttt{True} &
                 171,721,095 &
                 flux flag\\ \hline
      \texttt{NNPZ} &
                 \texttt{detect\_is\_primary} is \texttt{True} &
                 163,627,623 &
                 neighbor redshifts and weights \\
                     & 
                 objects with CModel fluxes in all five bands &
                     & 
                     \\ \hline \hline
    \end{tabular*}
  \end{center}
  \caption{
    Target selection applied by each code. The number of objects
    that satisfy the selection is shown.
    Details of other  quantities available in the catalog can be found
    in Section \ref{sec:products}.
    
    \label{tab:photoz_depend}
  }
\end{table*}

\section{Discussion and Summary}
\label{sec:summary}

We have presented the photo-$z$'s computed with several independent codes using
the data from HSC-SSP.  We have constructed the training sample by combining spec-$z$,
grism-$z$, and high accuracy photo-$z$ and applied a weight to each object to
reproduce the color-magnitude distribution of galaxies in the Wide layer.
The codes are trained, validated, and tested using this training sample.
We also use the COSMOS wide-depth stacks, in which the photometry is quasi-independent
from the training sample, in order to evaluate the seeing and depth dependence
of our photo-$z$ performance.

We have compared the performance between the codes in Section~\ref{sec:performance}.
There are trends common to all the codes such as, (1) our photo-$z$'s are most accurate
at $0.2\lesssim z \lesssim1.5$ where we can straddle the 4000\AA\ break with our filter set, and
(2) accuracy is nearly constant at $i\lesssim23$ and becomes worse at fainter magnitudes.
We use a few different algorithms in our machine-learning codes (i.e., neural network,
nearest-neighbor, self-organizing map), but all the machine-learning codes perform
better than the classical template fitting code (\texttt{Mizuki}).  Although this may not be
a firm, general conclusion because we have only one template-fitting code (and
it was trained against an old version of the training sample with problematic weights),
this may have implications for our future photo-$z$ strategy.

It is not a surprising result that machine-learning outperforms the classical
template-fitting.  There are multiple reasons for this.  One of them would
be that template-fitting codes suffer directly from systematic effects in
the photometry such as less accurate CModel photometry at bright magnitudes
(see \cite{aihara17}), while machine-learning codes make the empirical
mapping between the photometry and redshift including such systematic effects.  Machine-learning
codes are thus less prone to systematic effects.  

However, in order to train a machine-learning code, we need an unbiased training sample.
This is a fundamentally difficult problem because photometry always goes deeper than
spectroscopy (at least with the current detector technology) and there is no complete
spectroscopic sample down to faint enough (e.g., $i=25$) magnitudes.
There are on-going efforts to mitigate the problem and that will be useful for weak-lensing
science, in which only relatively bright galaxies are used.  However, in the UltraDeep
layers of HSC-SSP for instance, we reach deeper than $i=27$, where we have few spectroscopic
redshifts.

While further spectroscopic efforts are definitely needed, another way to mitigate
the problem would be to combine the template-fitting and machine-learning.  We can first
use the template-fitting technique with photometry in many filters.  If our understanding
of galaxy SEDs is reasonable, we can {\it assume} that these many-band photo-$z$'s are relatively
accurate even beyond the depth of the spectroscopic limit.  We can then train
machine-learning codes against these many-band photo-$z$'s using much fewer filters to compute photo-$z$'s over a wide area.

In fact, this is exactly what we did in our photo-$z$ training; we trained our 5-band
photo-$z$'s against the COSMOS many-band photo-$z$ catalog \citep{laigle16}.
However, there are problems in the current dataset.  First, the current optical data
in COSMOS used in the photo-$z$ calculation is not very deep, roughly 30-60~min integration,
and it is not quite deep enough to train our codes for the Wide survey with 20~min integration.
Fortunately, the UltraDeep COSMOS data from HSC-SSP is much deeper and that will solve this problem.  Another problem is
that COSMOS is currently the only wide enough field observed in many filters and high accuracy photo-$z$'s
are available.  As discussed in Section~\ref{ssec:dndz}, there are significant large-scale
structures even in COSMOS with multiple redshift peaks.  We have re-weighted the training sample
to reproduce the multi-color distribution of galaxies in the Wide layer and that largely
reduces the effects of large-scale structures in COSMOS.  But, it will still be very useful to
have multiple COSMOS-like fields to suppress any field-specific systematics.  UDS may be
the next COSMOS field given its deep optical to IR data over the wide area, although intensive
spectroscopic efforts are unfortunately missing in the field.
There are also very narrow spikes in the $N(z)$ distribution of COSMOS, which are likely
introduced by attractor solutions in the photo-$z$ code and are not accounted for by the re-weighting.
We need to run multiple template-fitting codes, not just one, to suppress such systematics.

We should also resort to clustering techniques to circumvent the problem.  There are on-going efforts on
clustering-based $N(z)$ estimations in HSC and we hope to report on that in our future
paper.  The technique does not suffer from any problems with photometry as it only
requires positional information.  A dense spectroscopic sample over the entire redshift
range is needed, but SDSS already offers it at least for tests of $N(z)$ reconstruction
in the Wide layer.  We could also apply the technique to validate the many-band photo-$z$'s
at very faint magnitudes, where no spectroscopic data is available, to check how reliable
many-band photo-$z$'s are beyond the reach of spectroscopic sensitivities.
It is an open question how to handle the evolution of the galaxy bias, but the clustering-based
redshift inference is certainly a promising way forward.

We have focused on redshifts in this paper, but there are other information we would need for science
such as stellar mass and star formation rates of galaxies.  A template-fitting code
delivers such information, but we could also train machine-learning codes to compute these
physical properties.  The training sample will again come from COSMOS-like fields and
we probably need to run multiple codes with templates from multiple stellar population
synthesis codes in order to have a sense for systematics in the physical properties.
That will also be our next step.

Aside from the problem of the training sample, there is another question of whether we should
'synthesize' photo-$z$ estimates from all the codes into one, master photo-$z$.
We probably should do so since the photo-$z$ synthesis hopefully reduces uncertainties
in each photo-$z$ estimates under the assumption that not all the codes make the same mistake.
It is also good for users to
have just one photo-$z$ for each object.  Our preliminary analysis performed in an earlier
photo-$z$ production run suggests that, when there is a code that performs significantly
better than the others, that code tends to dominate the master photo-$z$.  However,
in this release, most of the codes perform equally well and it is probably worth testing
the photo-$z$ synthesis again.  This is another future task of the HSC photo-$z$ group.

Finally, we remind the readers once again that the photo-$z$ products discussed in this paper are publicly available.
The photo-$z$ point estimates, confidence and risk parameters, as well as other ancillary
information are all stored in the database.  A full P(z) for each object is available in the fits
format and can be downloaded from the photo-$z$ page on the data release site.
Some of our codes suffered from sub-optimal weights used in the training and
also from over-training.  We hope to mitigate these issues and release
improved versions of our photo-$z$ products in a future incremental release
of HSC-SSP.

\section*{Acknowledgment}
The Hyper Suprime-Cam (HSC) collaboration includes the astronomical communities of Japan and Taiwan,
and Princeton University.  The HSC instrumentation and software were developed by the National
Astronomical Observatory of Japan (NAOJ), the Kavli Institute for the Physics and Mathematics of
the Universe (Kavli IPMU), the University of Tokyo, the High Energy Accelerator Research Organization (KEK),
the Academia Sinica Institute for Astronomy and Astrophysics in Taiwan (ASIAA), and Princeton University.
Funding was contributed by the FIRST program from Japanese Cabinet Office, the Ministry of Education,
Culture, Sports, Science and Technology (MEXT), the Japan Society for the Promotion of Science (JSPS),
Japan Science and Technology Agency  (JST),  the Toray Science  Foundation, NAOJ, Kavli IPMU, KEK, ASIAA,
and Princeton University.

This paper makes use of software developed for the Large Synoptic Survey Telescope. We thank the LSST
Project for making their code available as free software at http://dm.lsst.org.

The Pan-STARRS1 Surveys (PS1) have been made possible through contributions of the Institute for Astronomy,
the University of Hawaii, the Pan-STARRS Project Office, the Max-Planck Society and its participating institutes, the Max Planck Institute for Astronomy,
Heidelberg and the Max Planck Institute for Extraterrestrial Physics, Garching, The Johns Hopkins University, Durham University, the University of Edinburgh,
Queen's University Belfast, the Harvard-Smithsonian Center for Astrophysics, the Las Cumbres Observatory Global Telescope Network Incorporated,
the National Central University of Taiwan, the Space Telescope Science Institute, the National Aeronautics and Space Administration under Grant No.
NNX08AR22G issued through the Planetary Science Division of the NASA Science Mission Directorate, the National Science Foundation under Grant No. AST-1238877,
the University of Maryland, and Eotvos Lorand University (ELTE) and the Los Alamos National Laboratory.

This paper is based on data collected at the Subaru Telescope and retrieved from the HSC data archive system,
which is operated by Subaru Telescope and Astronomy Data Center at National Astronomical Observatory of Japan.
We thank the COSMOS team for making their private spectroscopic redshift catalog available for our calibrations.
MT acknowledges supported by JSPS KAKENHI Grant Number JP15K17617. AN is supported in part by MEXT KAKENHI
Grant Number 16H01096. JSS is supported by the National Science Foundation Graduate Research Fellowship under Grant No. 2016222625.
Any opinion, findings, and conclusions or recommendations expressed in this material are 
those of the authors(s) and do not necessarily reflect the views of the National Science Foundation.
HM was supported by the U.S. DOE under Contract DE-AC02-05CH11231, and by
the NSF under grants PHY-1316783 and PHY-1638509.  HM was also
supported by the JSPS Grant-in-Aid for Scientific Research (C)
(No.~17K05409), Scientific Research on Innovative Areas (No.~15H05887), and by WPI, MEXT, Japan.  
We thank Jun Nakano for discussions on the training of machine-learning codes.
We thank the anonymous referee for a very useful report, which helped improve the paper.

\appendix

\section{Previous Internal Photo-$z$ Releases}
\label{sec:previous}

As summarized in \citet{aihara17}, we have made 5 internal data releases.
For each release, the HSC photo-$z$ working group computed photo-$z$'s using several
independent codes and released the photo-$z$ products to the collaboration.  As these
internal photo-$z$ products are used in our science papers, we briefly summarize them here.

This paper is based on our photo-$z$ products in the S16A internal data release (i.e.,
latest release at the time of writing).  In the current release, we have used 6 codes.
But, we started with 4 codes (\texttt{DEmP}, \texttt{MLZ}, \texttt{Mizuki}, and \texttt{LePhare})
in the first data release (S14A0).
\texttt{FRANKEN-Z} was included in S15B and \texttt{EPHOR} in S16A.  For \texttt{MLZ}, random-forest
was used to compute photo-$z$'s until S15B and it changed to SOM in S16A.  In the early runs,
we used a template-fitting code, \texttt{LePhare}, but it was later replaced with \texttt{NNPZ}, which performs better.
There have been incremental updates in all the codes in each release, which helped steadily
improve our photo-$z$ performance over the years. But, the performance in the earlier runs
is not drastically different from that presented in this paper.  Thus,
the accuracy quoted in this paper can be used as a rough reference to our previous releases.
Once again, the photo-$z$'s for PDR1 are based on the S16A internal release.

Our calibration strategy in earlier releases were similar to the one presented in this paper,
but we almost exclusively relied on the many-band photo-$z$'s from COSMOS.  We cross-matched the HSC
objects with the COSMOS photo-$z$ catalog by position and split it into two: training+validation
and test.  Each photo-$z$ runner used the first sample to train and validate the code
and applied the trained code to the second sample to test the performance.  While this approach
worked well for faint objects, bright nearby objects were under-represented in COSMOS
and we discovered problems with low-$z$ objects in the Wide area.  This led to the combined
sample of bright spec-$z$ sample and faint photo-$z$ sample used in the training in this paper.
Also, the best point estimator and the risk parameter were first introduced in S16A and in this paper
and were not used in our previous releases.  Most papers based on our previous photo-$z$
products use $z_{median}$ and $C(z_{median})$ instead.


\section{Biases and scatter in the physical parameter estimates by \texttt{Mizuki}}
\label{sec:phys_mizuki}

\texttt{Mizuki} infers physical properties of galaxies such as stellar mass and
star formation rates(SFRs) self-consistently in addition to redshifts.
This section evaluates how accurate the physical parameter estimates are.
For this goal, we use data from the Newfirm Medium Band Survey (NMBS; \cite{whitaker11}).
We here focus on the AEGIS field and use the stellar mass and SFR estimates
by \citet{whitaker11} based on the NMBS and multi-wavelength data available in AEGIS.

\begin{figure*}
  \begin{center}
    \includegraphics[width=6cm]{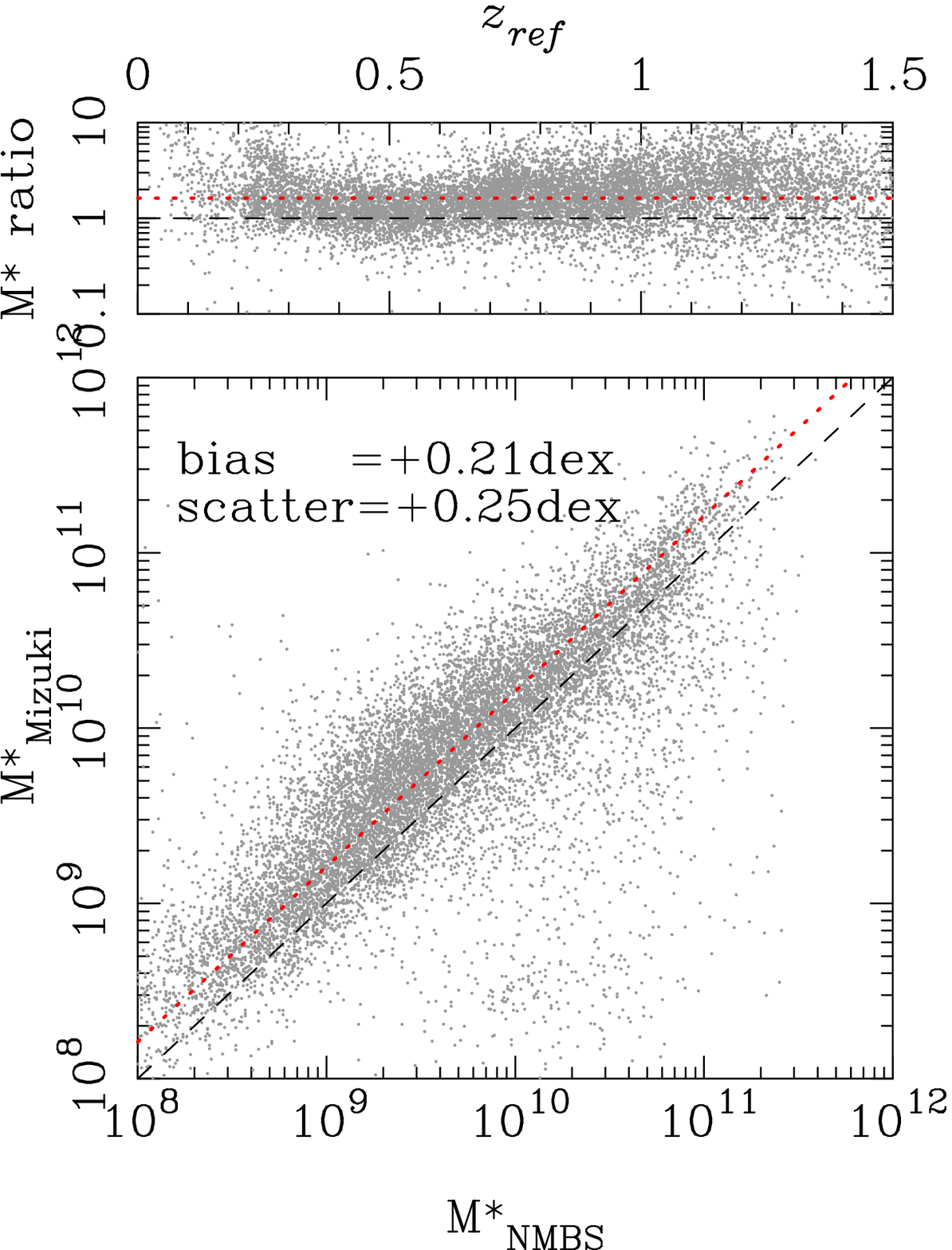}\hspace{0.5cm}
    \includegraphics[width=6cm]{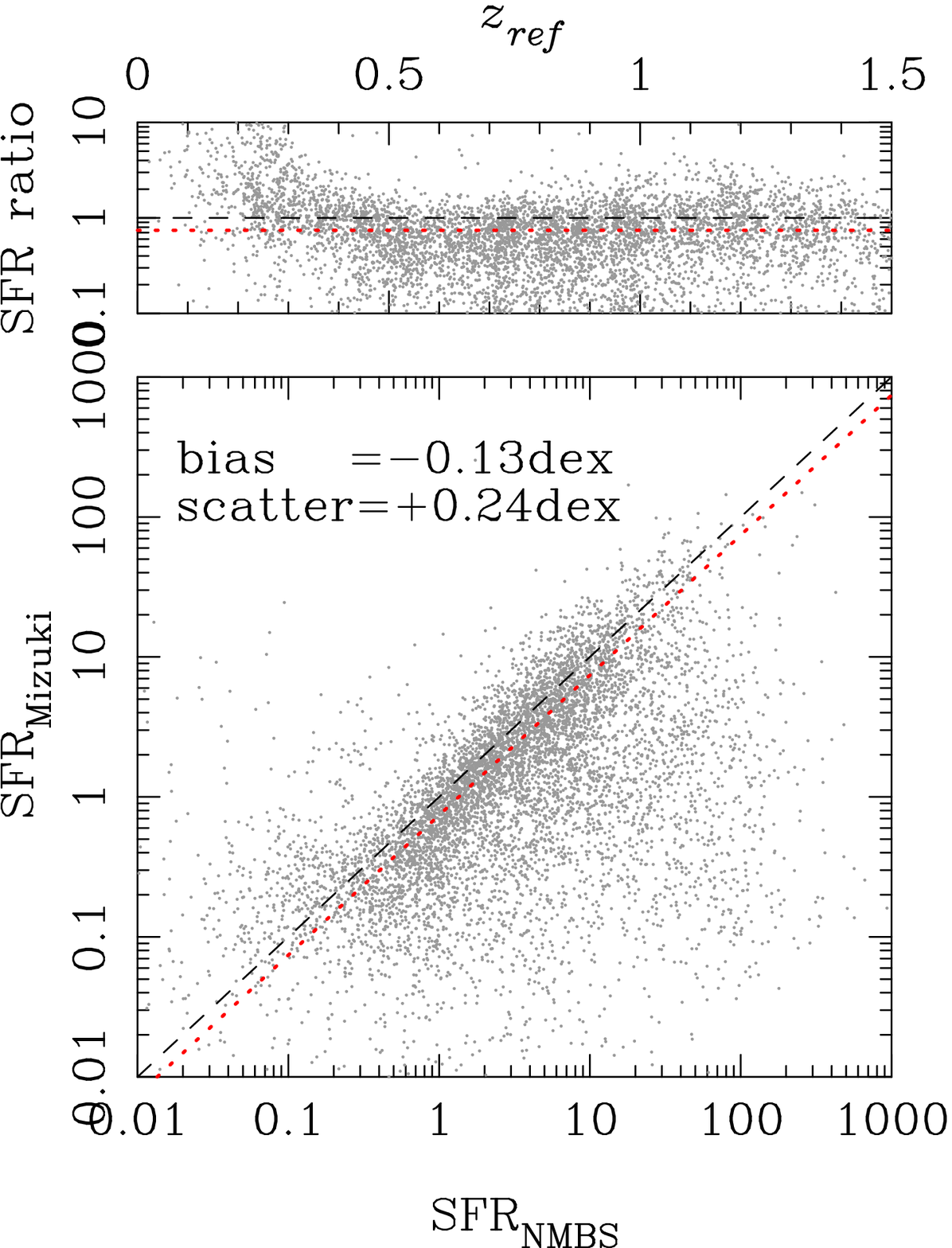}
  \end{center}
  \caption{
    Stellar mass (left) and SFR (right) from \texttt{Mizuki} plotted against
    those from NMBS.  The top panel in each plot shows the ratio between
    \texttt{Mizuki} and NMBS as a function of redshift.
    The dashed lines show the perfect correspondence and the dotted lines sow
    the mean bias.
 }
 \label{fig:phys_mizuki}
\end{figure*}

Fig~\ref{fig:phys_mizuki} compares stellar mass and SFR from \texttt{Mizuki} against NMBS.
As shown in the main body of the paper, our photo-$z$'s are not very accurate at $z\gtrsim1.5$,
where we lose the 4000\AA\ break, and we focus on galaxies at $z<1.5$ here.
Note that redshift is not fixed to those from NMBS but left as a free parameter.
Overall,
our stellar mass and SFR agree well with those from NMBS over the entire plotted range
with a scatter of about 0.25~dex, including photo-$z$ errors.
However, there is a systematic bias; stellar mass is over-estimated by 0.2 dex and SFR underestimated by 0.1 dex.
These biases in the physical properties are a function of redshift as shown in the top panels.
The biases are likely due to combination of template error functions and physical priors applied
\citep{tanaka15}. Work is in progress to reduce the systematic biases, but we note that
a level of 0.3~dex biases are relatively common in this field; \citet{vandokkum14}
found a relatively large stellar mass offset of $0.2\sim-0.3$~dex between 3D-HST and UltraVISTA
catalogs even though both catalogs have deep photometry in many filters.  Part of the bias we observe
here might come from systematics in the data (either in HSC or NMBS).

\end{document}